\documentclass[twocolumn,tighten]{aastex63}

\shorttitle{Black Hole Inner Shadows}
\shortauthors{Chael, Johnson, and Lupsasca}
\graphicspath{{./}{figures/}}

\usepackage{newtxmath,newtxtext,xspace}

\def\spin{{a^\ast}}
\def\m87{{M87$^{\ast}\xspace$}}
\def\sgra{{Sgr~A$^{\ast}\xspace$}}
\DeclareMathOperator\sign{sign}
\DeclareMathOperator\sn{sn}

\newcommand{\ab}[1]{\left|#1\right|}
\newcommand{\br}[1]{\left[#1\right]}
\newcommand{\cu}[1]{\left\{#1\right\}}
\newcommand{\pa}[1]{\left(#1\right)}
\newcommand{\ed}{\mathop{}\!\mathrm{d}}
\renewcommand{\deg}{\mathop{}\!\text{deg}}
\renewcommand{\O}[1]{\mathcal{O}\pa{#1}}

\defcitealias{PaperI}{Paper~I}
\defcitealias{PaperII}{Paper~II}
\defcitealias{PaperIII}{Paper~III}
\defcitealias{PaperIV}{Paper~IV}
\defcitealias{PaperV}{Paper~V}
\defcitealias{PaperVI}{Paper~VI}
\defcitealias{PaperVII}{Paper~VII}

\begin{document}

\title{Observing the Inner Shadow of a Black Hole:
A Direct View of the Event Horizon}

\correspondingauthor{Andrew Chael}
\email{achael@princeton.edu}

\author{Andrew Chael}
\altaffiliation{NASA Hubble Fellowship Program Einstein Fellow}
\affiliation{Princeton Center for Theoretical Science, Princeton University, Jadwin Hall, Princeton, NJ 08544, USA}

\author{Michael D.~Johnson}
\affiliation{Center for Astrophysics | Harvard \& Smithsonian, 60 Garden St, Cambridge, MA 02138, USA}

\author{Alexandru Lupsasca} 
\affiliation{Princeton Gravity Initiative, Princeton University, Princeton, NJ 08544, USA}

\begin{abstract}
Simulated images of a black hole surrounded by optically thin emission typically display two main features: a central brightness depression and a narrow, bright ``photon ring'' consisting of strongly lensed images superposed on top of the direct emission.
The photon ring closely tracks a theoretical curve on the image plane corresponding to light rays that asymptote to unstably bound photon orbits around the black hole. This critical curve has a size and shape that are purely governed by the Kerr geometry; in contrast, the size, shape, and depth of the observed brightness depression all depend on the details of the emission region.  
For instance, images of spherical accretion models display a distinctive dark region---the ``black hole shadow''---that completely fills the photon ring.
By contrast, in models of equatorial disks extending to the black hole's event horizon, the darkest region in the image is restricted to a much smaller area---an \emph{inner shadow}---whose edge lies near the direct lensed image of the equatorial horizon.
Using both semi-analytic models and general relativistic magnetohydrodynamic (GRMHD) simulations, we demonstrate that the photon ring and inner shadow may be simultaneously visible in submillimeter images of \m87, where magnetically arrested disk (MAD) simulations predict that the emission arises in a thin region near the equatorial plane.
We show that the relative size, shape, and centroid of the photon ring and inner shadow can be used to estimate the black hole mass and spin, breaking degeneracies in measurements of these quantities that rely on the photon ring alone. Both features may be accessible to direct observation via high-dynamic-range images with a next-generation Event Horizon Telescope.
\end{abstract}

\nocite{PaperI}
\nocite{PaperII}
\nocite{PaperIII}
\nocite{PaperIV}
\nocite{PaperV}
\nocite{PaperVI}
\nocite{PaperVII}
\nocite{PaperVIII}

\section{Introduction}
\label{sec:intro}

\begin{figure*}[ht]
\centering
\includegraphics[width=\textwidth]{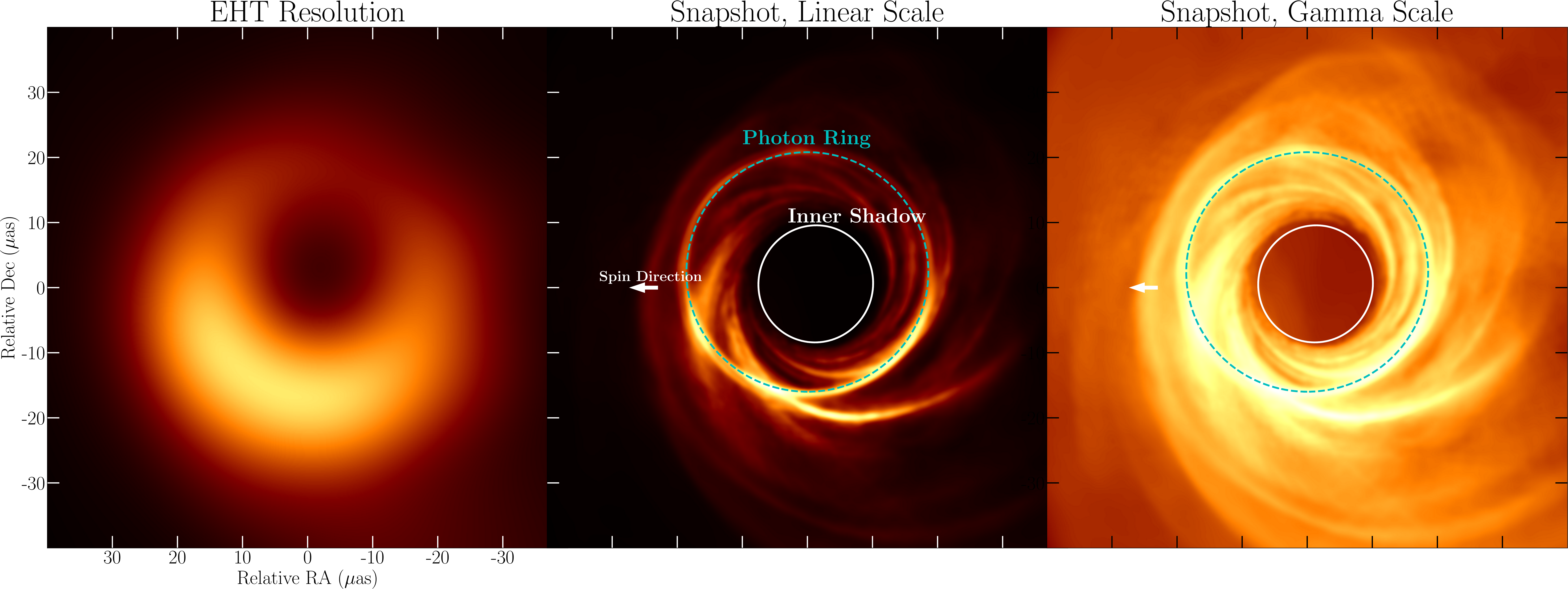}
\caption{
(Left) Snapshot image from a magnetically arrested radiative GRMHD simulation of \m87 \citep[Model R17;][]{Chael_19}, convolved with a circular Gaussian blurring kernel with a full width at half maximum (FWHM) of $15\,\mu$as.
The features of the simulated image at this resolution qualitatively match those seen in the first images of \m87 from the EHT \citep{PaperIV}.
(Middle) The simulation snapshot at native resolution. The simulation is viewed at an inclination $\theta_{\rm o}=163\deg$ \citep{Mertens2016,PaperV}; the black hole spin vector is oriented to the left and into the page.
The snapshot image shows filamentary, turbulent structures, a central brightness depression, and a narrow, bright photon ring that closely tracks the theoretical critical curve (cyan curve).
(Right) The same simulation snapshot in a gamma color scale that accentuates low-brightness features.
In this scale, the central brightness depression corresponds to the black hole's \emph{inner shadow}, or the direct lensed image of the equatorial event horizon (white curve).
The EHT images released thus far do not resolve the inner shadow of \m87, as they lack the requisite resolution and dynamic range.
These requirements may be met with a next-generation EHT.
\label{fig:Summary}
}
\end{figure*}

The Event Horizon Telescope (EHT) has recently produced the first resolved images of a black hole \citep{PaperI,PaperII,PaperIII,PaperIV,PaperV,PaperVI,PaperVII,PaperVIII}.
These 230~GHz images resolve the emission surrounding the supermassive black hole \m87 \citep[$M=6.5\pm 0.7\times10^9 M_\odot$;][]{PaperVI} at the center of the giant elliptical galaxy M87.
The EHT resolution of $\approx\!20\,\mu$as ($\approx\!5\,GM/Dc^2$ for \m87 at a distance $D\approx\!16.8$~Mpc) only just reveals the horizon-scale structure in \m87.
The EHT images display a ring with a diameter of $\approx\!40\,\mu$as with a North-South brightness asymmetry and a relatively dim interior.

In models where the accretion flow onto a Kerr black hole is spherically symmetric and the emission is optically thin, the central brightness depression in the observed image coincides precisely with those light rays that terminate on the event horizon when traced backwards from the observer's image plane into the black hole spacetime \citep{Falcke2000,Narayan_Shadow}.
This dark region---the ``black hole shadow''--- is bounded by a ``critical curve'' consisting of light rays that asymptote to unstably bound photon orbits around the black hole \citep{Bardeen1973a}.
Motivated by these models, the critical curve is sometimes also called the ``shadow edge.''
Approaching the shadow edge, the path length through the emission region diverges logarithmically as null geodesics wrap around the black hole multiple times \citep{Luminet1979a,Ohanian_1987,GHW,Johnson_Ring,GrallaLupsasca2020}.
Hence, in models featuring a spherically symmetric and optically thin emission region, the image brightness also diverges logarithmically at the critical curve, resulting in a bright ``photon ring'' encircling the black hole shadow.

By contrast, in models where the emission region is confined to an equatorial disk that extends down to the event horizon, the edge of the observed central brightness depression does not generically correspond to the critical curve \citep[e.g.,][]{Beckwith,Broderick06,GHW}.
Nevertheless, as long as the emission is optically thin, these models still feature a photon ring with logarithmically divergent brightness at the critical curve.
Contrary to the case of spherical accretion, however, the brightness increase is not continuous; rather, it is broken up into a sequence of strongly lensed images of the disk stacked on top of each other.
These images arise from rays with deflection angles $>180\deg$ that execute an increasing number of half-orbits around the black hole \citep{Luminet1979a,GHW,Johnson_Ring}.

In reality, the hot ($T>10^{10}$~K), collisionless plasma that produces the submillimeter emission in \m87 is expected to be turbulent, with a more complex structure than can be captured in either of these simple geometric pictures (\autoref{fig:Summary}).
The primary numerical tools for investigating the structure and dynamics of hot accretion flows are general relativistic magnetohydrodynamic (GRMHD) simulations \citep[e.g.,][]{Komissarov99,Gammie03}.
To constrain the properties of \m87, analyses of EHT images in both total intensity \citep{PaperV,PaperVI} and in polarization \citep{PaperVIII} made use of a library of these GRMHD simulations spanning a range of different parameters, including the black hole spin, accumulated magnetic flux on the black hole, and ion-to-electron temperature ratio.
Significantly, \citet{PaperVIII} found that, among the GRMHD simulation models in the EHT library, the currently favored models for \m87 all fall into the class of magnetically arrested disks \citep[MADs;][]{NarayanMAD,Igumenschchev2003}.  
In addition to producing images that are consistent with those observed by the EHT, MAD simulations naturally produce powerful jets \citep[e.g.,][]{Tchekhovskoy11,Chael_19} similar in both observed shape and total power to the prominent jet in \m87 \citep[e.g.,][]{Junor99,Stawarz06,Abramowski2012,Hada2016,Walker18,Mwl}.

Analyses of GRMHD simulation images have generally focused on the mathematical shadow edge, i.e., the critical curve \citep[e.g.,][]{Dexter2012,Psaltis_2015,Moscibrodzka_16,Bronzwaer_2021}.
Because this curve only depends on the black hole mass and spin vector, inferring its size and shape would provide information about the black hole's intrinsic parameters and enable tests of the validity of the Kerr metric \citep[e.g.,][]{Takahashi,JohannsenPsaltis,PaperVI}.
However, in performing these tests with limited-resolution observations, it is critical to account for the systematic uncertainty in relating observed image features such as the emission ring and central brightness depression to gravitational properties such as the size and shape of the critical curve \citep[e.g.,][]{PaperVI,Bronzwaer_2021}.
These systematic uncertainties may be dramatically reduced via future observations using an enhanced ground or space-based array capable of distinguishing lensed subrings within the photon ring \citep[e.g.,][]{Johnson_Ring,Astro2020Ground,Astro2020Space,Astro2020Space2,GLM_20,Broderick21}.

\begin{figure*}[t]
\raisebox{3ex}{\includegraphics[height=0.425\textwidth,trim=0 0 0 0]{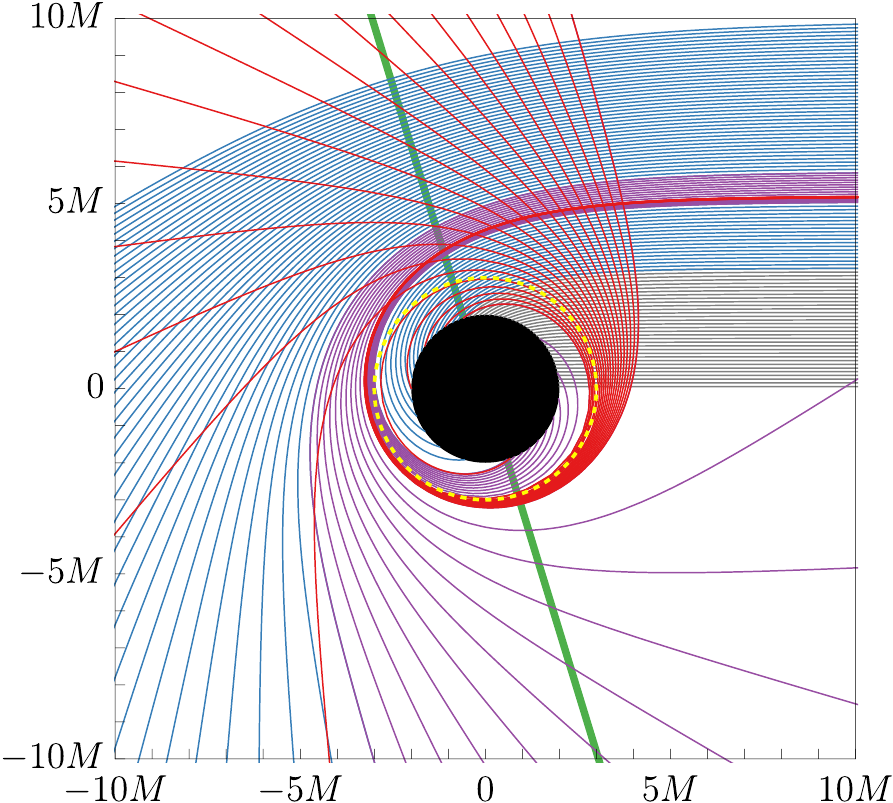}}\hfill 
\includegraphics[height=0.45\textwidth]{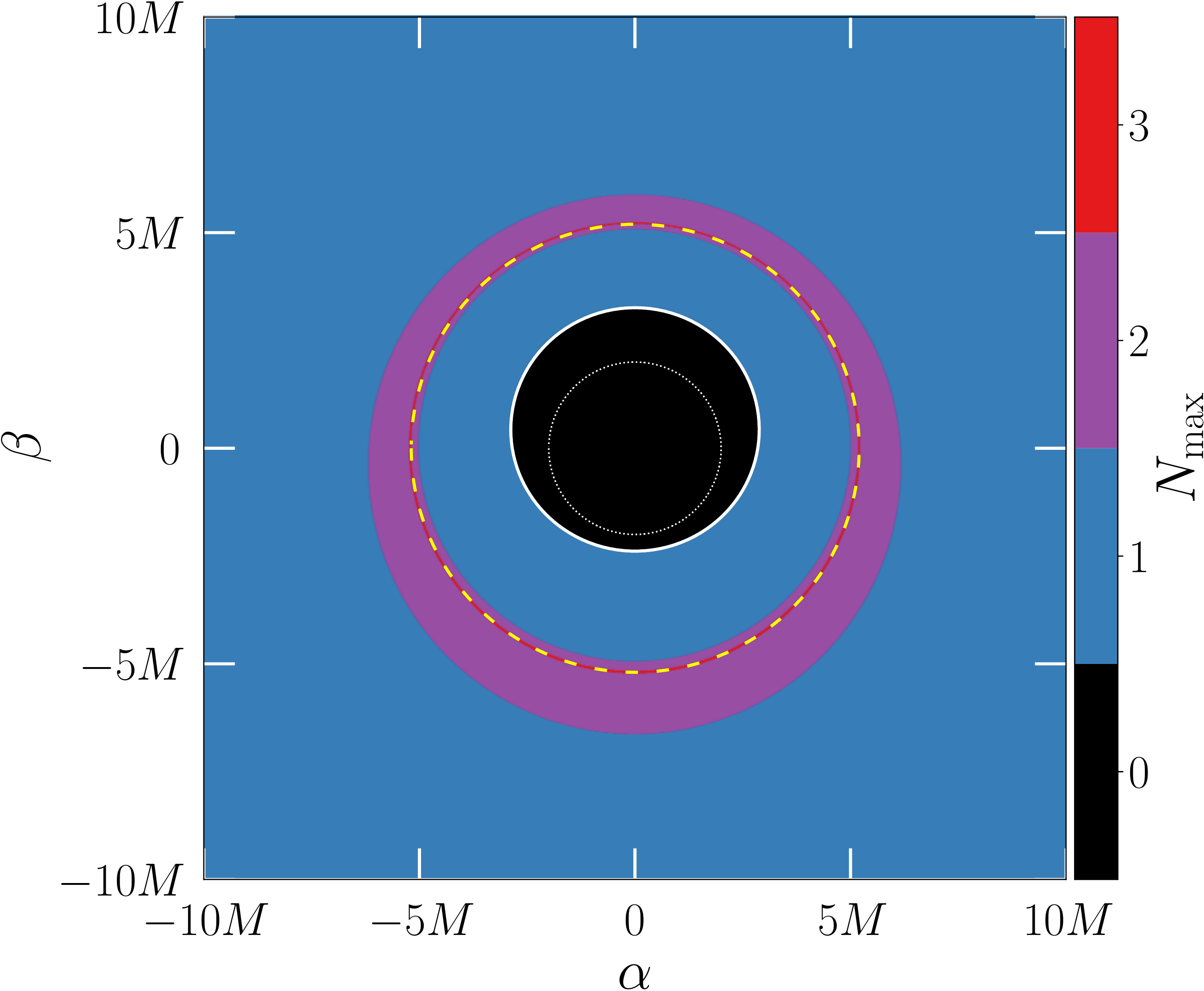}
\caption{
(Left) Photon trajectories around the black hole that reach a distant observer located to the far right \citep[see also][]{JohannsenPsaltis,GHW}.
The black hole is nonrotating ($\spin=0$), with an event horizon at $r_+=2M$ (black disk) and a photon sphere at $r_{\rm c}=3M$ (dashed yellow circle).
Photon trajectories are colored according to the number of times they cross the equatorial plane (green line), which is inclined at $\theta_{\rm o}=17\deg$ from the observer.
Most trajectories cross the equatorial plane once (blue), but photons that appear close to the critical curve on the image plane wrap around the black hole and cross the plane twice (purple), three times (red), or more, with photons appearing exactly on the critical curve describing trajectories that asymptote to unstably bound orbits ruling the photon sphere.
The inner shadow is defined by the trajectories that do not cross the equatorial plane ($N_{\rm max}=0$) before intersecting the event horizon (black disk).
(Right) The maximum number of equatorial crossings $N_{\rm max}$ for null geodesics as a function of the coordinates $(\alpha,\beta)$ in the image plane for a black hole of spin $\spin=0$ observed at an inclination angle of $\theta_{\rm o}=17\deg$.
Rings with an increasing number of equatorial crossings $N_{\rm max}$ become increasingly narrow and exponentially approach the critical curve (dashed yellow circle).
Inside the contour of the lensed equatorial horizon (solid white line), light rays do not cross the equatorial plane even once.
In models of equatorial emission extending to the horizon, these rays are therefore dark, resulting in an ``inner shadow'' feature.
The ``unlensed'' outline of the event horizon $r_+=2M$ is indicated with the dotted white line.
}
\label{fig:Rings}
\end{figure*}

In this paper, we show that MAD models of \m87 naturally exhibit a deep flux depression whose edge is contained well within the photon ring and critical curve.
This darkest region in a MAD simulation image corresponds to rays that terminate on the event horizon before crossing the equatorial plane even once (\autoref{fig:Rings}).
We refer to this feature as the ``inner shadow'' of the black hole.
This lensing feature was previously studied by \citet{DN19b,DN20,DN20b}.
As long as the emission is equatorial and extends all the way to the horizon, the darkest region in the observed image will correspond to the inner shadow, with a boundary defined by the direct, lensed image of the event horizon's intersection with the equatorial plane.
The MAD GRMHD models that we consider satisfy these criteria, with their submillimeter emission originating in the equatorial plane close to the event horizon \citep[as seen in][]{PaperV}.
Due to the effects of increasing gravitational redshift, the image brightness falls off rapidly near the edge of the inner shadow.
As a result, the correspondence between the lensed image of the equatorial horizon and the edge of the central brightness depression in an image is only apparent in faint image features viewed at high dynamic range.  

The inner shadow of a Kerr black hole has a significantly different dependence on its parameters than the critical curve \citep{Takahashi}.
For instance, the photon ring and critical curve of a Schwarzschild black hole are circular and independent of the viewing inclination, while the inner shadow is only circular when viewed face-on and has a size, shape, and centroid that are highly sensitive to the viewing inclination.
The photon ring and inner shadow provide complementary information.
When considered independently, each is subject to degeneracies in its size and shape as a function of black hole mass, spin, and viewing angle, yet these degeneracies can be broken via simultaneous observations of both features.

In simple toy models, spherical accretion flows produce a central brightness depression that completely fills the critical curve, but they do not give rise to an inner shadow \citep[e.g.,][]{Falcke2000,Narayan_Shadow}.
By contrast, thin-disk accretion models with emission extending to the horizon and a large optical depth present a precisely observable inner shadow, but they do not display any visible feature near the critical curve, since the lensed images that would produce a photon ring are blocked by the optically thick disk \citep[e.g.,][Figure 5.]{Beckwith}.
As a result, past work has generally analyzed these two features independently under the expectation that only one or the other will be relevant to the observed image \citep[see, e.g.,][]{Takahashi,DN20b}. 
Remarkably, we find that in both GRMHD simulations with strong magnetic fields and in semi-analytic, optically thin disk models with a radially dependent emissivity, the photon ring and the inner shadow are both prominent as potentially observable features (\autoref{fig:Summary}).
Thus, in the future, it may become possible to simultaneously measure both features in images of a black hole and thereby derive tighter, joint constraints on its parameters.

In this paper, we explore how the inner shadow may appear in images from realistic simulations and models of \m87, we assess the information contained in the relative size and shape of the inner shadow compared to the critical curve, and we discuss the prospects for direct observation of this feature using submillimeter very-long-baseline interferometry (VLBI).
In \autoref{sec:BlackHoleImages}, we review the basic properties of null geodesics and radiative transfer in the Kerr spacetime that give rise to the photon ring and inner shadow.
In \autoref{sec:Models}, we discuss the appearance of the inner shadow in images simulated from GRMHD and semi-analytic models.
In \autoref{sec:GeometricDescription}, we discuss geometric properties of both the critical curve and inner shadow, including their relative size, shape, and centroid positions, and we provide convenient analytic approximations for these quantities.
Throughout the paper, we only consider the inner shadow arising from equatorial emission near the event horizon of a Kerr black hole; in \autoref{sec:discussion} we discuss some of the factors---including jet emission, disk thickness and tilt, and alternative spacetime geometries---that could affect whether or how this feature appears in black hole images.
We summarize our conclusions in \autoref{sec:Conclusions}.

\section{Black hole images}
\label{sec:BlackHoleImages}

In this section, we review key features of the Kerr metric and the multiple lensed images of emission surrounding a black hole.
We argue that the curve marking the direct image of the equatorial event horizon should be visible as the edge of an ``inner shadow'' if the emission region is sufficiently equatorial and extends down to the event horizon.
From here on, we work in units normalized such that $G=c=1$.

\subsection{Kerr metric}

In Boyer-Lindquist coordinates $(t,r,\theta,\phi)$, the metric of a Kerr black hole of mass $M$ and angular momentum $J=aM$ ($0\leq a\leq M$) is
\begin{align}
	ds^2&=-\frac{\Delta}{\Sigma}\pa{\ed t-a\sin^2{\theta}\ed\phi}^2+\frac{\Sigma}{\Delta}\ed r^2\notag\\
	&\quad+\Sigma\ed\theta^2+\frac{\sin^2{\theta}}{\Sigma}\br{\pa{r^2+a^2}\ed\phi-a\ed t}^2, 
\end{align}
where
\begin{align}
	\Delta\equiv r^2-2Mr+a^2,\quad
	\Sigma\equiv r^2+a^2\cos^2{\theta}.
\end{align}
We frequently use the dimensionless spin $0\leq\spin\equiv a/M\leq1$.

The (outer) event horizon is located at radius
\begin{align}
	r_+=M+\sqrt{M^2-a^2}.
\end{align}
Unstable bound null geodesics, which neither escape to infinity nor intersect the event horizon, form a ``photon shell'' \citep{Bardeen1973a,Teo_2003,Johnson_Ring} outside of the outer event horizon.
Each bound orbit exists at a fixed Boyer-Lindquist radius $r_{\rm c}$ in the range $r_{\rm c,-}\leq r_{\rm c}\leq r_{\rm c,+}$, where  
\begin{align}
	r_{\rm c,\pm}=2M\br{1+\cos\pa{\frac{2}{3}\arccos\pa{\pm\spin}}}.
\end{align}
The bound orbits at $r=r_{\rm c,\pm}$ are confined to the equatorial plane ($\theta=\pi/2$).
At intermediate radii $r_{\rm c,-}<r_{\rm c}<r_{\rm c,+}$, the bound orbits oscillate between two fixed polar angles $\theta_\pm$ (see \autoref{eq:TurningPoints}).
In the case of a nonrotating Schwarzschild black hole ($a=0$), the photon shell reduces to a single ``photon sphere'' at $r_{\rm c}=3M$.

There exist timelike, equatorial geodesics forming stable prograde circular orbits around the black hole for all radii $r\geq r_{\rm ISCO}$, where $r_{\rm ISCO}$ denotes the radius of the  ``Innermost Stable Circular Orbit,''
\begin{align}
	r_{\rm ISCO}=M\br{3+Z_2-\sqrt{(3-Z_1)(3+Z_1+2Z_2)}},
\end{align}
with
\begin{subequations}
\begin{align}
	Z_1&=1+\pa{1-\spin^2}^{1/3}\br{\pa{1+\spin}^{1/3}+\pa{1-\spin}^{1/3}},\\
	Z_2&=\sqrt{3\spin^2+Z_1^2}.
\end{align}
\end{subequations}
For Schwarzschild, $r_{\rm ISCO}=6M$.

\subsection{Lensed images and the critical curve} 
\label{sec:LensedImages}

\begin{figure*}[ht]
\centering
\includegraphics[width=0.9\textwidth]{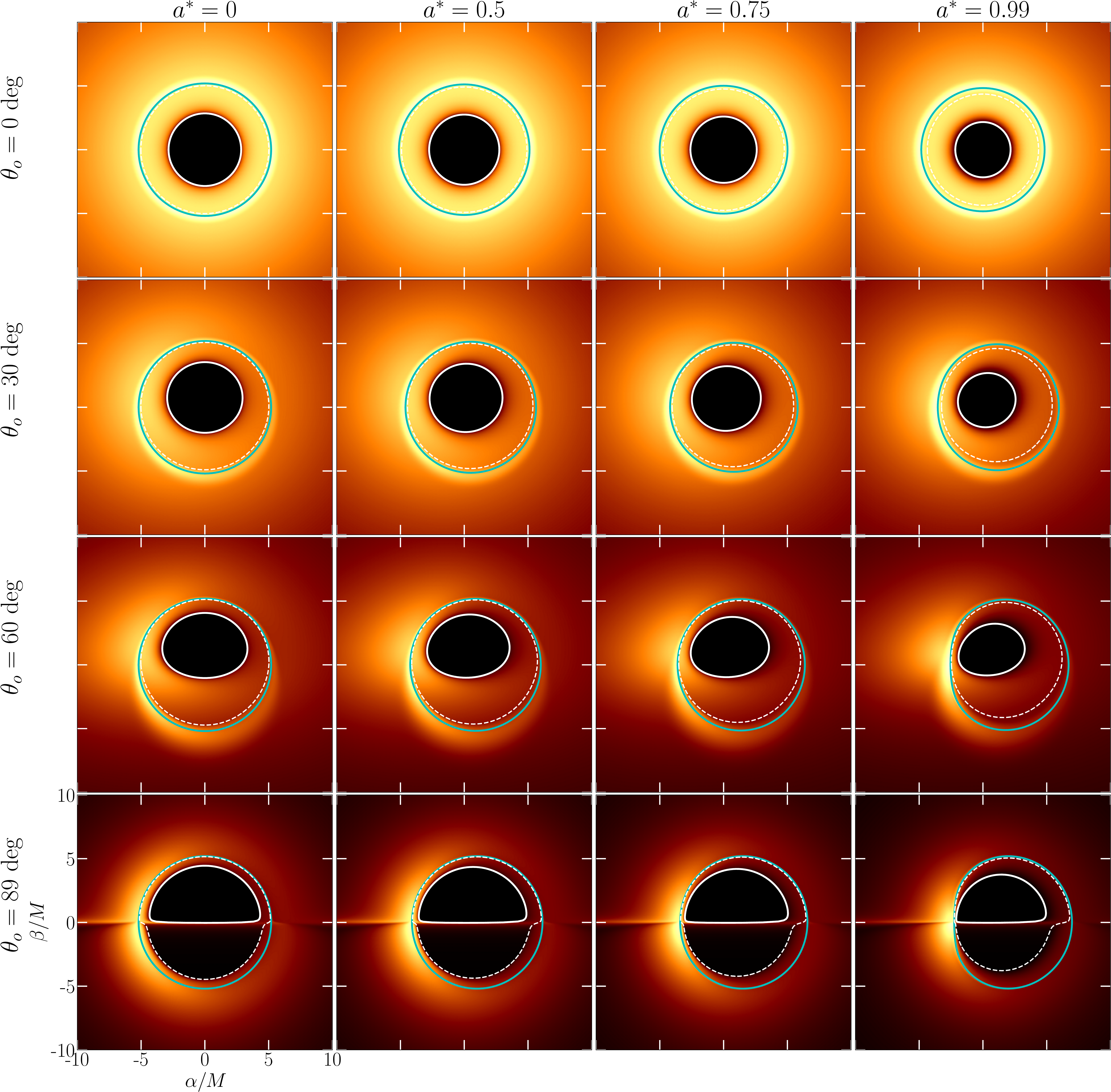}
\caption{Analytic models of emission from a Kerr black hole's equatorial plane covering a range of black hole spins and observer inclinations.
Columns from left to right display models with dimensionless spins $\spin=(0,0.5,0.75,0.99)$ and rows from top to bottom display inclinations $\theta_{\rm o}=(0,30,60,89)\deg$.
The intensity in each panel is normalized independently: all images are plotted in a gamma color scale with index $\gamma=1/4$.
For each model, we show the critical curve (cyan), the direct (primary, $n=0$) lensed image of the equatorial horizon (solid white) and the backside (secondary, $n=1$) lensed image of the equatorial horizon (dashed white).
\label{fig:imgrid}
}
\end{figure*}

We consider a distant observer ($r_{\rm o}\to\infty$) viewing the black hole at an inclination angle $0\leq\theta_{\rm o}<\pi$ with respect to its spin axis.
We parameterize the observer's image plane using ``Bardeen coordinates'' $(\alpha,\beta)$, given in units of $M$, defined such that the $\beta$ axis corresponds to the black hole spin axis projected onto the plane perpendicular to the ``line of sight.''

Each point in the image plane is associated with a null geodesic extending into the Kerr spacetime and labeled by two conserved quantities: the energy-rescaled angular momentum $\lambda$ and Carter constant $\eta$.
For a point $(\alpha,\beta)$ in the image plane, these constants are
\begin{subequations}
\label{eq:ConservedQuantities}
\begin{align}
	\lambda&=-\alpha\sin{\theta_{\rm o}},\\
	\eta&=\pa{\alpha^2-a^2}\cos^2{\theta_{\rm o}}+\beta^2.
\end{align}
\end{subequations}
The covariant four-momentum $k_\mu$ of the null geodesic at any point in the spacetime is given in terms of $\lambda$, $\eta$ and the photon energy-at-infinity $E$ as
\begin{subequations}
\label{eq:NullGeodesics}
\begin{align}
	k_t&=-E,\quad
	k_\phi=E\lambda,\\
	k_r&=\pm E\sqrt{\mathcal{R}}/\Delta,\\
	k_\theta&=\pm E\sqrt{\Theta},
\end{align}
\end{subequations}
where $\mathcal R(r)$ and $\Theta(\theta)$ are the radial and angular potentials
\begin{align}
	\label{eq:RadialPotential}
	\mathcal{R}(r)&=\pa{r^2+a^2-a\lambda}^2-\Delta\br{\eta+\pa{\lambda-a}^2},\\
	\Theta(\theta)&=\eta+a^2\cos^2{\theta}-\lambda^2\cot^2{\theta}.
\end{align}
By integrating the null geodesic equation \ref{eq:NullGeodesics}, we can solve for the trajectory $x^\mu(\tau)$ through the Kerr spacetime of a photon shot back from position $(\alpha,\beta)$ on the observer's image plane.

Such trajectories can be divided into three classes: those that eventually cross the event horizon (photon capture), those that are deflected by the black hole but return to infinity (photon escape), and those that asymptote to unstable bound orbits around the black hole.
The latter form a closed curve $(\alpha_{\rm c},\beta_{\rm c})$ in the image plane---the \emph{critical curve}---delineating the region of photon capture (the curve's interior) from that of photon escape (its exterior).
Critical photons have conserved quantities $(\lambda_{\rm c},\eta_{\rm c})$ equal to those of a photon on a bound orbit. For a given photon orbit radius $r_{\rm c,-}\leq r_{\rm c}\leq r_{\rm c,+}$, these are
\begin{subequations}
\begin{align}
	\lambda_{\rm c}&=a+\frac{r_{\rm c}}{a}\br{r_{\rm c} -\frac{2\Delta(r_{\rm c})}{r_{\rm c}-M}},\\
	\eta_{\rm c}&=\frac{r_{\rm c}^3}{a^2}\br{\frac{4M\Delta(r_{\rm c})}{(r_{\rm c}-M)^2}-r_{\rm c}}.
\end{align}
\end{subequations}
The critical curve in the image plane is obtained by inverting \autoref{eq:ConservedQuantities} to find $(\alpha,\beta)$ for all $r_{\rm c,-}\leq r_{\rm c}\leq r_{\rm c,+}$.
Each bound photon orbit at constant radius $r_{\rm c}$ maps to \emph{two} points in the image plane corresponding to the two signs $\pm\beta$ allowed for a given pair $(\lambda,\eta)$.
As a result, the critical curve is symmetric about the $\alpha$ axis perpendicular to the projected spin. 

The interior of the critical curve corresponds to geodesics that connect the observer to the event horizon and is often referred to as the ``black hole shadow.''
This name is motivated by the observation that, for a black hole that is immersed within an optically thin accretion flow with a spherically symmetric emissivity, light rays inside the critical curve (which terminate on the horizon) have a shorter path length along which to accumulate brightness than those in the exterior (which extend to infinity and can pick up more photons as they pass through the emission region); as a result, in such configurations, the critical curve's interior displays a brightness depression \citep{Falcke2000,Narayan_Shadow}.

Tracing back from the image plane, light rays that originate very near the critical curve approach the photon shell of bound orbits and execute many oscillations in $\theta$ between the turning points $\theta_\pm$ (\autoref{eq:TurningPoints}) before either terminating on the event horizon or escaping to infinity.
The number of oscillations (and the path length of the null geodesic) diverge logarithmically as the image-plane coordinate approaches a point on the critical curve.
If the black hole is surrounded by a uniform, optically thin emission region, this divergence in path length manifests as a logarithmic increase in the image brightness surrounding the critical curve: the ``photon ring'' \citep{Johnson_Ring,GrallaLupsasca2020}.

If instead the black hole has an optically thin emission region that does not fully surround it (e.g., one concentrated near the equatorial plane, in a tilted plane, or in a ``jet sheath'' region), then each oscillation in $\theta$ corresponds to an additional pass of the null geodesic through the emission region.
In this case, the photon ring is still present but exhibits additional substructure: its brightness profile increases in steps, forming exponentially narrow subrings that converge to the critical curve, with each ring assigned a label $n$ corresponding to the number $n-1$ of passes its light rays execute through the emission region \citep{Johnson_Ring}.\footnote{
Following \citet{Johnson_Ring}, we assign a number $n$ to each subring such that light rays appearing on that ring describe at least $n$ librations in $\theta$, or at least $n+1$ passes through the emission region.
Thus, $n=0$ refers to the ``direct image'' formed by rays passing through the emission region once.
}

\subsection{Equatorial images and the lensed horizon}
\label{sec:ComputingCurves}

\begin{figure*}[t]
\centering
\includegraphics[width=0.9\textwidth]{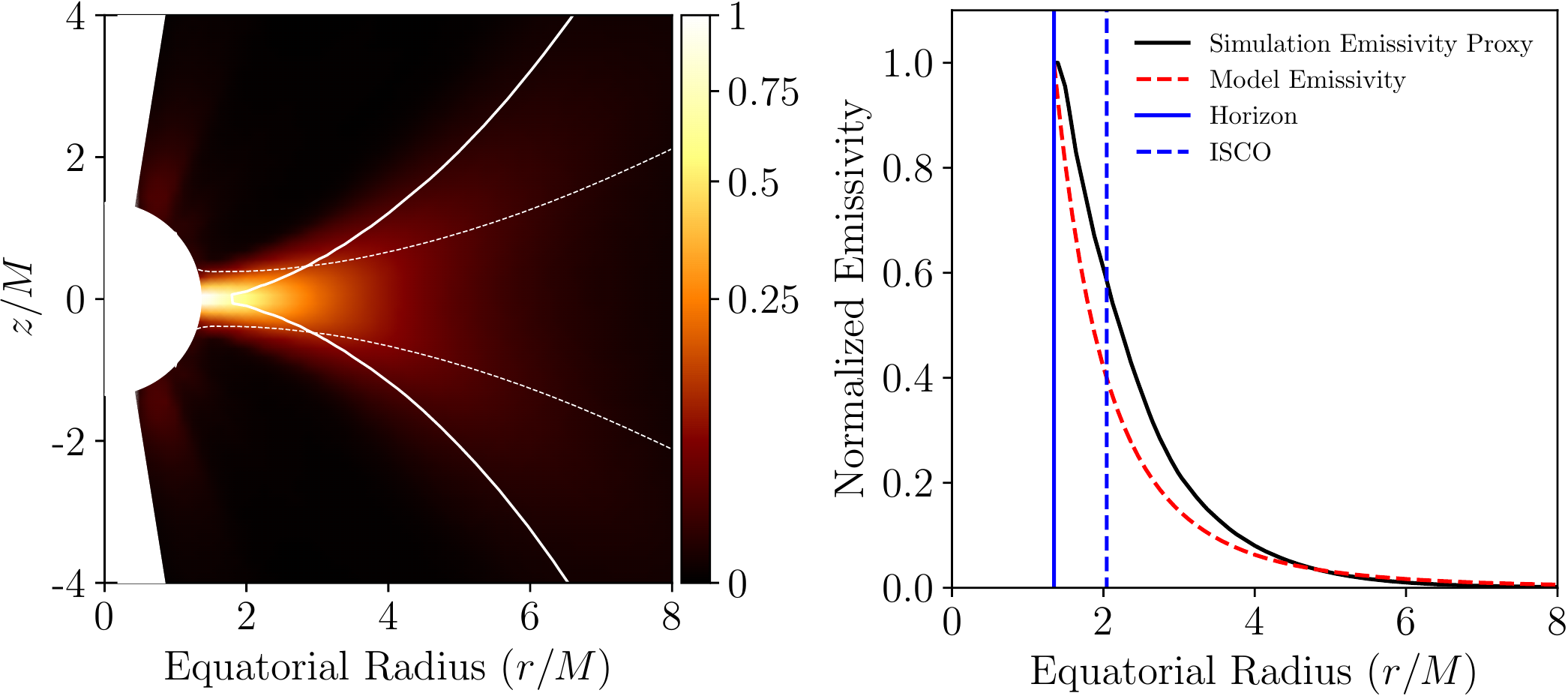}
\caption{
(Left) Map of 230~GHz synchrotron emissivity proxy (\autoref{eq:EmissivityProxy}) in the poloidal plane for time- and azimuth-averaged data from a MAD radiative GRMHD simulation of \m87.
(Right) Equatorial slice of the simulation emissivity proxy (solid), compared with the emissivity profile used in the analytic model (dashed), both normalized to unity at the horizon.
The simulation emissivity proxy is concentrated in the equatorial plane and does not truncate at the ISCO, but rather continues increasing with decreasing radius all the way to the event horizon.
}
\label{fig:EmissivityProfile}
\end{figure*}

We now focus on emission that is concentrated in the black hole's equatorial plane ($\theta=\pi/2$).
A geodesic ending at position ($\alpha,\beta$) in the image plane crosses the equatorial plane a maximum number of times $N_{\rm max}(\alpha,\beta)$ outside of the event horizon (an analytic procedure from \citet{GrallaLupsasca2020} for calculating the equatorial crossings is reviewed in \autoref{app:RayTracing}; in particular, see \autoref{eq:Nmax}).
In most of the image plane, $N_{\rm max}=1$; that is, geodesics cross the equator only once and project a direct (but still lensed) image of the equatorial emission on the observer sky.
In parts of the image plane that form increasingly narrow rings around the black hole, we instead have $N_{\rm max}=2,3,\ldots$
These concentric regions are the ``lensed subrings'' carrying contributions from geodesics that wrap around the black hole and cross its equatorial plane multiple times.
\autoref{fig:Rings} shows how $N_\mathrm{max}$ varies across the image plane for the case of a Schwarzschild black hole viewed at $\theta_{\rm o}=17\deg$.

For each $0\leq n<N_{\rm max}(\alpha,\beta)$, one can calculate the radius $r_{\rm eq}(\alpha,\beta;n)$ where the geodesic impinging on the observer's image plane at position $(\alpha,\beta)$ crosses the equatorial plane for the $(n+1)^\text{th}$ time.
This computation can be done either analytically (e.g., using the analytic method described in \cite{GrallaLupsasca2020} and reviewed in \autoref{app:RayTracing}) or numerically (e.g., using a GR ray tracing code like \texttt{grtrans} \citep{Dexter16} or \texttt{ipole} \citep{MoscibrodzkaIPole}). 

One can also invert $r_{\rm eq}(\alpha,\beta;n)$ to determine the successive lensed images of equatorial circles of constant source radius $r_{\rm s}=r_{\rm eq}$.
These contours are convex curves in the image plane and can be described in image-plane polar coordinates $(\rho,\varphi)$ as curves $\rho(\varphi;r_{\rm s},n)$ with $-\pi\leq\varphi<\pi$ defined by\footnote{
We take the polar angle $\varphi=0$ in the image plane to lie along the $+\alpha$ axis: $\alpha=\rho\cos{\varphi}$, $\beta=\rho\sin{\varphi}$.
Because the lensed images of equatorial rings are convex curves containing the origin, there is a unique $\rho$ satisfying this equation for each $\varphi\in[-\pi,\pi)$, so the curves $\rho(\varphi;r_{\rm s},n)$ are well-defined.
}
\begin{align}
    \label{eq:Contours}
	r_{\rm eq}\pa{\alpha=\rho\cos{\varphi},\beta=\rho\sin{\varphi};n=0}=r_{\rm s},
\end{align}
For any fixed radius $r_{\rm s}\ge r_+$, the curves $\rho(\varphi;r_{\rm s},n)$ approach the critical curve exponentially fast with increasing $n$.
For small observing angles $\theta_{\rm o}\approx0$, the $n=0$ image of an equatorial ring of constant radius $r_{\rm s}$ is lensed by approximately one gravitational radius; that is, $\rho\approx r_{\rm s}+M$ \citep{GrallaLupsasca2020,Gates2020}.

While most of the image plane has $N_{\rm max}\geq1$, it also has a small region with $N_{\rm max}=0$ wherein geodesics do not cross the equatorial plane even once, but instead pierce the event horizon before ever reaching $\theta=\pi/2$ (right panel of \autoref{fig:Rings}).
This $N_{\rm max}=0$ region corresponds exactly to the interior of the direct ($n=0$) lensed image of the equatorial event horizon, and is therefore bounded by the curve
\begin{align}
    \rho_{\rm h}(\varphi)=\rho(\varphi;r_+,0),
\end{align}
defined by \autoref{eq:Contours} with $r_{\rm s}=r_+$.
Like the critical curve, this curve divides the image plane into two qualitatively distinct regions.
Inside the critical curve, all geodesics terminate on the event horizon, while inside $\rho_{\rm h}(\varphi)$, all geodesics terminate on the horizon without crossing the equator (left panel of \autoref{fig:Rings}).
Thus, if the black hole is surrounded by an emission region that is predominantly equatorial and extends all the way down to the horizon, we should expect the interior of $\rho_{\rm h}(\varphi)$ to show up as a dark region in the image, thereby forming an ``inner shadow'' of low brightness. 

In \autoref{fig:imgrid}, we plot the critical curve $\rho_{\rm c}(\varphi)$ and the direct, lensed equatorial horizon image $\rho_{\rm h}(\varphi)$ for a range of black hole spins and observer inclinations, on top of an image generated from the analytic model described in \autoref{sec:EmissionModel} below.
In this model, the emission is purely equatorial and extends to the horizon; thus, the interior of $\rho_{\rm h}(\varphi)$---the black hole's ``inner shadow''--- is visible in each image as a deep brightness depression contained within the critical curve.

\section{Models for \texorpdfstring{\m87}{M87*}}
\label{sec:Models}

In this section, we investigate the appearance of the lensed equatorial horizon in images of synchrotron emission from a radiative GRMHD simulation of \m87.
We find that the lensed horizon image is visible in GRMHD simulation images of this magnetically arrested disk model for \m87 because its emission region is primarily equatorial.
We compare images from the simulation with images from an analytic model that assumes all emission originates in the equatorial plane. 

We scale all images of \m87 throughout this paper so that the angular gravitational size is \citep{PaperVI}
\begin{align}
	\frac{M}{D}=3.78\,\mu\text{as}.
\end{align} 
We also scale the total flux density at 230~GHz to 0.6~Jy \citep{PaperIII}.\footnote{
Note that the simulation images used here originally had an average flux density of $\approx\!1$~Jy based on observations of \m87 prior to 2017 \citep{Akiyama15}.
Here, we have scaled down the simulation's total flux density to match the updated value that better fits the 2017 EHT images.
}

\subsection{Radiative GRMHD simulation}

\begin{figure*}[t]
\centering
\includegraphics[height=0.475\textwidth]{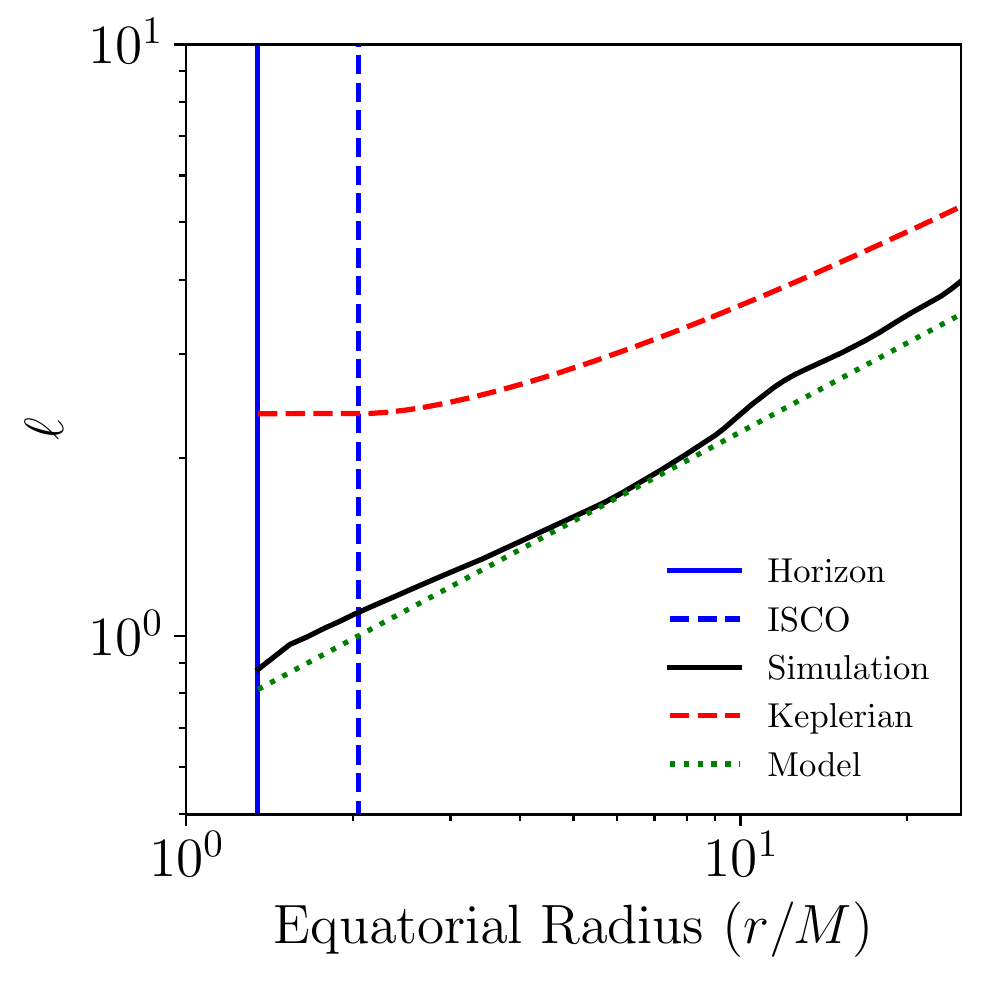}
\includegraphics[height=0.475\textwidth]{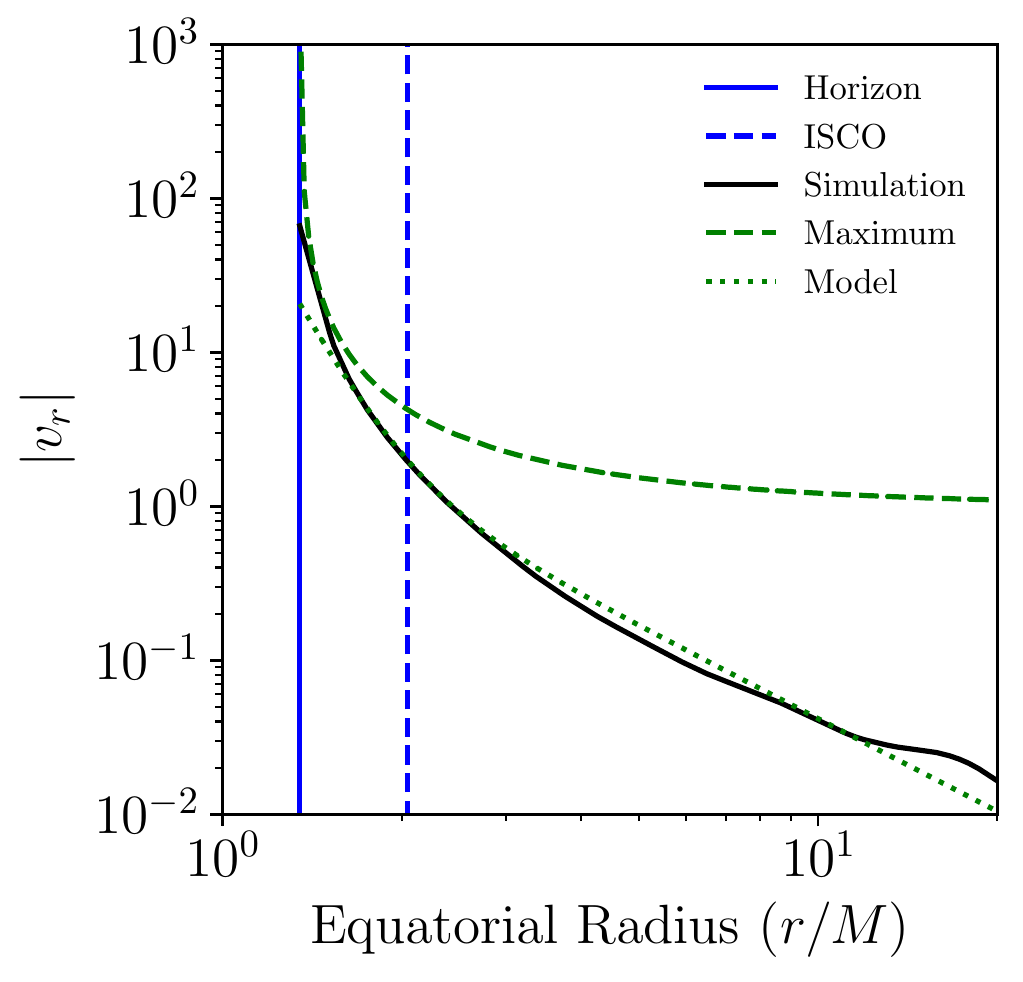}
\caption{
(Left) Equatorial value of the specific angular momentum $\ell\equiv u_\phi/u_t$ taken from the time- and azimuth-averaged simulation data (black curve).
At all radii, the averaged simulation angular momentum is well below the Keplerian value (red dashed curve).
We use a simple power-law fit to the GRMHD data in our analytic equatorial model (dotted green curve, \autoref{eq:Fit1}).
(Right) Equatorial value of the covariant infall velocity $v_r=u_r/u_t$ taken from the averaged simulation data (black curve).
We fit the average simulation data with a broken power law (dotted green curve, \autoref{eq:Fit2}).
This broken power-law fit is safely below the maximum infall velocity permitted for timelike geodesics by our power-law fit for $\ell$ (dashed green line).
}
\label{fig:SubKeplerian}
\end{figure*}

\begin{figure*}[t]
\centering
\includegraphics[width=\textwidth]{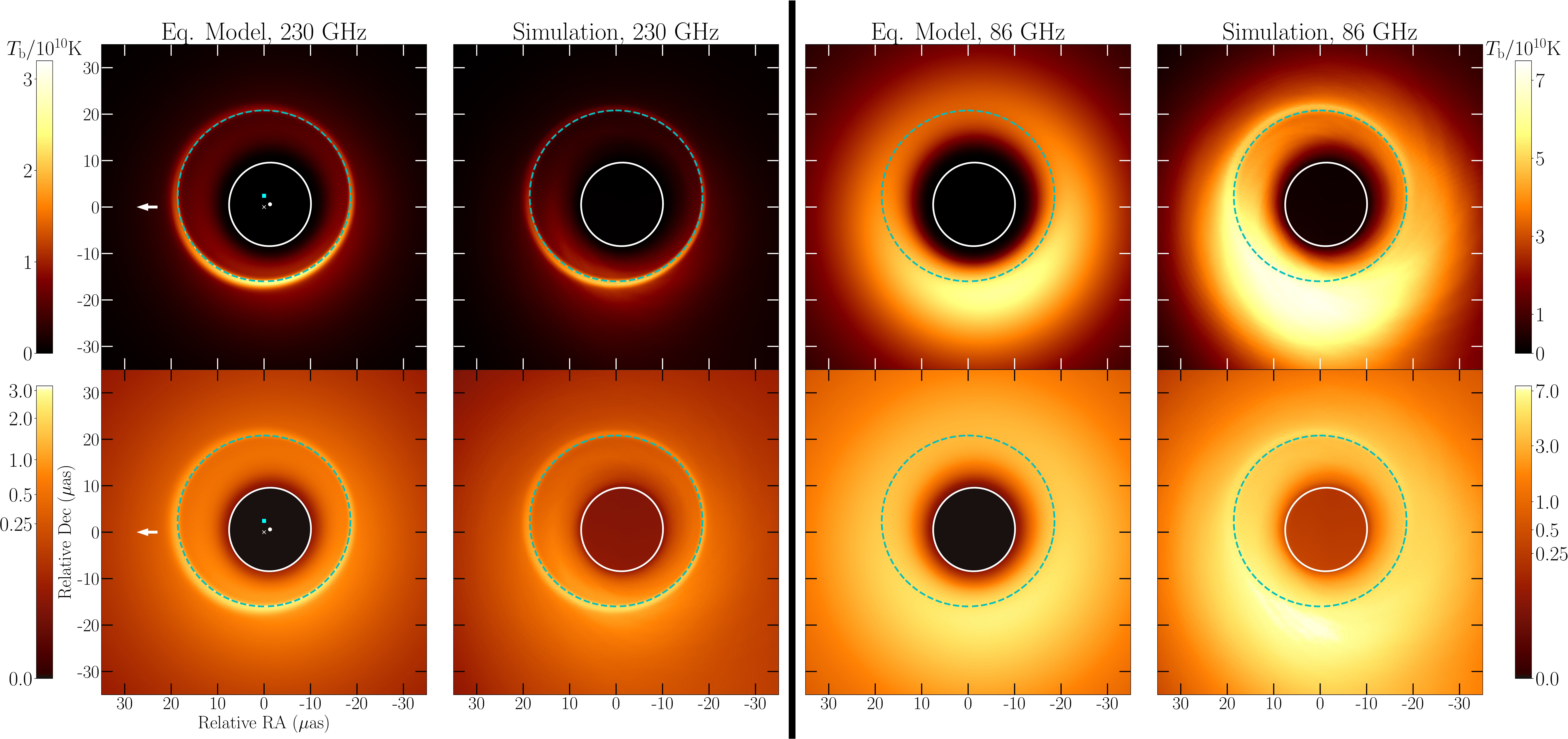}
\caption{
Top: (Left) 230~GHz image from an analytic, equatorial model for emission from \m87.
(Middle left) Time-averaged 230~GHz GRMHD image.
(Middle right) Equatorial model tuned to match the 86~GHz simulation image, with higher-order image subrings suppressed to mimic optical depth.
(Right) Time-averaged 86~GHz GRMHD image.
Bottom: The same images in a gamma color scale with index $\gamma=0.25$.
All images were generated with dimensionless black hole spin $\spin=0.9375$ and observer inclination $\theta_{\rm o}=163\deg$.
Both the 230~GHz and 86~GHz analytic model images had the parameters in their emissivity profile (\autoref{eq:EmissivityProfile}) separately fixed to best match the corresponding simulation images.
The black hole spin (positive $\beta$ axis) points to the left, as indicated by the arrow in the left column images.
Each panel displays the critical curve (cyan) and the direct image of the equatorial horizon (white line).
In the left column, we also indicate the centroid of the critical curve (cyan square marker), the centroid of the direct equatorial horizon image (white circular marker), and the origin ($\alpha=0$, $\beta=0$) of the Bardeen coordinate system (white cross).
}
\label{fig:ModelSimulationComparison}
\end{figure*}

\begin{figure*}[t]
\centering
\includegraphics[width=0.95\textwidth]{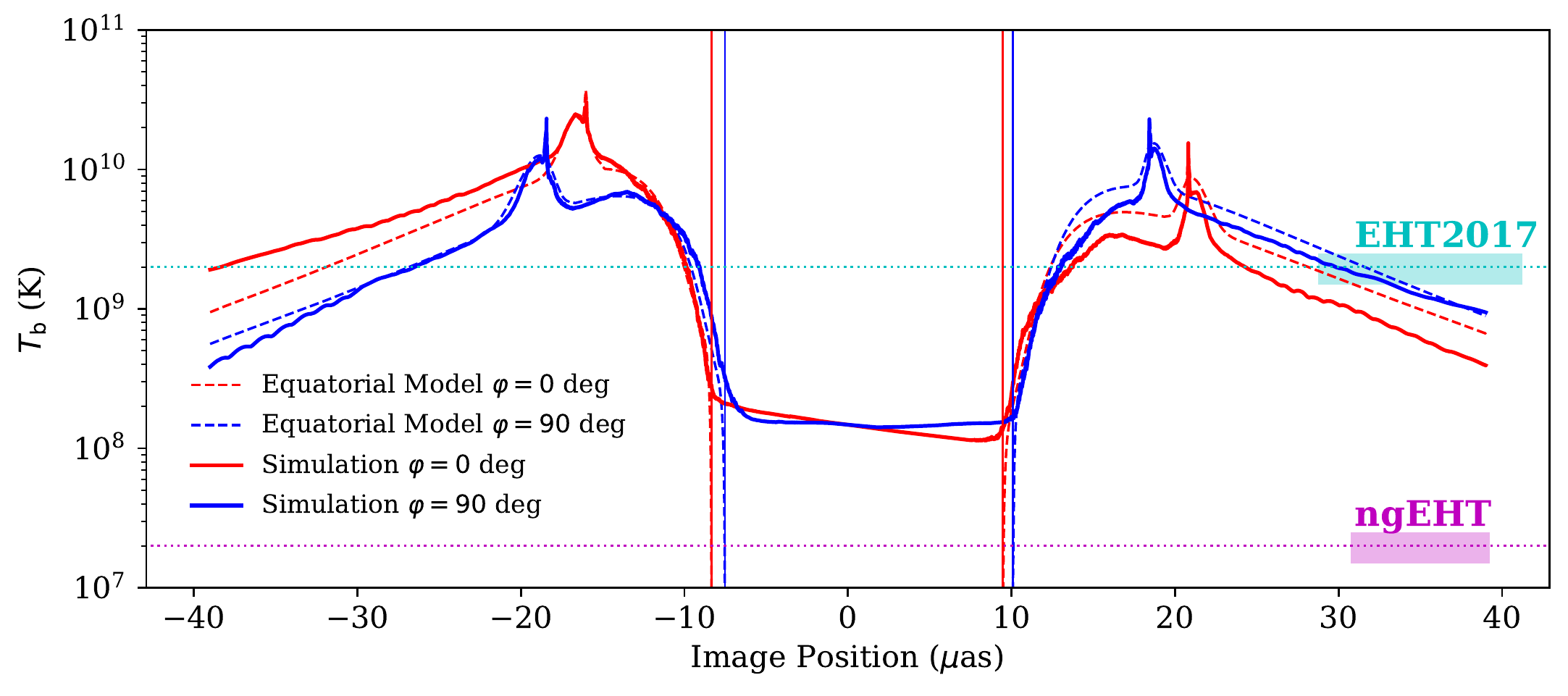}
\caption{
Slices along the $\beta=0$ (red) and $\alpha=0$ axes (blue) of the time-averaged 230~GHz images from the \m87 GRMHD simulation (solid curves) and the corresponding analytic equatorial disk model (dashed curves).
Solid vertical lines indicate the exact location of the direct image of the equatorial horizon on these slices.
An approximate dynamic range for the EHT2017 array is indicated by the cyan horizontal line, while the width of the cyan rectangle shows one-half the nominal resolution of the 2017 array.
Likewise, the dynamic range for an ngEHT concept array is indicated by the magenta line, and the magenta rectangle indicates one-half of the ngEHT concept's nominal resolution.
In the time-averaged simulation image, the brightness inside the lensed horizon contour levels off at a finite floor value produce by foreground jet emission.
In this simulation, the position of this dark depression inside the main emission ring---the ``inner shadow''---coincides with the image of the lensed equatorial horizon to within a microarcsecond.
In this model, the photon ring contains $\approx\!10\%$ of the total flux in the image.
}
\label{fig:Slices}
\end{figure*}

In this paper, we consider images of a radiative GRMHD simulation of \m87; specifically, we use 
simulation \texttt{R17} from \citet{Chael_19}.
This simulation was performed using the radiative GRMHD code \texttt{KORAL} \citep{KORAL13,KORAL14,KORAL16}.
Unlike most GRMHD codes, which evolve a single combined electron-ion fluid and must apply a model for the electron-to-ion temperature ratio in post-processing, the \texttt{KORAL} code directly evolves the temperature of the emitting electrons under radiative cooling and dissipative heating.
The primary cooling mechanism in the simulation is from synchrotron radiation at submillimeter wavelengths.
The electron heating fraction in the simulation is provided by the \citet{Rowan17} subgrid prescription fit to simulations of collisionless, transrelativistic magnetic reconnection.

The simulation used a black hole mass of $M=6.2\times10^9\,M_\odot$ and spin $\spin=0.9375$.
The initial magnetic field was set up so that the magnetic flux saturates on the black hole, putting the system in the magnetically arrested (MAD) accretion state \citep{Igumenschchev2003,NarayanMAD,Tchekhovskoy11}.
Polarimetric EHT observations of \m87 favor this accretion state over one with weaker, turbulent magnetic fields \citep{PaperVIII}.
The simulation produces a relativistic jet of power $\approx\!10^{43}$~erg~s$^{-1}$, satisfying measurements of the jet power from M87 \citep[e.g.,][]{Stawarz06}.
Furthermore, the jet opening angle in 43~GHz images from this simulation is large.
When observed at an inclination angle $\theta_{\rm o}=163\deg$ \citep{Mertens2016}, the apparent opening angle is $\approx50\deg$, similar to that observed in VLBI images of M87 \citep{Walker18}.
The extended jet in the simulation is in steady-state out to $\approx\!2500\,M\approx1$~pc, while the disk in the midplane is in steady-state out to $\approx\!40\,M$.

The GRMHD simulation is turbulent and time-variable.
To investigate the persistent features of the GRMHD fluid data, we computed profiles of the key plasma quantities (e.g., the density $\rho$, electron temperature $T_{\rm e}$, magnetic field $B^i$, and velocity $u^\mu$) in the poloidal $(r,\theta)$ plane after averaging in time and azimuth.
We also generated images of the 230~GHz and 86~GHz synchrotron emission from this simulation using the GR ray tracing and radiative transfer code \texttt{grtrans} \citep{Dexter16}.
The images were generated at an observer inclination angle of $\theta_{\rm o}=163\deg$ and rotated so that the black hole spin points to the East, opposite to the direction of the approaching jet \citep{PaperV}.
The snapshot images from this simulation exhibit small-scale structure from plasma turbulence and magnetic filaments (\autoref{fig:Summary}).
In this paper, we focus on time-averaged images generated from the collection of snapshot images of the simulation; both the time-averaged images and the time-averaged simulation data were produced from simulation snapshots spanning $5000\,M$ in time at a cadence of $10\,M$.

Radiatively inefficient simulations with weak magnetic flux form geometrically thick disks supported by the gas pressure.
By contrast, in the high-magnetic-flux MAD state, the magnetic pressure exceeds the gas pressure in the ``disk'' near the black hole.
In the time-averaged simulation data, the near-horizon material forms a thin, highly magnetized structure in the equatorial plane; this thin structure is the source of the observed 230~GHz emission.
Note that the thickness of the equatorial ``disk'' in these simulations is limited by the resolution; in very-high-resolution simulations, the emission region is even thinner, and it occasionally collapses to a current sheet that may source very high energy emission \citep[e.g.,][]{Ripperda20}.

In \autoref{fig:EmissivityProfile}, we investigate the 230~GHz emissivity from the time- and azimuth-averaged \texttt{R17} simulation data.
The true rest frame emissivity used in the radiative transfer (e.g., that given in Appendix A1 of \citealt{Dexter16}) depends on the combined special-relativistic and gravitational redshift of the geodesic at the source, as well as the orientation of the magnetic field with respect to the wavevector in the fluid rest frame.
As a result, it is nontrivial to directly extract an emissivity profile from the time-averaged simulation data that would correspond meaningfully to the time-averaged images generated by \texttt{grtrans}.
Here, we use the proxy for the 230~GHz  emissivity defined in the EHT GRMHD code comparison project \citep{CodeComparison}.
This function follows the characteristic behavior of the true synchrotron emissivity \citep[e.g.,][]{Leung11} with the density $\rho$, electron pressure $p_{\rm e}$, and magnetic field strength $|B|$.
The emissivity proxy is
\begin{align}
	\label{eq:EmissivityProxy}
	j_{\rm sim}=\frac{\rho^3}{p_{\rm e}^2}\exp\br{-C\pa{\frac{\rho^2}{|B|p_{\rm_e}^2}}^{1/3}}.
\end{align}
We follow \citet{CodeComparison} in setting the free constant $C=0.2$ so that the 230~GHz emission is contained within a characteristic radius $r_{\rm em}\leq5\,M$.

From the left panel of \autoref{fig:EmissivityProfile}, it is apparent that near the black hole, the emissivity proxy predicts that emission from the \citet{Chael_19} simulation is predominately located in the equatorial plane.
In the right panel, we extract the simulation emissivity in the equatorial plane $(\theta=\pi/2)$ and compare with the emissivity function we use in the analytic model described in the next section, \autoref{eq:EmissivityProfile}.
The simulation emissivity satisfies two criteria necessary for the lensed equatorial event horizon, or black hole inner shadow, to be visible as an image feature at 230~GHz:
\begin{enumerate}
	\item The simulation emissivity is predominately concentrated in the equatorial plane (\autoref{fig:EmissivityProfile}, left panel).
	\item The simulation emissivity extends to the event horizon, and is not truncated at any earlier radius such as the innermost stable circular orbit (\autoref{fig:EmissivityProfile}, right panel). 
\end{enumerate}

We also investigate the time-averaged simulation velocity profile in the equatorial plane. \autoref{fig:SubKeplerian} shows profiles of the specific angular momentum $\ell\equiv u_\phi/u_t$ and covariant infall velocity $v_r=u_r/u_t$ computed from the average simulation data.
Notably, the average angular momentum in the equatorial plane is significantly below the Keplerian value at all radii \citep[as also seen in, e.g.,][]{NarayanMAD}.
These sub-Keplerian velocities significantly reduce the total Doppler+gravitational redshift factor for emission close to the event horizon.
As a result of this reduced redshift factor, the brightness of the emission falls off less severely near the lensed horizon curve than it would in a Keplerian model with infall inside the ISCO \citep[e.g,][]{Cunningham_75}.

\subsection{Equatorial emission model}
\label{sec:EmissionModel}

Because the time-averaged emissivity of the GRMHD simulation is predominantly equatorial (\autoref{fig:EmissivityProfile}), it is reasonable to compare time-averaged images from this simulation with those from a simple model with the emission confined to the equatorial plane.
\citet{GLM_20} introduced a convenient, analytic model for computing images of equatorial emission around a black hole.
These images are specified by the black hole spin $\spin$ and observer inclination $\theta_{\rm o}$ (which determine the lensed subring structure), 
the four-velocity of the emitting material in the equatorial plane $u^\mu(r)$ (which determines the redshift of the emission), and
the rest-frame emissivity in the equatorial plane $j_{\rm model}(r)$.
The emissivity and velocity are assumed to be constant in azimuth.

In this model, the observed intensity at a point $(\alpha,\beta)$ on the image plane is
\begin{align}
	\label{eq:EquatorialModel}
	I(\alpha,\beta)=\sum_{n=0}^{N_{\rm max}-1}f_n\,j_{\rm model}(r_n)\,g^3(r_n,\alpha,\beta),
\end{align}
where $r_n=r_\mathrm{eq}(\alpha,\beta;n)$ is the radius at which the geodesic crosses the equatorial plane for the $(n+1)^\text{th}$ time (see \autoref{sec:ComputingCurves}), $N_{\rm max}=N_{\rm max}(\alpha,\beta)$ is the maximum number of equatorial crossings, $j_{\rm model}(r_n)$ is the equatorial emissivity at $r_n$, and $g$ is the redshift factor computed from the emitted photon wavevector $k_\mu$ and the four-velocity $u^\mu$ of the emitting material at radius $r_n$.
The factor $f_n$ is a ``fudge'' that can enhance or diminish the brightness of higher-order rings: we set $f_0=1$ and $f_n=2/3$ for $n>0$ to best match the time-averaged images from the radiative GRMHD simulation. 

Note that while \autoref{eq:EquatorialModel} is of the same general form introduced in \citet{GLM_20}, we use a factor of $g^3$ to represent the redshift of the specific 230~GHz intensity (assuming a flat emssion spectrum in \autoref{eq:EmissivityProfile}) rather than the $g^4$ redshift factor they use for bolometric intensity.
We also use a ``fudge'' factor $f_n<1$ for $n>0$, while \citet{GLM_20} uses $f_n=1.5$, as we find it necessary to slightly suppress the contributions from higher-order subrings to match our model images to the time-averaged simulation images used here.

For the emissivity $j_{\rm model}(r)$, we use a second-order polynomial in log-space; that is,
\begin{align}
	\label{eq:EmissivityProfile}
	\log\br{j_{\rm model}(r)}=p_1\log[r/r_+]+p_2\pa{\log[r/r_+]}^2.
\end{align}
For the 230~GHz images shown throughout this paper, we set $p_1=-2$, and $p_2=-1/2$.\footnote{
Note that for the analytic model to match the change in the image structure with frequency observed in the GRMHD simulation, the emissivity profile parameters must change with the observation frequency; in the 86~GHz images in \autoref{fig:ModelSimulationComparison}, we set $p_1=0,p_2=-3/4$.
}
The overall scale of the emissivity in \autoref{eq:EmissivityProfile} is arbitrary; in computing images of \m87, we normalize the emission so that the 230~GHz flux density is 0.6~Jy \citep{PaperIII}. The right panel of \autoref{fig:EmissivityProfile} compares the parametrization from \autoref{eq:EmissivityProfile} to the equatorial emissivity profile from the time-averaged GRMHD simulation (\autoref{eq:EmissivityProxy}).

The redshift factor is given by
\begin{align}
	g=\frac{-k_t}{k_\mu u^\mu}
	=\frac{1}{u^t-\lambda u^\phi\pm u^r\sqrt{\mathcal{R}}/\Delta},
\end{align}
where we assume $u^\theta=0$.
The sign of the $\pm\sqrt{\mathcal{R}}$ term is equal to the sign of the radial component of the null wavevector, $k^r$.
The factor of $g^3$ in \autoref{eq:EquatorialModel} suppresses the $n=0$ emission rapidly with decreasing radius toward the lensed horizon image.
Different models for the velocity $u^\mu$ will feature different rates of suppression, with different implications for how close the observable brightness depression on the sky is to the analytic solution for the inner shadow edge $\rho_{\rm h}(\varphi)$.

While \citet{GLM_20} follow \citet{Cunningham_75} and define the velocity $u^\mu$ to be on Keplerian circular orbits for $r>r_{\rm ISCO}$ and infalling for $r<r_{\rm ISCO}$, we instead model $u^\mu$ with sub-Keplerian angular velocities so as to mimic the characteristic behavior of magnetically arrested disks in GRMHD simulations.
In particular, we use a simple power-law fitting function to the covariant angular momentum $\ell\equiv u_\phi/u_t$, and a broken power-law fitting function to the infall velocity $v_r\equiv u_r/u_t$ derived from the GRMHD simulation:
\begin{align}
	\label{eq:Fit1}
	\ell&=\ell_{\rm ISCO}\pa{\frac{r}{r_{\rm ISCO}}}^{1/2},\\
	\label{eq:Fit2}
	v_r&=-V_{\rm ISCO}\pa{\frac{r}{r_{\rm ISCO}}}^{q_1}\br{\frac{1}{2}+\frac{1}{2}\pa{\frac{r}{r_{\rm ISCO}}}^{1/\delta}}^{\delta(q_2-q_1)}.
\end{align}
We set $\ell_{\rm ISCO}=1$, $V_{\rm ISCO}=2$, $q_1=-6$, $q_2=-2$, and $\delta=1/5$.
\autoref{fig:SubKeplerian} compares these fitting functions to the values obtained from the time-averaged GRMHD data. 

\subsection{230~GHz and 86~GHz images}

\autoref{fig:ModelSimulationComparison} compares time-averaged images from the \m87 simulation at 230 and 86~GHz with images generated using the modified analytic model described in \autoref{sec:EmissionModel}.
We see that the direct lensed image of the equatorial event horizon is apparent as a deep central brightness depression---the inner shadow---in both the time-averaged images from the simulation and the equatorial model.
The brightness of the emission surrounding the horizon image is suppressed by the gravitational redshift; nonetheless, when looking at the image in a gamma color scale\footnote{
In the gamma scale, $I^\gamma$ is plotted in the linear color scale instead of $I$, where $I$ is the image brightness and we set $\gamma=1/4$.
}
(bottom row of \autoref{fig:ModelSimulationComparison}), the apparent edge of the central brightness depression in the simulation and model image approaches the exact location of the lensed equatorial horizon contour within a microarcsecond.  

\begin{figure*}[t]
\centering
\includegraphics[width=\textwidth]{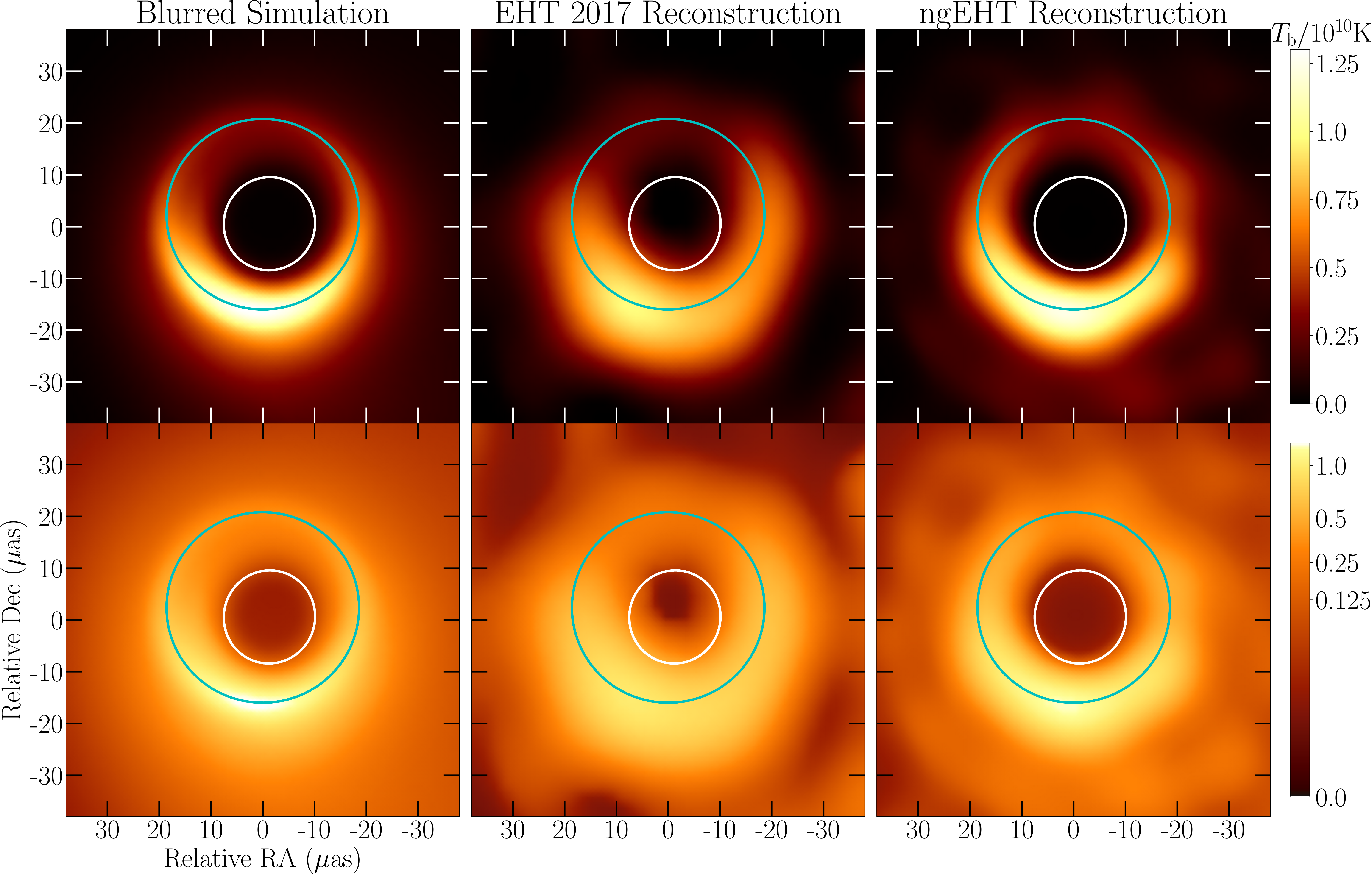}
\caption{
(Left) Time-averaged GRMHD images blurred to an approximate ngEHT imaging resolution of $10\,\mu$as.
(Middle) Reconstruction of the simulation model from synthetic data generated on EHT2017 baselines.
(Right) Reconstruction of the simulation model from synthetic data generated from an example ngEHT array.
The top row shows images in a linear color scale and the bottom row shows the same images in gamma scale.
In all images, the white curve corresponds to the lensed equatorial horizon, while the cyan contour is the critical curve.
}
\label{fig:Reconstructions}
\end{figure*}

At 86~GHz, the increasing optical depth of the accretion flow washes out the images of the higher-order $(n=1,2,\ldots)$ subrings in the simulation image, except for part of the $n=1$ ring on the north half of the image.
We mimic this effect in the image from the analytic equatorial model by suppressing the higher-order rings and only showing the direct, $n=0$ emission.
Despite the optical depth suppressing the appearance of the lensed subrings, the central ``inner shadow'' depression is still visible at 86~GHz.
This is because in the simulation, the emitting material contributing to the increased total optical depth is still contained within the equatorial plane; the optical depth through the jet material in front of the event horizon remains low.
As a result, the direct geometrical effect of the equatorial emission being truncated at the event horizon is still visible at this frequency.
At lower frequencies ($<\!40$~GHz in this simulation), the jet material becomes optically thick and obscures both the equatorial emission and the inner shadow.
The transition between the optically thick opaque jet and optically thin transparent jet regimes occurs at the frequency above which the image ``core'' no longer moves along the jet, but rather stabilizes at the location of the black hole \citep[][Figure 11]{Hada2016,Chael_19}.

\subsection{Observability with the EHT}
\label{sec:observability}

In \autoref{fig:Slices}, we show profiles from the simulation and analytic model 230~GHz images in \autoref{fig:ModelSimulationComparison} extracted along the $\beta=0$ (red; North-South) and $\alpha=0$ (blue; East-West) axes.
The time-averaged simulation image and the equatorial model image show the same characteristic features: a ring of direct $n=0$ emission that peaks at a radius of $\approx\!15-20\,\mu$as from the origin, $n=1$ and $n=2$ subring images that approach the critical curve at a radius of $\approx\!20\,\mu$as, and a central brightness depression corresponding to the lensed image of the equatorial horizon, i.e., an inner shadow.
The exact position of the horizon image on these slices is indicated by the vertical lines.
The equatorial analytic model has no emission outside the equatorial plane; its brightness plunges toward zero with increasing redshift as the projected radius approaches the direct lensed image $\rho_{\rm h}(\varphi)$ of the equatorial horizon on the sky.
The simulation image features faint foreground emission from the approaching relativistic jet which lies in front of the bulk of the emission in the equatorial plane.
The approaching jet provides a finite brightness ``floor'' inside the main $n=0$ emission ring.
In this simulation, the edges of the floor correspond to the analytic location of the horizon image to within about a microarcsecond.
In other simulations, the exact location of the emission floor will depend on the equatorial emissivity profile, the velocity/redshift of the equatorial fluid, and the intensity of the foreground emission. 

The cyan line on \autoref{fig:Slices} indicates the dynamic range of the EHT in 2017; the limited interferometric $(u,v)$ coverage of the array in this first observation of \m87 makes it impossible to extract dim features below $\approx\!10\%$ of the peak brightness \citep{PaperIV}.
The magenta line is an approximate forecast for the dynamic range of the next-generation EHT (ngEHT) array \citep{Astro2020Ground,Raymond21}.
With the addition of new sites and short interferometric baselines, the dynamic range of the ngEHT array should improve to be sensitive to emission that is a factor $10^{-3}$ dimmer than the beam emission.
In this simulation, the emission ``floor'' that fills the lensed horizon image is a factor of $10^{-2}$ dimmer than the peak of the emission.
As a result, in this scenario, we would expect an ngEHT array with improved coverage to be able to directly image the inner shadow feature down to the floor set by the foreground emission.

In \autoref{fig:Reconstructions}, we investigate the ability of the EHT and ngEHT arrays to recover the inner shadow feature with simulated image reconstructions.
We generate synthetic VLBI data from the time-averaged 230~GHz simulation image using the $(u,v)$ coverage on 2017 April 11 \citep{PaperIV}.
We also generate a synthetic ngEHT observation using an example array explored in \citet{Raymond21}.
This ngEHT concept array adds 12 telescopes to the current EHT, dramatically filling in the EHT's $(u,v)$ coverage and increasing its imaging dynamic range.
In both cases, we generated synthetic data including thermal noise and completely randomized station phases from atmospheric turbulence.
We did not include the time-variable amplitude gain errors that complicate real EHT imaging \citep{PaperIII,PaperIV}.

The left column of \autoref{fig:Reconstructions} shows the simulation image blurred to half of the nominal ngEHT resolution at 230~GHz (using a circular Gaussian blurring kernel of $10\,\mu$as FWHM).
The middle column shows the reconstruction from EHT2017 synthetic data, and the right column shows the ngEHT reconstruction.
Both reconstructions were performed using the \texttt{eht-imaging} library \citep{Chael_18}; in particular, the settings used in imaging the 2017 data were the same as those used in \texttt{eht-imaging} in the first publication of the M87 results in \citet{PaperIV}.
While the EHT2017 reconstruction shows a central brightness depression, its size and brightness contrast cannot be constrained or associated with the inner shadow.
However, the increased $(u,v)$ coverage of the ngEHT array dramatically increases the dynamic range, and the image reconstruction recovers the position and size of the high-dynamic-range ``inner shadow'' depression that is visible in the simulation image blurred to the equivalent resolution.

This imaging test is idealized.
First, neither the ngEHT nor EHT2017 directly image the time-averaged structure in \m87, so an imaging test using a GRMHD snapshot would be more realistic.
However, the inner shadow is prominent in simulation snapshots as well as in the time-averaged image (\autoref{fig:Summary}).
Furthermore, we neglect realistic station amplitude gains and polarimetric leakage factors that complicate image inversion from EHT data.
However, \m87 is weakly polarized, making accurate recovery of the total intensity image possible with no leakage correction \citep{PaperIV,PaperVII}, and image reconstruction of EHT data with even very large amplitude gain factors is possible with a relatively small degradation of the reconstruction quality using \texttt{eht-imaging} \citep{Chael_18}.

This example demonstrates that the candidate ngEHT array from \citet{Raymond21} \emph{could} constrain the presence of an inner shadow in \m87 if it is indeed present in the image.
In particular, detecting this feature does not require dramatic increases in imaging \emph{resolution} (which, in the absence of a 230~GHz VLBI satellite, is limited by the size of the Earth) but by the imaging \emph{dynamic range}, which is limited by the sparse number of baselines in the EHT array.
Once its presence is established via imaging, parametric visibility domain modeling could recover the size and shape of the inner shadow to higher accuracy than is possible from imaging alone \citep[e.g.,][]{PaperVI}.

\section{Geometric description of the lensed horizon image}
\label{sec:GeometricDescription}

\begin{figure}[t]
\centering
\includegraphics[width=\columnwidth]{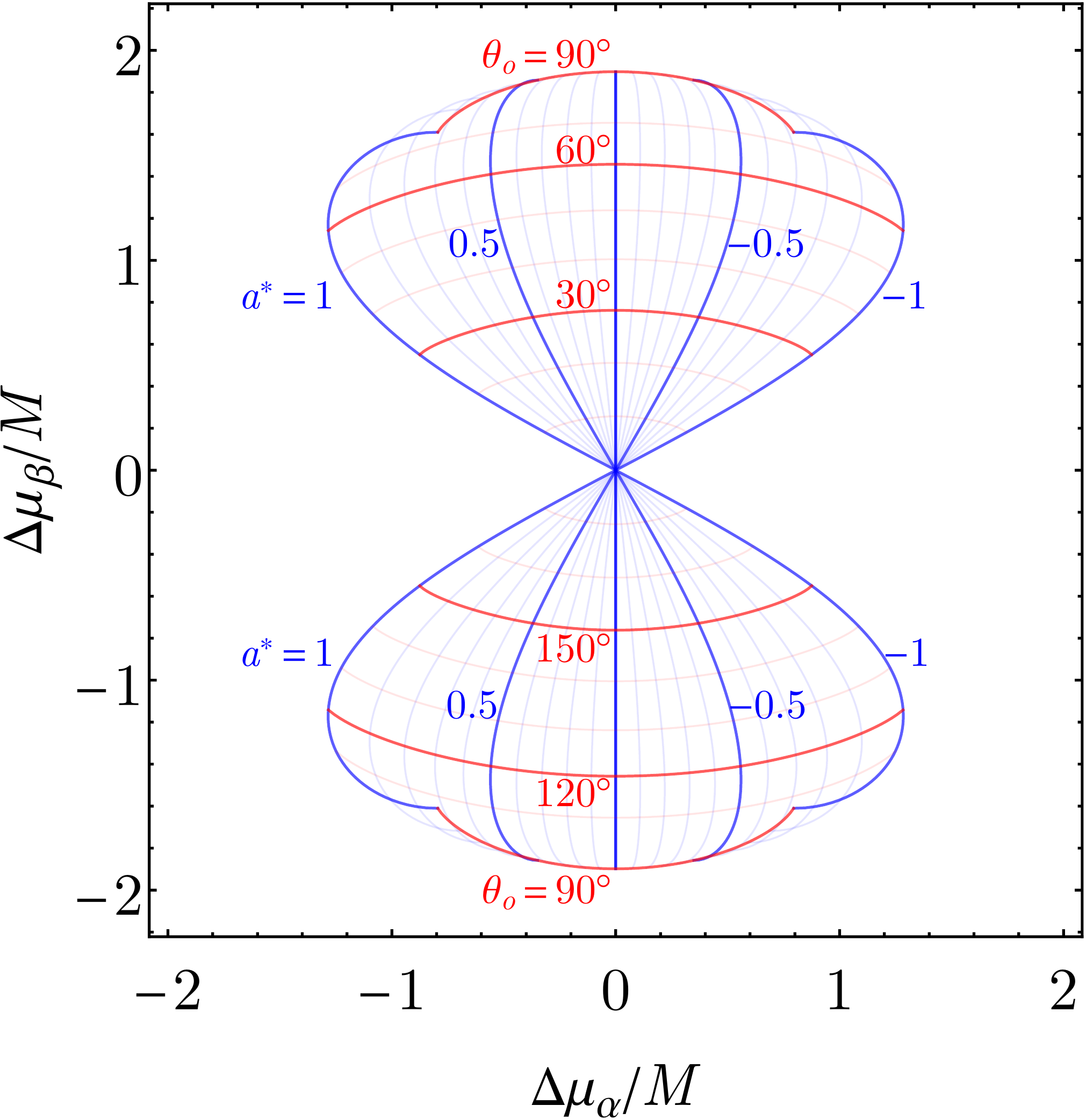} 
\caption{
Relative centroid displacement $(\Delta\mu_\alpha,\Delta\mu_\beta)$ of the direct equatorial horizon image with respect to the critical curve.
For all values of black hole spin and inclination, $\sign(\Delta\mu_\alpha)=\sign(\spin)$ and $\sign(\Delta\mu_\beta)=\sign(\cos{\theta_{\rm o}})$.
An abrupt transition occurs at $\theta_{\rm o}=90\deg$, where the $n=0$ and $n=1$ equatorial horizon images are degenerate.
The mapping $(\spin,\theta_{\rm o})\rightarrow (\Delta\mu_\alpha,\Delta\mu_\beta)$ is one-to-one and fairly linear up to high spin and nearly edge-on inclination.
}
\label{fig:Centroid}
\end{figure}

\begin{figure*}[t]
\centering
\includegraphics[width=0.32\textwidth]{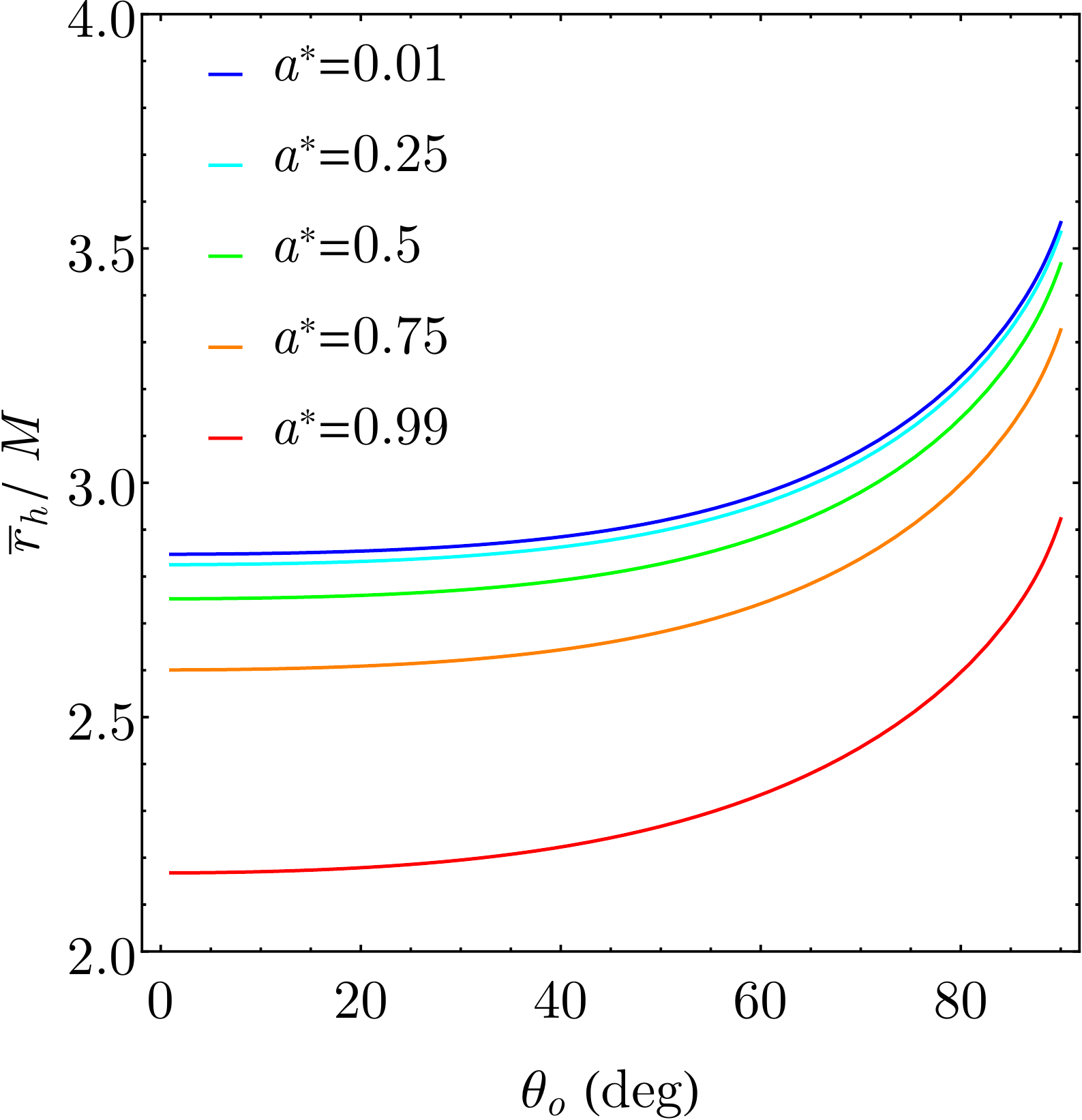}
\includegraphics[width=0.32\textwidth]{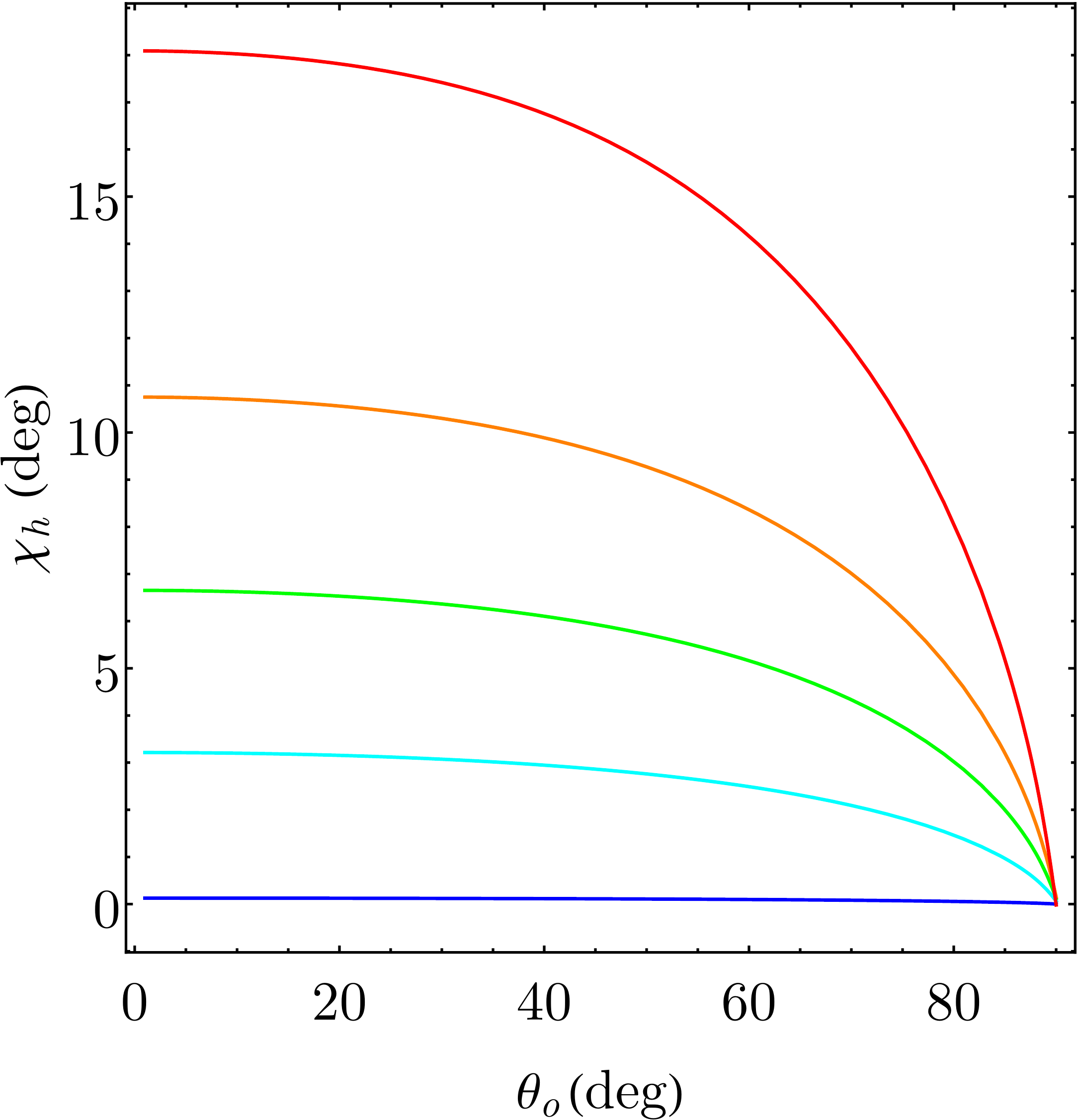}
\includegraphics[width=0.32\textwidth]{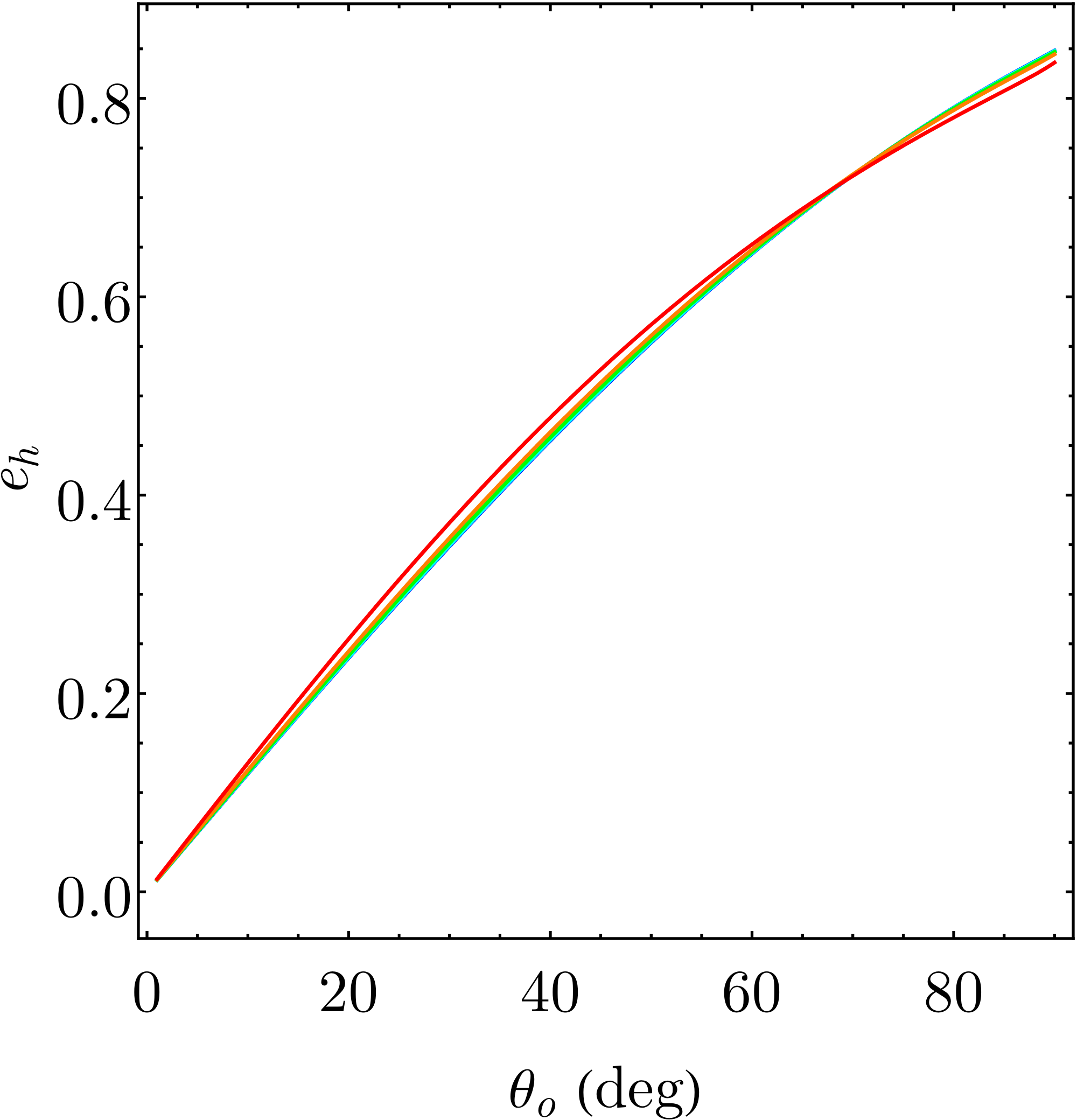}
\caption{
(Left) The mean radius $\bar{r}_{\rm h}$ of the inner shadow as a function of inclination $\theta_{\rm o}$ for several values of the black hole spin: from top to bottom, $\spin=(0.01,0.25,0.5,0.75,0.99)$.
(Middle) The orientation angle $\chi_{\rm h}$ with respect to the $+\alpha$ axis as a function of inclination for the same spin values; the orientation angle increases with spin at low and moderate inclination.
(Right) The eccentricity $e_{\rm h}$ for the same spin values; the eccentricity of the inner shadow is nearly independent of spin and depends primarily on the inclination.
}
\label{fig:2ndMoment}
\end{figure*}

In this section, we describe the behavior of the lensed equatorial horizon contour as a function of black hole spin and observer inclination using image moments.
While not a complete description of the horizon image shape (particularly at high inclination), the moment description captures important properties of the horizon image that may be observable by the EHT or future VLBI experiments. 

In the procedure outlined in \autoref{sec:ComputingCurves}, we compute the $n=0$ lensed horizon image as a closed curve $\rho_{\rm h}(\varphi)$ in the $(\alpha,\beta)$ image plane.
Given $\rho_{\rm h}$, we can compute image moments in a standard way (explicitly described in \autoref{app:Moments}).
The zeroth moment is the area $A_{\rm h}$ of the inner shadow.
The first moment is the centroid vector $\boldsymbol{\mu}_{\rm h}$ defined with respect to the $(\alpha,\beta)$ axes.
The second central moment is the covariance matrix $\boldsymbol\Sigma_{\rm h}$.
By diagonalizing $\boldsymbol\Sigma_{\rm h}$, we can compute the lengths $a_{\rm h}$ and $b_{\rm h}$ of the principal axes (where $a_{\rm h}\geq b_{\rm h}$), as well as the orientation angle $\chi_{\rm h}$ between the first principal axis and the positive $\alpha$ axis.
We can then define the mean radius $\bar{r}_{\rm h}$ and the eccentricity $e_{\rm h}$ of the lensed horizon as
\begin{align}
	\label{eq:MeanRadius}
	\bar{r}_{\rm h}=\sqrt{\frac{a_{\rm h}^2+b_{\rm h}^2}{2}},\quad
	e_{\rm h}=\sqrt{1-\frac{b_{\rm h}^2}{a_{\rm h}^2}}.
\end{align}

In addition to computing these image moments for the lensed horizon, we also compute the area, centroid, average radius, eccentricity, and orientation angle of the critical curve ($A_{\rm c},\boldsymbol\mu_{\rm c},\bar{r}_{\rm c},e_{\rm c},\chi_{\rm c}$).
Note that our definition of the average critical curve radius $\bar{r}_{\rm c}$ differs from that introduced in \citet{JohannsenPsaltis}.
In \autoref{app:Moments}, \autoref{fig:RadiusComparison}, we compare results for the average critical curve radius from these two methods and find that they agree within one percent for all values of black hole spin and observer inclination.

\subsection{Centroid}
\label{sec:Centroid}

In \autoref{fig:Centroid}, we plot the \emph{relative} centroid displacement between the lensed horizon and the critical curve $\Delta\boldsymbol\mu=\boldsymbol\mu_{\rm h}-\boldsymbol\mu_{\rm c}$.
Measuring the absolute centroid displacement of either the lensed horizon or the critical curve would require precise prior knowledge of the location of the black hole on the sky; by contrast, the relative centroid displacement $\Delta\boldsymbol\mu$ could in principle be observed by simply measuring the two curves and determining the direction of the black hole spin to set the orientation of the $+\beta$ axis (in M87, for instance, these axes can be inferred from the direction of the large-scale jet). 

The critical curve is symmetric about the $\beta=0$ axis for all values of spin and inclination, so the vertical displacement $\Delta\boldsymbol\mu_\beta$ is purely due to the offset of the inner shadow's centroid.
The direction of the vertical offset is set by the hemisphere that the observer lies in: $\sign(\Delta\mu_\beta)=\sign(\cos{\theta_{\rm o}})$.
Both the critical curve and the lensed horizon image have a horizontal displacement $\Delta\boldsymbol\mu_\alpha$ that is approximately linear with spin.
The sign of this displacement follows that of the spin: $\sign(\Delta\mu_\alpha)=\sign(\spin)$.\footnote{
The projected spin direction is along the $\beta$ axis: a positive spin is aligned with the $+\beta$ axis and a negative spin with the $-\beta$ axis.
} 
In general, the mapping $(\spin,\theta_{\rm o})\rightarrow(\Delta\mu_\alpha,\Delta\mu_\beta)$ is one-to-one and fairly linear up to high spin and nearly edge-on inclination.
There is an abrupt transition in $\Delta\boldsymbol\mu$ at $\theta_{\rm o}=90\deg$, where the $n=0$ and $n=1$ images are degenerate.

The geometric centroids of the inner shadow and the critical curve are well approximated by
\begin{align}
	\mu_{\alpha}&\approx
	\begin{cases} 
		2M\spin\sin{\theta_{\rm o}}
		&\text{critical curve},\\
		\frac{1}{2}M\spin\sin{\theta_{\rm o}}
		&\text{equatorial horizon},
	\end{cases}\\
	\mu_{\beta}&\approx
	\begin{cases} 
		0
		&\text{critical curve},\\
		\pm\frac{3}{2}M\pa{1-\frac{1}{4}\spin^2}\sin{\theta_{\rm o}}
		&\text{equatorial horizon},
\end{cases}
\end{align}
where $\sign(\mu_\beta)=\sign(\cos{\theta_{\rm o}})$.
For $\sin{\theta_{\rm o}}<1/2$ and $\ab{\spin}<1/2$, these centroid approximations have a maximum absolute error less than $0.03\,M$. 

If the inclination and mass are known a~priori, then it is possible to estimate the spin by  
\begin{align}
	\spin\approx-\frac{2}{3\sin{\theta_{\rm o}}}\frac{\Delta\mu_\alpha}{M}.
\end{align}
For instance, if the spin of \m87 is aligned with its large-scale jet, then $\spin\approx-2.3\Delta\mu_\alpha/M\approx -\frac{\Delta\mu_\alpha}{0.6\,\mu\text{as}}$.
Indeed, based on the jet inclination, we expect that $\ab{\Delta\mu_\alpha}<0.53\,M$ and $0.31\,M<\ab{\Delta\mu_\beta}<0.44\,M$ for \m87.
The narrow range in allowed $\Delta \mu_\beta$ is useful to assess whether features detected in the image can be associated with the equatorial horizon image or critical curve.  

Likewise, if the inclination is unknown but there is an a priori spin estimate, then it is possible to estimate the inclination using
\begin{align}
	\sin{\theta_{\rm o}}&\approx\frac{2}{3\pa{1-\frac{1}{4}\spin^2}}\frac{\ab{\Delta\mu_\beta}}{M}.
\end{align}
Hence, a measured centroid offset along the spin direction $\Delta\mu_\beta$ determines a narrow range of possible inclinations: $\sin{\theta_{\rm o}}\in\br{\frac{2}{3}, \frac{8}{9}}\ab{\Delta \mu_\beta}/M$.
For instance, measuring an offset $\Delta\mu_\beta=1.5\,\mu$as in \m87 would give a constraint $15\deg<\theta_{\rm o}<21\deg$.

\subsection{Radius, orientation, eccentricity}

\autoref{fig:2ndMoment} shows the variation of the quantities that define the second moment of the lensed horizon image---the mean radius $\bar{r}_{\rm h}$, the orientation angle $\chi_{\rm h}$, and the eccentricity $e_{\rm h}$---with the inclination $\theta_{\rm o}$ (in the range $0\leq\theta_{\rm o}\leq\pi/2$) for several values of the black hole spin $\spin$.
At low inclinations ($\theta_{\rm o}\lesssim30\deg$), the mean radius $r_{\rm h}$ and image orientation angle $\chi_{\rm h}$ are approximately independent of the inclination and hence directly probe the spin.
By contrast, the eccentricity $e_{\rm h}$ is almost entirely independent of spin over the whole range and thus provides a direct probe of the inclination.
Measuring $e_{\rm h}$ or $\chi_{\rm h}$ at inclinations $\theta_{\rm o}\lesssim30\deg$ would require extremely high precision measurements of the lensed horizon shape; since the eccentricity $e_{\rm h}<0.4$ for these inclinations, the relative sizes of the major and minor image axes differ by $\lesssim 8$\%.

\begin{figure}[t]
\includegraphics[width=0.4\textwidth]{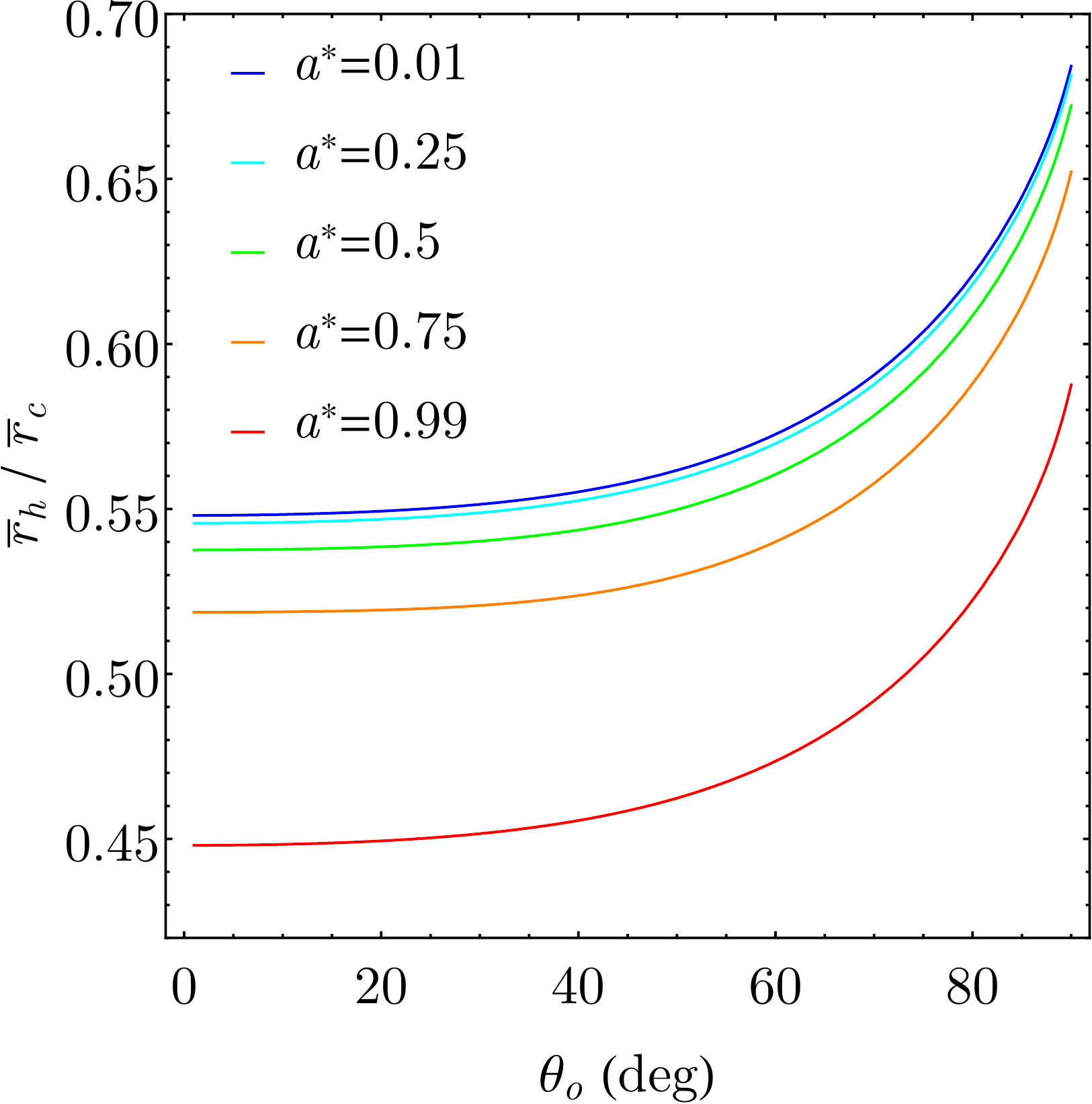}
\caption{
The ratio of the mean radius of the lensed horizon $\bar{r}_{\rm h}$ to the mean radius of the critical curve $\bar{r}_{\rm c}$. In the low-inclination case, $\bar{r}_{\rm h}/\bar{r}_{\rm c}$ shrinks from $\approx\!55$\% at zero spin to $\approx\!45$\% at maximal spin.
}
\label{fig:RelativeRadius}
\end{figure}

At these low inclinations, the mean image radius varies by $\approx\!20$\% from zero to maximal spin.
\autoref{fig:RelativeRadius} shows, for several fixed values of the black hole spin, the dependence on inclination of the ratio $\bar{r}_{\rm h}/\bar{r}_{\rm c}$ of the lensed horizon mean radius to the critical curve mean radius.
Again, for $\theta_{\rm o}\lesssim30\deg$, $\bar{r}_{\rm h}/\bar{r}_{\rm c}$ is approximately independent of the inclination and hence provides a direct measurement of the spin.
In the low inclination case, $\bar{r}_{\rm h}/\bar{r}_{\rm c}$ shrinks from $\approx\!55$\% at zero spin to $\approx\!45$\% at maximal spin.
Importantly, measuring $\bar{r}_{\rm h}/\bar{r}_{\rm c}$ for an astrophysical black hole would \emph{not} require accurate measurements of the black hole mass $M$ or distance $D$.

Referencing the lensed horizon image directly to the critical curve would require detecting a lensed subring of order $n\gtrsim1$, which for \m87 only becomes visible on very high resolution baselines $\gtrsim\!20$~G$\lambda$ \citep{Johnson_Ring}.
However, it should be possible to constrain the location of the critical curve with measurements of the $n=0$ and $n=1$ rings by determining systematic calibration factors (and associated systematic uncertanties) that relate size of the EHT image to the critical curve size in a library of astrophysical models, as was done in \citet{PaperV,PaperVI} to measure the mass of \m87.
Alternatively, parametric modeling with priors on the image structure may constrain the $n=1$ and higher subring images from measurements at lower spatial frequencies rings using parametric model fits \citep{Broderick20}.

For all spins at low and moderate inclinations, the lensed horizon image is well approximated by an ellipse, and the first three image moments give a fairly complete description of the curve.
At higher inclinations, the structure becomes more complex, with more information in the full curve shape than is captured in the first three moments.
In particular, at $\spin=0$ and $\theta_{\rm o}=90\deg$, the horizon image becomes a semicircle, degenerate with the $n=1$ horizon image semicircle in the other half plane (at higher spins, the image is not a perfect semicircle, but is still a mirror image of the degenerate $n=1$ image; see \autoref{fig:imgrid}).
In this case, the curve is probably better described by the single radius of the combined $n=0$ and $n=1$ circle, rather than the second moment parameters of just the $n=0$ image.

\begin{figure*}[t]
\centering
\includegraphics[width=0.45\textwidth]{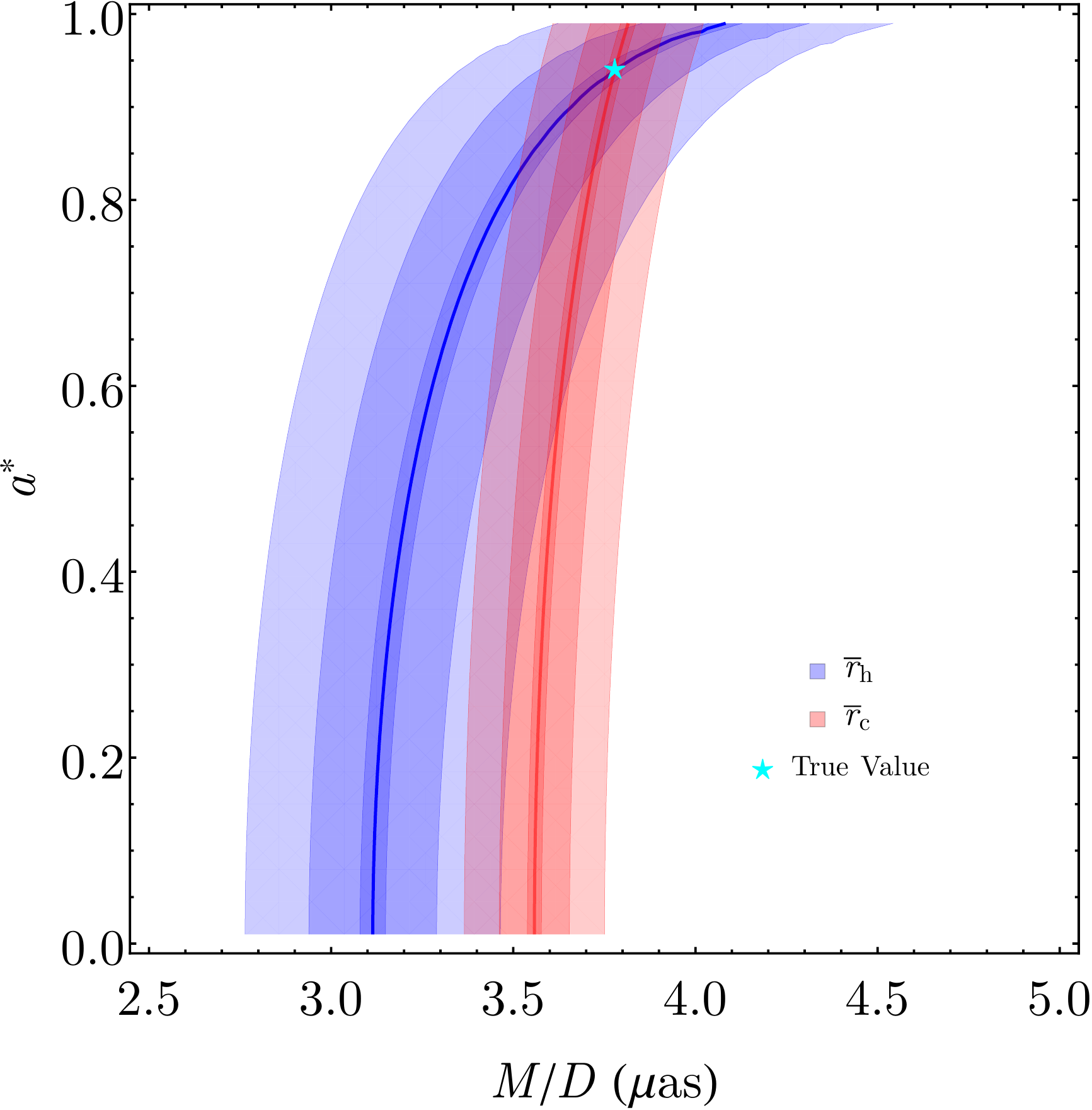}\hfill
\includegraphics[width=0.45\textwidth]{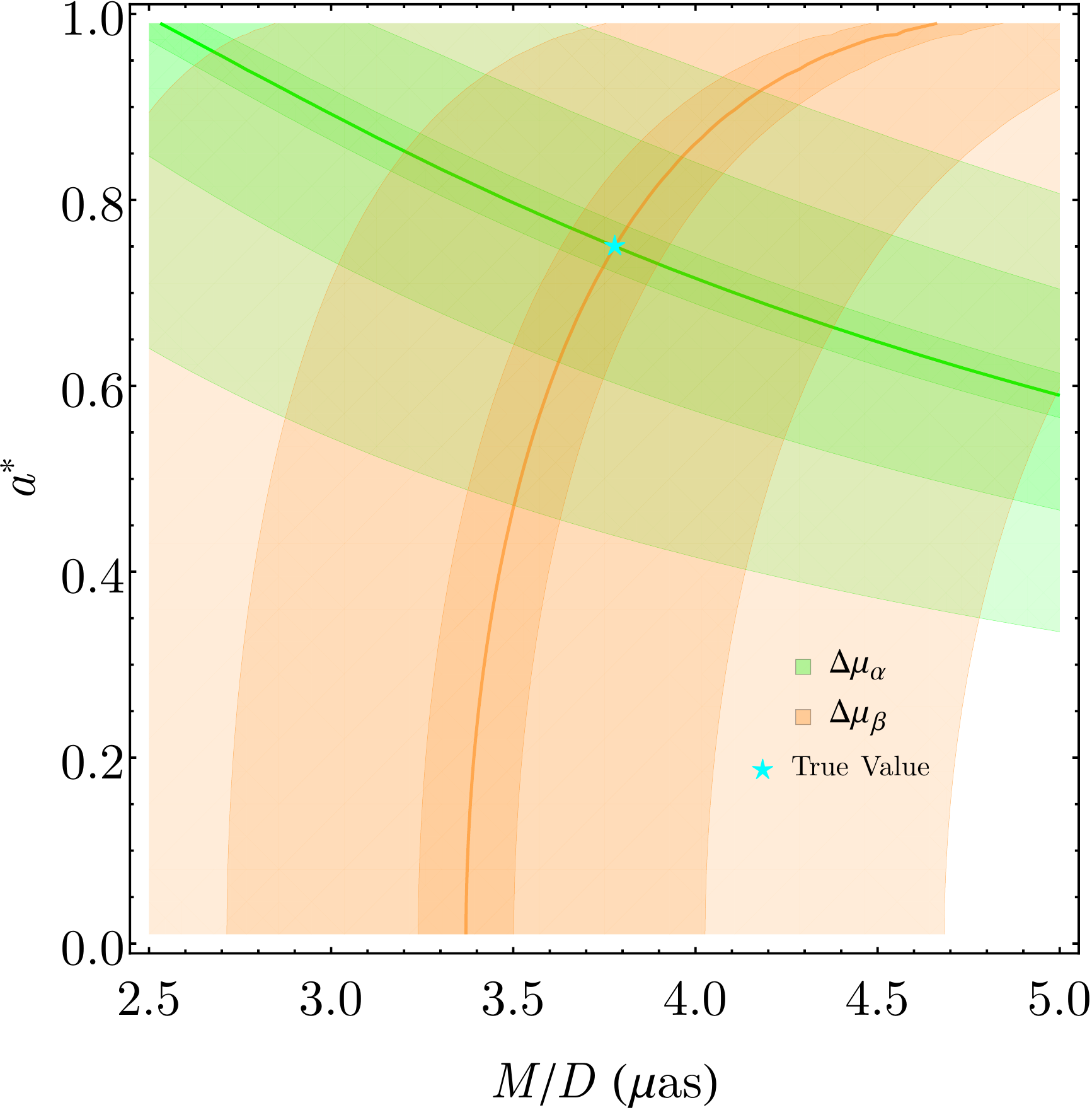} \\
\caption{
(Left) Simultaneous constraints on the black hole mass-to-distance ratio $M/D$ and spin $\spin$ enabled by measuring the mean radius of the lensed horizon (blue, $\bar{r}_{\rm h}$) and critical curve (red $\bar{r}_{\rm c}$) in \m87, when the inclination is fixed $\theta_{\rm o}=17\deg$ \citep{Mertens2016}.
Without fixing the mass, multiple values of $\spin$ provide the same result for the size of each feature, but combining a measurement of both features breaks the degeneracy.
The shaded regions show errors on the radius measurement of $0.1$, $0.5$, and $1\,\mu$as.
The input mass scale and spin are $M/D=3.78\,\mu$as and $\spin=0.94$.
(Right) Simultaneous constraints on $M/D$ and $\spin$ from measurements of the centroid offset $\boldsymbol\mu_{\rm h}-\boldsymbol\mu_{\rm c}$ in the $\alpha$ direction (green) and the $\beta$ direction (orange).
}
\label{fig:SpinMassConstraint}

\bigskip
\centering
\includegraphics[width=0.45\textwidth]{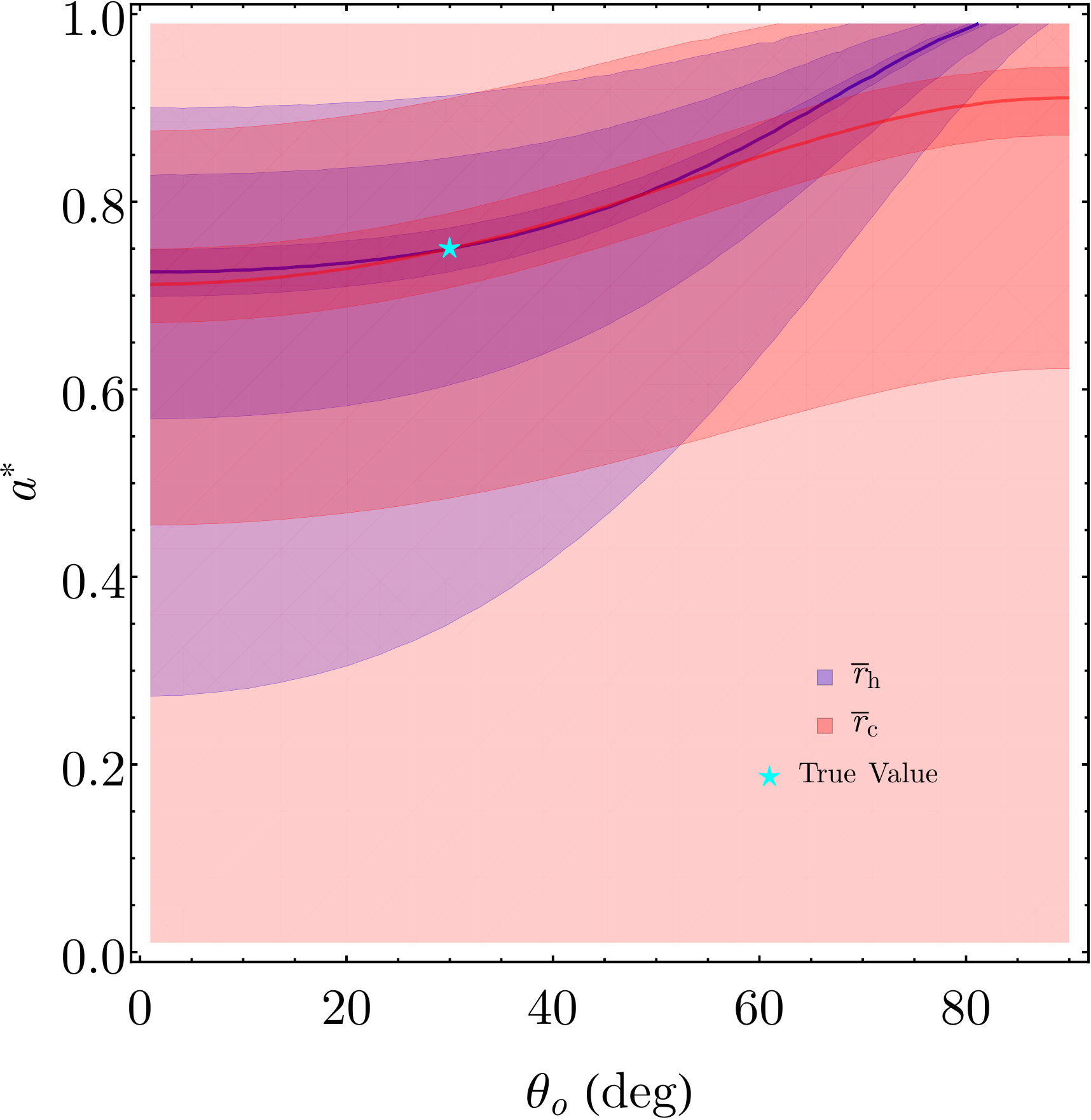}\hfill
\includegraphics[width=0.45\textwidth]{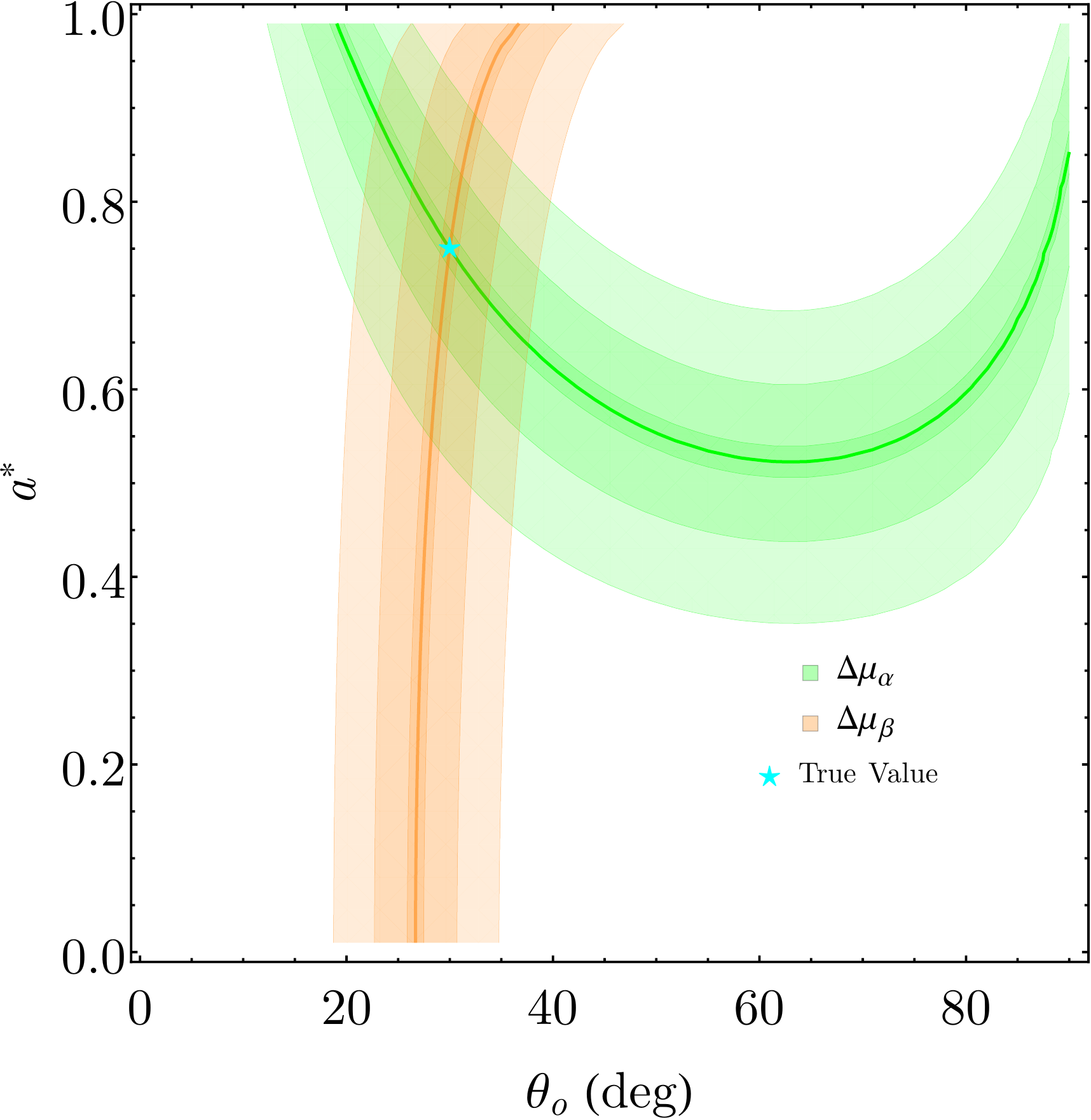}
\caption{
(Left) Simultaneous constraints on the black hole spin $\spin$ and inclination $\theta_{\rm o}$ enabled by measuring the mean radius of the lensed horizon (blue, $\bar{r}_{\rm h}$) and critical curve (red $\bar{r}_{\rm c}$) in \sgra, fixing the mass-to-distance ratio $M/D=5.01\,\mu$as \citep{Gravity2019}.
The shaded regions show errors on the radius measurement of $0.1$, $0.5$, and $1\,\mu$as.
The input spin and inclination are $\spin=0.75$ and $\theta_{\rm o}=30\deg$.
(Right) Simultaneous constraints on $\spin$ and $\theta_{\rm o}$ from measurements of the centroid offset $\boldsymbol\mu_{\rm h}-\boldsymbol\mu_{\rm c}$ in the $\alpha$ direction (green) and the $\beta$ direction (orange).
The absolute errors in the centroid measurement depicted by the shaded regions are the same as in the left panel.
When $M/D$ is fixed, the relative sizes of the lensed horizon and critical curve poorly constrain $\theta_{\rm o}$, but the centroid offset strongly constrains both $\spin$ and $\theta_{\rm o}$.
}
\label{fig:SpinInclinationConstraint}
\end{figure*}

\autoref{fig:SpinMassConstraint} demonstrates how a simultaneous measurement of the radius of the critical curve and the lensed horizon could be used to constrain the mass and spin in \m87 when the inclination is fixed at $\theta_{\rm o}=17\deg$ \citep{Mertens2016}.
These simultaneous constraints are analogous to those discussed in \citet{Broderick21}, which considers constraints from measuring multiple lensed images from a single face-on emitting ring.
The blue line shows the space of mass-to-distance ratios $M/D$ and spins $\spin$ that give the same mean lensed horizon radius for an image of \m87; the red line shows the same for the critical curve.
The red and blue lines intersect in only one location corresponding to the input black hole mass $M/D=3.78\,\mu$as and spin $\spin=0.94$.
The shaded bands around the intersecting lines show absolute errors in the radius measurements of $0.1,0.5,1\,\mu$as.
Given a reported EHT radius measurement uncertainty of $1.5\,\mu$as from geometric modeling of the EHT2017 data in  \citet{PaperVI}, measurements of the ring and inner shadow radius and centroid locations at $\lesssim\!1\,\mu$as precision may be feasible with the ngEHT.
In addition to reducing uncertainty in the image size measurement itself, precisely constraining both features will depend on reducing systematic uncertainty in the relationship between the gravitational features and images from a set of plausible astrophysical models \citep[e.g.,][]{PaperVI}.

The right panel of \autoref{fig:SpinMassConstraint} shows a similar figure for a simultaneous measurement of the centroid offset along the $\alpha$ ($\Delta\mu_\alpha$, in green) and $\beta$ axes ($\Delta\mu_\beta$, in orange).
Because the centroid offsets are relatively small, an absolute error of $1\,\mu$as in the measurement of the centroid offsets is less constraining than the corresponding radius measurement.
However, measuring the offset $\Delta\mu_\alpha$ to $1\,\mu$as precision could put a lower limit on the spin $\spin\gtrsim0.5$ independent of the mass.
Measurements that jointly constrain the image size and eccentricity could even more precisely constrain the mass and spin in \m87. 

In \sgra, the mass-to-distance ratio is known to high precision, $M/D=5.011\pm0.016\,\mu$as \citep{Gravity2019}, but the inclination is unconstrained.
In \autoref{fig:SpinInclinationConstraint}, we show similar simultaneous constraints on $\theta_{\rm o}$ and $\spin$ for \sgra from measurements of the inner shadow and critical curve radii (left panel) and centroid offset (right panel).
The inner shadow and critical curve radii poorly constrain the inclination, and constraining the spin with these radii and an unknown inclination requires a measurement precision finer than $1\,\mu$as.
By contrast, as discussed in \autoref{sec:Centroid}, the centroid offset is highly constraining of both $\theta_{\rm o}$ and $\spin$ if $M/D$ is known (right panel of \autoref{fig:SpinInclinationConstraint}).
Note that the constraints shown in the right panels of \autoref{fig:SpinMassConstraint} and \autoref{fig:SpinInclinationConstraint} require the $\alpha$ and $\beta$ axes to be known a priori; in practice, this could be informed with reference to the jet at large scales in \m87, or by reference to the location of the brightness asymmetry from Doppler beaming in 230~GHz images of \m87 or \sgra.
\autoref{fig:SpinMassConstraint} and \autoref{fig:SpinInclinationConstraint} are idealized examples; in practice, the orientation of the $\alpha$ and $\beta$ axes would likely have to be fit to observations along with $\spin$, $\theta_{\rm o}$, and other model parameters that relate the critical curve and lensed equatorial horizon shapes to their corresponding features in the observed image (e.g., the emissivity and redshift profiles in the model described in \autoref{sec:EmissionModel}). 

\subsection{Semi-analytic description of the lensed horizon}

\begin{figure*}[t!]
\centering
\includegraphics[width=0.33\textwidth]{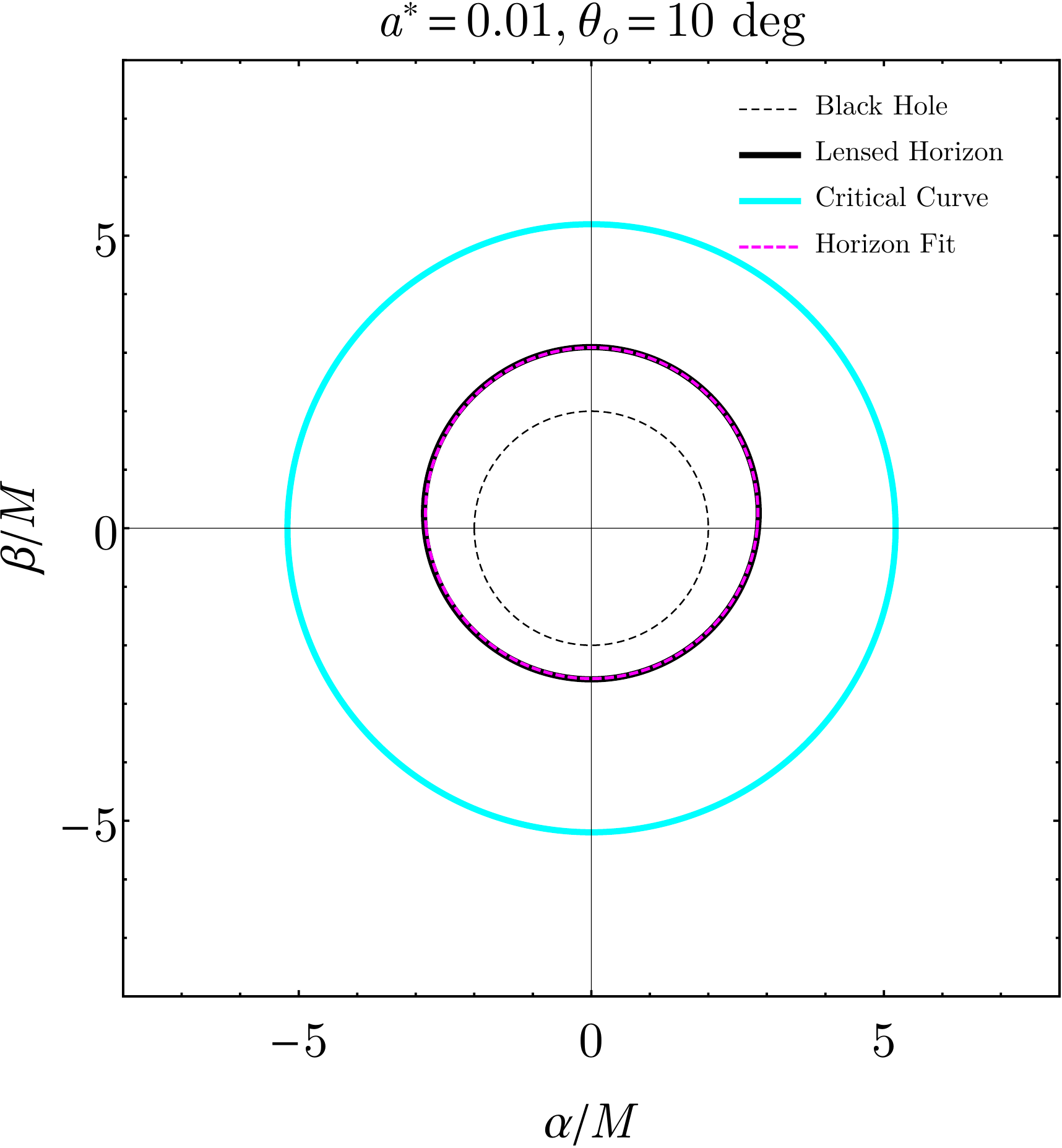}
\includegraphics[width=0.33\textwidth]{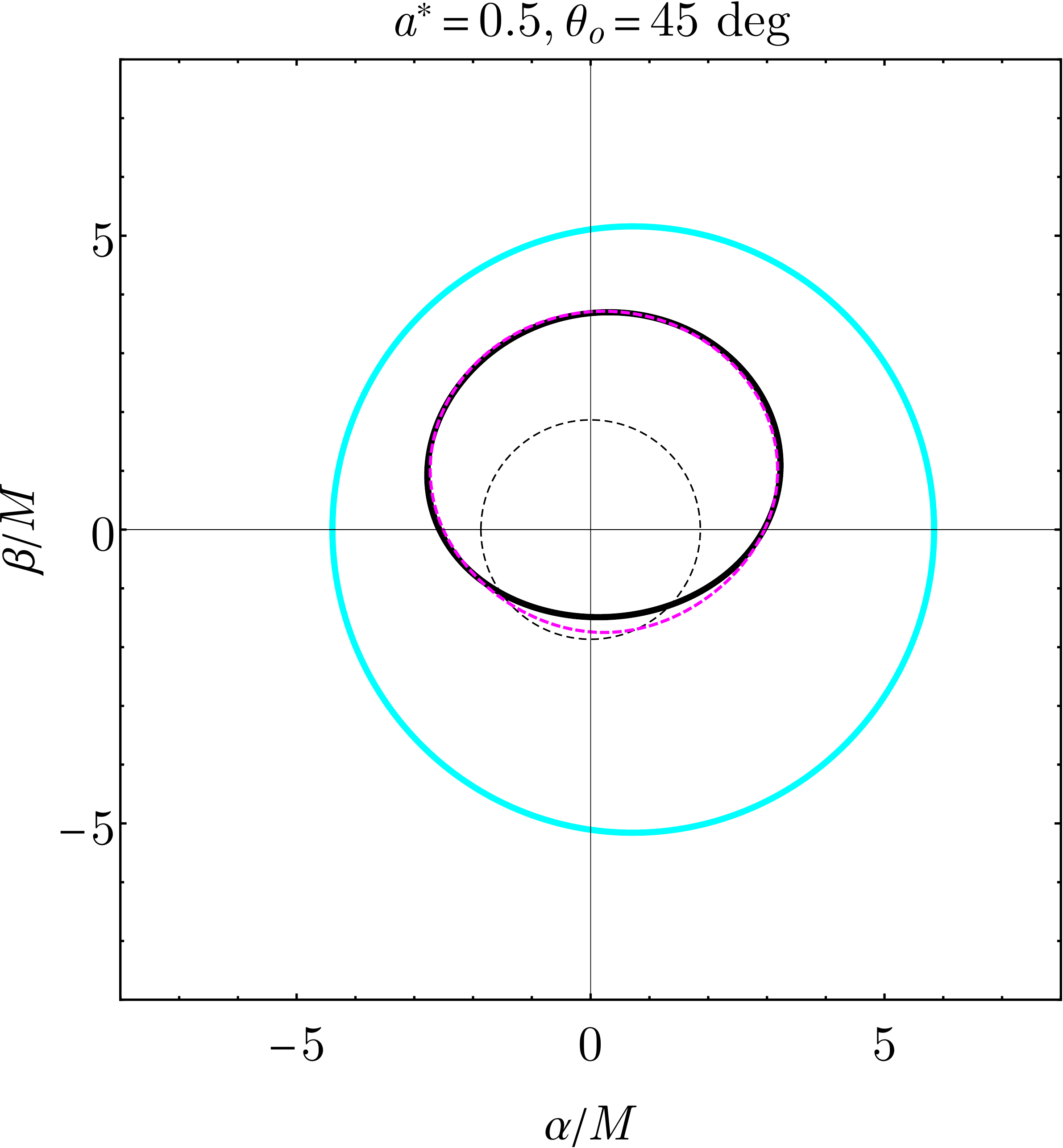}
\includegraphics[width=0.33\textwidth]{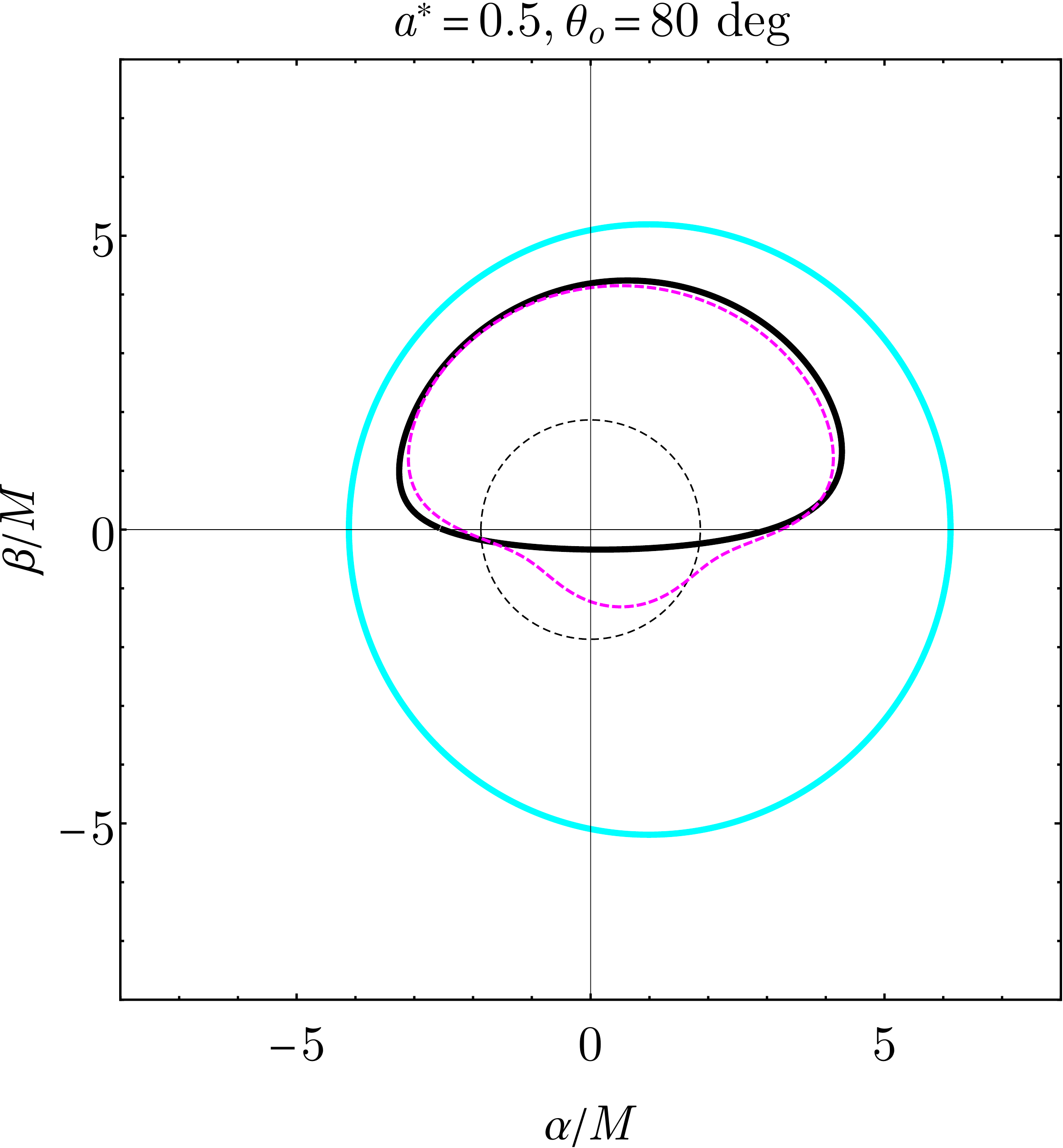}
\caption{
Examples of the semi-analytic fitting function $\rho_{\rm h, fit}(x)$ (\autoref{eq:FittingFunction}) to the lensed equatorial horizon curve for a black hole with spin $\spin=0.5$ viewed at inclination $\theta_{\rm o}=10\deg$ (left), $45\deg$ (middle), and $80\deg$ (right).
In each panel, we indicate the ``unlensed'' image of the black hole event horizon $r=r_+$ (black dashed line), the direct lensed image of the equatorial horizon (black solid line), the critical curve (cyan solid line), and the approximate parameterization of the horizon image $\rho_{\rm h, fit}(x)$ (dashed magenta line).
The approximate parameterization matches the horizon image well at low inclination, but deviates from the truth in the $\beta<0$ part of the curve at high inclination.
}
\label{fig:FitCurve}
\end{figure*}

Here, we attempt to derive an approximate parameterization for the general shape of the lensed equatorial event horizon, starting with a nonrotating ($\spin=0$) Schwarzschild black hole.
As is the case for the critical curve, such approximations are useful for easily exploring the behavior of the shape with spin and inclination and to identify parameter degeneracies \citep[e.g.,][]{devries_2003,Cunha_2018,Farah_2020,GL20}.
\cite{Gates2020} provide an analytic formula for the equatorial radius $r_{\rm eq}(\rho,\varphi,\theta_{\rm o};n)$ that a light ray shot back from position $(\rho,\varphi)$ on the image plane will cross on its $(n+1)^\text{th}$ equatorial crossing.
While the dependence of this transfer function on the impact radius $\rho$ is rather complicated, its dependence on the impact angle $\varphi$ and observer inclination $\theta_{\rm o}$ only enter through the particular combination
\begin{align}
	x=\arctan\br{\sin\pa{\varphi}\tan\pa{\theta_{\rm o}}}.
\end{align}
In this paper, we are only interested in the direct image $n=0$.\footnote{
Lensed images with higher $n$ lie in the photon ring, and one may set up a large-$n$ expansion as in \cite{GrallaLupsasca2020} to obtain a simple asymptotic formula for $r_{\rm eq}(\rho,\varphi;n)$ given in App.~A of \cite{Hadar2020}.}
It is tempting to expand
\begin{align}
	\label{eq:EquatorialRadius}
	r_{\rm eq}(\rho,x;n=0)=Mf_0(\rho)-Mf_1(\rho)x+\O{x^2},
\end{align}
where $Mf_0(\rho)=r_{\rm eq}(\rho,x=0;n=0)$ is the exact, axisymmetric transfer function for a polar observer at $x=\theta_{\rm o}=0$.
\cite{Gates2020} derive its asymptotic expansion in large impact radius $\rho\to\infty$,
\begin{align}
	\label{eq:f0}
	f_0(\rho)=\frac{\rho}{M}-1+\frac{1}{2}\frac{M}{\rho}+\frac{3(5\pi-16)}{4}\frac{M^2}{\rho^2}+\O{\frac{M^3}{\rho^3}}.
\end{align}
\autoref{eq:f0} is already an excellent approximation even when truncated after the first two terms \citep{GrallaLupsasca2020}.
Likewise, $f_1(\rho)$ admits the expansion
\begin{align}
	f_1(\rho)=2+\frac{(15\pi-64)}{8}\frac{M}{\rho}-\frac{(15\pi-50)}{2}\frac{M^2}{\rho^2}+\O{\frac{M^3}{\rho^3}}.
\end{align}
Inverting Eq.~\eqref{eq:EquatorialRadius} results in an expression of the form
\begin{align}
	\rho(x; r_{\rm eq},n=0)=g_0(r_{\rm eq})+g_1(r_{\rm eq})x+\O{x^2},
\end{align}
which unfortunately is not as good an approximation as its inverse \eqref{eq:EquatorialRadius}.
In particular, this expansion breaks down at large inclination, where the higher-order terms in $x$ grow more relevant.
Nonetheless, we empirically observe that an excellent semi-analytic fit to the lensed equatorial horizon ($r_{\rm eq}=r_+$) for Schwarzschild ($r_+=2M$) is provided by the expression
\begin{align}
	\rho_{\mathrm{h},\spin=0}(x)\approx M\br{2\sqrt{2}+\pa{1+\frac{1}{2}\cos^2{\theta_{\rm o}}}x}.
\end{align}
Note that the Schwarzschild approximation $g_0(r_+)=2\sqrt{2}M$ follows from an analytic approximation for Schwarzschild geodesics given by \cite{Beloborodov2002}.

Finally, we note that this expression can be simply extended to the rotating Kerr case with nonzero spin $\spin>0$ to obtain a general fitting function $\rho_{\rm h, fit}(\varphi)$ for the shape of the direct lensed equatorial horizon image: 
\begin{align}
	\label{eq:FittingFunction}
	\rho_{\rm h,fit}(x)=M\left[2\sqrt{\frac{r_+}{M}}+\pa{1+\frac{1}{2}\cos^2{\theta_{\rm o}}}x\right].
\end{align}
In particular, for a face-on observer, the radius of the lensed equatorial horizon is well fit by 
\begin{align}
    \rho_{\rm h}(\theta_{\rm o}=0)\approx2M\sqrt{r_+/M}.
\end{align}
\autoref{fig:FitCurve} shows the quality of the approximation $\rho_{\rm h, fit}(x)$ (\autoref{eq:FittingFunction}) to the lensed equatorial horizon image for a black hole of spin $\spin=0.5$ at several inclinations.
The fitting function breaks down at high inclinations, where it develops an artifical ``bump'' on the part of the curve below the $\alpha$ axis. 

\section{Discussion}
\label{sec:discussion}

\begin{figure*}[t]
\centering
\begin{tabular}{ccc}
\includegraphics[width=\textwidth]{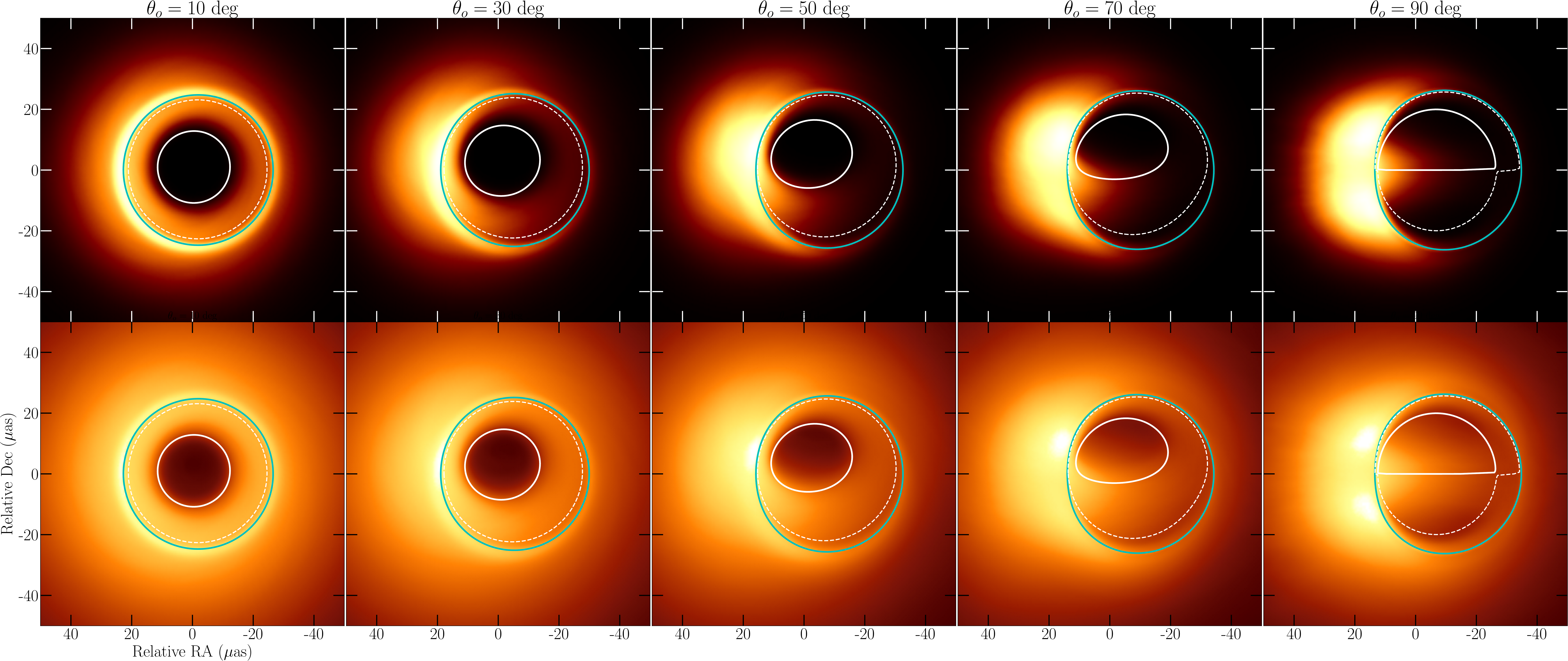} 
\end{tabular}
\caption{
Time-averaged 230~GHz images of a \sgra simulation with weak magnetic flux and $\spin=0.9375$ at different inclinations: from left to right, $\theta_{\rm o}=(10,30,50,70,90)\deg$.
The top row shows images in a linear scale, and the bottom row displays images in a gamma scale with $\gamma=1/4$.
The different panels are each normalized to their own peak brightness, since the total flux density increases with increasing inclination.
The critical curve is indicated in cyan, the direct image of the equatorial horizon is indicated by the solid white curve, and the $n=1$ image of the equatorial horizon is displayed with the dashed white curve.
In this low-magnetic-flux simulation, the 230~GHz emission is concentrated in a geometrially thick disk extending to the horizon.
The lensed equatorial horizon image is visible at low inclinations, but is blocked by direct emission from the disk when viewed at edge-on inclinations.
}
\label{fig:SANE}
\end{figure*}

Images from the radiative GRMHD simulations of \m87 from \citet{Chael_19} display a deep central brightness depression.
For these simulations of synchrotron-emitting plasma around a Kerr black hole, the edge of this image feature is the lensed image of the event horizon's intersection with the equatorial plane; the brightness depression is the ``inner shadow'' of the black hole.
Because the gravitational redshift diverges for emission approaching the event horizon, this inner shadow is only visible at high dynamic range in the simulation images.
Nonetheless, if it is present in \m87 images, this feature could be observable with the next-generation Event Horizon Telescope, which will increase the dynamic range of the current EHT by a factor of $\approx\!100$.

An inner shadow bounded by the image of the equatorial event horizon is only visible in images of a black hole if the emission region
\begin{enumerate}
	\item is concentrated in the equatorial plane,
	\item is not obscured by foreground emission (e.g., from the approaching jet), and
	\item extends to the event horizon.
\end{enumerate}
This scenario is realized by 230~GHz emission from the MAD simulations of \m87 considered in this paper; however, there are other plausible scenarios for the distribution of emitting plasma.

If the emission is spherically symmetric and extends to the horizon, the direct outline of the horizon is not visible; the only brightness depression in this case occurs inside the critical curve \citep[e.g.,][]{Falcke2000,Narayan_Shadow}.
Because GRMHD simulations of hot accretion flows produce geometrically thick accretion disks, one might think that these simulations are better approximated by a spherically symmetric emissivity distribution than an equatorial model.
On the contrary, however, even geometrically thick accretion disks in GRMHD simulations are better approximated by the equatorial emission model than a spherically symmetric one.
\autoref{fig:SANE} shows time-averaged images from a GRMHD simulation of \sgra in the low-magnetic-flux accretion state at several observer inclinations $\theta_{\rm o}$.
At low inclination, the images from this simulation look qualitatively similar to the magnetically arrested \m87 simulation images in \autoref{fig:ModelSimulationComparison}; they display both a bright photon ring near the critical curve, and an inner shadow marking the equatorial event horizon.
Despite the effects of significant disk thickness, the horizon image is still visible up to moderate inclinations ($\theta_{\rm o}\approx50\deg$) in this simulation.
At higher inclinations, emission from the disk in front of the black hole blocks the appearance of the horizon inner shadow, and it is not visible.

Some  GRMHD simulations produce 230~GHz emission predominantly in the black hole jet or along the ``jet sheath'' \citep[e.g.,][Figure 4]{PaperV}.
Because the observed emission in these simulations is not predominantly equatorial, the inner shadow seen in the simulations presented in this paper will not be present in this scenario.
Funnel emission in \citet{PaperV} is most often seen in simulations with weak magnetic flux, while \citet{PaperVIII} shows that polarimetric EHT observations strongly prefer MAD models with strong magnetic flux for \m87.
While the MAD models in this paper and in \citet{PaperV} show 230~GHz emission that originates predominantly from the equatorial plane, a larger survey over different GRMHD images should be performed to assess how generic this behavior is with respect to different simulation parameters and electron heating/acceleration models.  

Even in the \citet{Chael_19} MAD models explored here, in which the 230~GHz emission is predominantly equatorial, nonzero brightness from the forward jet adds a finite ``floor'' to the inner shadow (\autoref{fig:Slices}).
While this emission is dim at 230~GHz and even 86~GHz in these simulations, at even lower frequencies, the forward jet becomes optically thick and eventually obscures the equatorial emission, and hence the inner shadow.
Furthermore, the precise size of the observed brightness depression in these models depends on where the foreground emission becomes as bright as emission from the equatorial plane, which falls off rapidly due to the increasing gravitational redshift incurred as the emission radius approaches the horizon.
As a result, in these models, the area of the observed brightness depression is larger than the inner shadow of the equatorial emission.
By contrast, emission from a thick disk without forward jet emission could produce a central brightness depression with a slightly smaller area than the equatorial inner shadow, as the emission region intersects with the event horizon at a higher latitude above the equatorial plane.

The GRMHD simulation considered in this paper assumes that the plasma angular momentum is aligned with the black hole spin; however, alignment may not be generic in black hole accretion flows.
Relatively few ``tilted'' or misaligned disk GRMHD simulations have been conducted \citep[e.g.,][]{DexFrag,White19,Chaterjee20}.
Both \citet{White19} and \citet{Chaterjee20} found that in misaligned simulations, new image features can emerge due to shocks and gravitational lensing that change the appearance of the 230~GHz image relative to the images seen from aligned-disk simulations.
In particular, tilted simulations can show peak brightness contours that are farther from the black hole (and at shifted azimuthal angles) compared to what is typically seen in aligned simulations.
Because our definition of the inner shadow in this paper requires equatorial emission, the feature as discussed in this paper would not be visible if the accretion disk feeding the black hole is tilted.
However, if the emission extends to the horizon, a similar inner shadow---corresponding to the intersection of the event horizon with an inclined plane---may be present in images from tilted disks.
\citet{Chaterjee20} show time-averaged simulation images at several observing frequencies from their tilted-disk MAD simulations (their figure 6).
In all of these images from \citet{Chaterjee20}, there is a prominent central brightness depression inside the photon ring that most likely marks the event horizon image from emission concentrated in an inclined plane.
In a forthcoming work, we will investigate the visibility of the lensed horizon image in tilted disk simulations, and explore the dependence of the shape and size of the horizon image when varying the orientation $\theta$ of the emission plane.  

While we have focused on the Kerr metric, exotic compact objects with horizons will also produce an analogous inner shadow.
For instance, \citet{Mizuno_2018} present GRMHD simulations of a dilaton black hole, which exhibits a prominent inner shadow (their figure 2).
The parametrized non-Kerr metrics explored in \citet{Medeiros20} all have well-defined horizons and so will likely display an inner shadow feature when illuminated by emitting plasma in the equatorial plane.
Thus, observing an inner shadow feature from \m87 or \sgra would not constrain the metric to be Kerr; nevertheless, if the properties of the emission region are well-constrained, then the relationship between the inner shadow and the photon ring can be used as a null hypothesis test of the Kerr metric.
For instance, in Kerr, an inner shadow must have diameter $0.45<\bar{r}_{\rm h}/\bar{r}_{\rm c}<0.7$ (see \autoref{fig:RelativeRadius}).
Non-Kerr metrics such as those explored in \citet{Medeiros20} may show different relationships between the inner shadow and critical curve sizes and shapes that depend on the deviation parameters, enabling these parameters to be constrained in fits to future observations.
In addition, horizonless compact objects may also exhibit a feature analogous to the inner shadow if their interior is evacuated \citep[see, e.g.,][]{Vincent_2016,Olivares_2020}; however, in this case, the properties of the inner shadow will be primarily determined by the structure of the emission region rather than the gravitational lensing of the compact object.

\section{Conclusions}
\label{sec:Conclusions}

In this paper, we have examined the appearance of the direct image of a Kerr black hole's equatorial event horizon. 
This lensed feature always lies within the critical curve and is sensitive to the black hole spin and viewing inclination \citep{Takahashi,DN20b}.
Using GRMHD and analytic models of the submillimeter emission from \m87, we have shown that:
\begin{itemize}
\item The direct ($n=0$) lensed image of the equatorial event horizon marks the boundary of an ``inner shadow'' that should be observable in future images of \m87 if the submillimeter or radio emission is both equatorial and extends to the horizon.
\item The radiative GRMHD simulations of \m87 from \citet{Chael_19} have emissivity profiles that extend to the horizon and sub-Keplerian flows that result in a relatively weak redshift near the horizon. The ``inner shadow'' is prominent in images of these simulations.
\item Analytic equatorial emission models show some of the main features we see in time-averaged images from these GRMHD simulations, including the photon ring structure and the inner shadow. 
\item The ngEHT should have the dynamic range necessary to observe the inner shadow in \m87, if it is present.
This feature could also be visible at other frequencies, even if the optical depth of the accretion disk is high at those frequencies (as it is in the GRMHD simulation we consider at 86~GHz).  
\item The radius and centroid offset of the direct lensed equatorial horizon image can be used to measure the black hole spin and inclination, and to break degeneracies in estimating both the black hole mass and spin from only one image feature.
\item The presence and observability of this feature in \m87 is contingent on the emission being predominantly equatorial and extending to the event horizon.
If instead the accretion disk is tilted, then there may be an analogous low-brightness feature corresponding to an image of the event horizon's intersection with an inclined plane.
Non-Kerr spacetimes may produce an inner shadow with a different size relative to the photon ring than predicted in Kerr, potentially enabling tests of the spacetime using both features.
\end{itemize}

\acknowledgments

The authors thank Pierre Christian for serving as the EHT collaboration internal referee for this paper; his comments significantly improved the manuscript. 
AC is supported by Hubble Fellowship grant HST-HF2-51431.001-A awarded by the Space Telescope Science Institute, which is operated by the Association of Universities for Research in Astronomy, Inc., for NASA, under contract NAS5-26555.
MDJ was supported by the National Science Foundation (AST-1716536, AST-1935980) and the Gordon and Betty Moore Foundation (GBMF-5278).
AL gratefully acknowledges support from the Jacob Goldfield Foundation and from Will and Kacie Snellings.
This work used the Extreme Science and Engineering Discovery Environment (XSEDE), supported by NSF grant ACI-1548562. XSEDE Stampede2 resource at TACC was allocated through grant TG-AST190053 and TG-AST080026N.

\appendix

\section{Analytic ray tracing in Kerr}
\label{app:RayTracing}

In this Appendix, we review the analytic ray tracing method we use in computing the shape of the lensed horizon image in \autoref{sec:LensedImages} and the emission from the analytic equatorial model used in \autoref{sec:EmissionModel}.
Our method is taken directly from \citet{GrallaLupsasca2020,GrallaLupsasca2020b}, and we only summarize the most important steps here.
We restrict our attention to geodesics with positive Carter constant $\eta>0$; the region of ``vortical'' geodesics with $\eta<0$ is always interior to the lensed equatorial horizon curve and thus within the inner shadow \citep{GrallaLupsasca2020}.

To solve the Kerr null geodesic equation \eqref{eq:NullGeodesics}, we parameterize the geodesic in terms of the Mino time $\tau$, such that
\begin{align}
	\frac{dx^\mu}{d\tau}=\frac{\Sigma}{E}k^\mu.
\end{align} 
The total Mino time $\tau$ elapsed between emission from a source at $(r_{\rm s},\theta_{\rm s})$ and detection by an observer at $(r_{\rm o},\theta_{\rm o})$ is
\begin{align}
	\label{eq:Mino}
	\tau=I_r
	=G_\theta,
\end{align}
where
\begin{align}
	I_r=\int_{r_{\rm s}}^{r_{\rm o}}\frac{\ed r}{\pm\sqrt{\mathcal{R}(r)}},\qquad
	G_\theta=\int_{\theta_{\rm s}}^{\theta_{\rm o}}\frac{\ed\theta}{\pm\sqrt{\Theta(\theta)}}.
\end{align}
The above integrals are along the photon trajectory, which can oscillate in both $r$ and $\theta$; the directions in $r$ and $\theta$ correspond to the $\pm$ factors in the denominator of the integrands.
In particular, the geodesic oscillates in $\theta$ between two turning points
\begin{align}
	\label{eq:TurningPoints}
	\theta_\pm=\arccos\pa{\mp\sqrt{u_+}},
\end{align}
where
\begin{align}
	u_\pm=\frac{1}{2}\pa{1-\frac{\eta+\lambda^2}{a^2}}\pm\frac{1}{2}\sqrt{\pa{1-\frac{\eta+\lambda^2}{a^2}}^2+\frac{4\eta}{a^2}}.
\end{align}

A geodesic shot back from a location $(\alpha,\beta)$ in the image plane with corresponding conserved quantities $(\lambda,\eta)$ (\autoref{eq:ConservedQuantities}) intersects the black hole's equatorial plane ($\theta=\pi/2$) for the $(n+1)^\text{th}$ time when the Mino time is \citep[][Equation 81]{GrallaLupsasca2020}
\begin{align}
	\label{eq:EquatorialGtheta}
	G_\theta=\frac{2mK-\sign(\beta)F_{\rm o}}{\sqrt{-u_-a^2}},
\end{align}
where
\begin{align}
	\label{eq:mn}
	m=n+H(\beta) 
\end{align}
is the number of turning points in $\theta$ that generate the image of order $n$.\footnote{
Note that \citet{GrallaLupsasca2020} use $\bar{m}$ to refer to the image order we call $n$, while we both use $m$ to refer to the number of angular turning points.
}
Here, $H(\beta)$ denotes the Heaviside step function; \autoref{eq:mn} indicates, for example, that photons emitted on the far side of the equatorial plane ($\beta>0$) must make one reversal in $\theta$ before reaching the observer in the direct, $n=0$ image, while direct photons emitted from the near side of the disk ($\beta<0$) make no reversals in $\theta$.
The $K$ and $F_{\rm o}$ factors in \autoref{eq:EquatorialGtheta} are given by elliptic integrals of the first kind $F(u|k)$:
\begin{align}
	K=F\pa{\left.\frac{\pi}{2}\right|\frac{u_+}{u_-}},\qquad
	F_0=F\pa{\left.\arcsin\pa{\frac{\cos{\theta_{\rm o}}}{\sqrt{u_+}}}\right|\frac{u_+}{u_-}}.
\end{align}

Given $G_\theta$ for the $(n+1)^\text{th}$ equatorial crossing from \autoref{eq:EquatorialGtheta}, we need to apply \autoref{eq:Mino} and find the equatorial radius $r_{\rm eq}(I_r)$ for which $I_r=G_\theta$.
The inversion $r_{\rm eq}(I_r)$ depends on the roots ${r_1,r_2,r_3,r_4}$ of the radial potential $\mathcal{R}(r)$ (\autoref{eq:RadialPotential}).
The character (real or complex) and number of unique roots is different in different regions of the $(\lambda,\eta)$ plane \citep[e.g.,][Figure 2]{GrallaLupsasca2020b}.
Nonetheless, \citet{GrallaLupsasca2020b} provide a unified inversion formula $r_{\rm eq}(I_r)$ that holds in all cases (their equation B119): 
\begin{align}
	r_{\rm eq}(I_r)=\frac{r_4r_{31}-r_3r_{41}\sn^2\pa{\left.\frac{1}{2}\sqrt{r_{31}r_{42}}I_r-\mathcal{F}_o\right|k}}{r_{31}-r_{41}\sn^2\pa{\left.\frac{1}{2}\sqrt{r_{31}r_{42}}I_r-\mathcal{F}_o\right|k}},
\end{align}
where $r_{ij}=r_i-r_j$ for $r_i$ in the set of (real or complex) radial roots $\cu{r_1,r_2,r_3,r_4}$, $\sn(u|k)$ is the Jacobi elliptic sine function, and
\begin{align}
	\label{eq:rIr}
	\mathcal{F}_o=F\pa{\left.\arcsin\sqrt{\frac{r_{31}}{r_{41}}}\right|k},\qquad
	k=\frac{r_{32}r_{41}}{r_{31}r_{42}}.
\end{align}
Note that, in practice, a numerical implementation of \autoref{eq:rIr} may give a complex result in certain regimes where some of the radial roots $\cu{r_1,r_2,r_3,r_4}$ are complex.
In such cases, it is more useful to use the formulae in \citet{GrallaLupsasca2020b}, Appendix B, which give manifestly real expressions for $r_{\rm eq}(I_r)$ in the different regimes of the radial roots' behavior. 

For completeness, the radial roots are computed with the following expressions \citep[][Section IV.A]{GrallaLupsasca2020b}: 
\begin{align}
	r_1=-z-\sqrt{-\frac{\mathcal{A}}{2}-z^2+\frac{\mathcal{B}}{4z}},\quad
	r_2=-z+\sqrt{-\frac{\mathcal{A}}{2}-z^2+\frac{\mathcal{B}}{4z}},\quad
	r_3=z-\sqrt{-\frac{\mathcal{A}}{2}-z^2-\frac{\mathcal{B}}{4z}},\quad
	r_4=z+\sqrt{-\frac{\mathcal{A}}{2}-z^2-\frac{\mathcal{B}}{4z}},
\end{align}
where
\begin{subequations}
\begin{gather}
	\mathcal{A}=a^2-\eta-\lambda^2,\qquad
	\mathcal{B}=2M\br{\eta+(\lambda-a)^2},\qquad
	\mathcal{C}=-a^2\eta,\\
	\mathcal{P}=-\frac{\mathcal{A}^2}{12}-\mathcal{C},\qquad
	\mathcal{Q}=-\frac{\mathcal{A}}{3}\br{\pa{\frac{\mathcal{A}}{6}}^2-\mathcal{C}}-\frac{\mathcal{B}^2}{8},\\
	\omega_\pm=\pa{-\frac{\mathcal{Q}}{2}\pm\sqrt{\frac{\mathcal{P}^3}{27}+\frac{\mathcal{Q}^2}{4}}}^{1/3},\qquad
	z=\sqrt{\frac{\omega_++\omega_--\mathcal{A}/3}{2}}.
\end{gather}
\end{subequations}

Note that \autoref{eq:rIr} only applies where the Mino time $\tau=I_r$ is less than its maximum value for a given geodesic: $0<I_r<I_r^{\rm total}$.
The total Mino time $I_r^{\rm total}$ elapsed along a trajectory is given by \citep[][Equation 29]{GrallaLupsasca2020}:
\begin{align}
	\label{eq:TotalMinoTime}
	I_r^{\rm total}=
	\begin{cases}
		\displaystyle2\int_{r_4}^\infty\frac{\ed r}{\sqrt{\mathcal{R}(r)}}
		&\text{if }r_+<r_4\in\mathbb{R},\vspace{2pt}\\
		\displaystyle\int_{r_+}^\infty\frac{\ed r}{\sqrt{\mathcal{R}(r)}}
		&\text{otherwise}.
	\end{cases}
\end{align}
The first case of \autoref{eq:TotalMinoTime} corresponds to light rays appearing outside the critical curve, which encounter a radial turning point at $r_4$ outside of the event horizon, while the second case corresponds to light rays appearing inside the critical curve, which terminate at the horizon.
Appendix B of \citet{GrallaLupsasca2020b} provides expressions for $I_r^{\rm total}$ in terms of elliptic integrals for the different regimes of the radial roots' behavior.

Combining \autoref{eq:EquatorialGtheta} and \autoref{eq:TotalMinoTime}, we can obtain an expression for $N_{\rm max}(\alpha,\beta)$, the maximal number of times a geodesic terminating at the image-plane position $(\alpha,\beta)$ crosses the equatorial plane: 
\begin{align}
	\label{eq:Nmax}
	N_{\rm max}=\left\lfloor\frac{I_r^{\rm total}\sqrt{-u_-a^2}+\sign(\beta)F_{\rm o}}{2K}\right\rfloor-H(\beta)+1.
\end{align}
Recall that $N_{\rm max}=0$ inside the direct lensed image of the equatorial horizon; geodesics inside this region do not cross the equatorial plane even once.

\section{Image moments}
\label{app:Moments}

In this Appendix, we define the second moment used to characterize both the lensed horizon and critical curve in \autoref{sec:GeometricDescription}.
For a closed convex curve $\rho(\varphi)$ enclosing the origin in the $(\alpha,\beta)$ coordinate system, the zeroth image moment is the area $A$: 
\begin{align}
	A=\frac{1}{2}\int\rho^2(\varphi)\ed\varphi.
\end{align}

The first moment is the centroid vector $\boldsymbol\mu=(\mu_\alpha,\mu_\beta)$:
\begin{subequations}
\begin{align}
	\mu_\alpha&=\frac{1}{3A}\int\rho^3(\varphi)\cos{\varphi}\ed\varphi,\\
	\mu_\beta&=\frac{1}{3A}\int\rho^3(\varphi)\sin{\varphi}\ed\varphi.
\end{align}
\end{subequations}

The second central moment is equivalent to the covariance matrix, or moment of inertia tensor $\boldsymbol\Sigma$.
In the $(\alpha,\beta)$ coordinate system, it has components $\Sigma_{\alpha\alpha}$, $\Sigma_{\beta\beta}$, and $\Sigma_{\alpha\beta}=\Sigma_{\beta \alpha}$, where
\begin{subequations}
\begin{align}
	\Sigma_{\alpha\alpha}&=\frac{1}{4A}\int\rho^4(\varphi)\cos^2{\varphi}\ed\varphi-\mu_\alpha^2,\\
	\Sigma_{\beta\beta}&=\frac{1}{4A}\int\rho^4(\varphi)\sin^2{\varphi}\ed\varphi-\mu_\beta^2,\\
	\Sigma_{\alpha\beta}&=\frac{1}{4A}\int\rho^4(\varphi)\cos{\varphi}\sin{\varphi}\ed\varphi-\mu_\alpha\mu_\beta.
\end{align}
\end{subequations}

We can diagonalize $\boldsymbol\Sigma$ to find the lengths of the principal axes $a,b$ and their orientation angle $\chi$ relative to the $+\alpha$ axis.
Explicitly:
\begin{subequations}
\begin{align}
	a&=\sqrt{2\pa{\Sigma_{\alpha\alpha}+\Sigma_{\beta\beta}+D_\Sigma}},\\ 
	b&=\sqrt{2\pa{\Sigma_{\alpha\alpha}+\Sigma_{\beta\beta}-D_\Sigma}},\\ 
	\chi&=\frac{1}{2}\arcsin\pa{\frac{2\Sigma_{\alpha \beta}}{D_\Sigma}},
\end{align}
\end{subequations}
where
\begin{align}
	D_\Sigma=\sqrt{\pa{\Sigma_{\alpha\alpha}-\Sigma_{\beta\beta}}^2+4\Sigma_{\alpha\beta}^2}.
\end{align}
The mean radius and eccentricity are then defined by \autoref{eq:MeanRadius} in the same manner as for an ellipse with semimajor axis $a$ and semiminor axis $b$. That is, the mean radius is $\bar{r}=\sqrt{\pa{a^2+b^2}/2}$ and the eccentricity is $e=\sqrt{1-b^2/a^2}$.

\begin{figure}[t]
\centering
\includegraphics[width=0.4\textwidth]{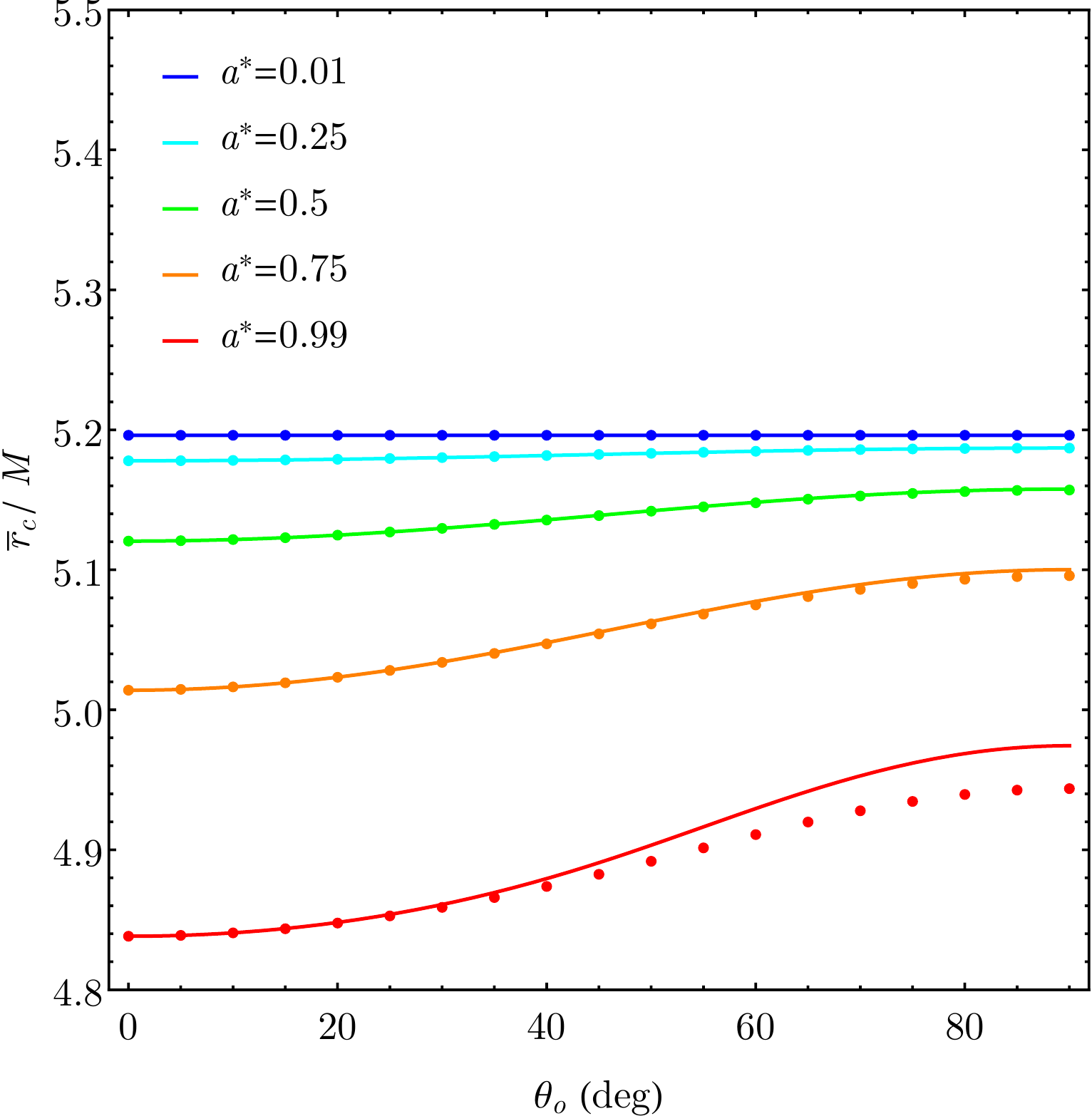}
\caption{
Comparison of the average radius of the critical curve $\bar{r}_\mathrm{c}$ computed using the second image moment following the method described in this Appendix (solid lines) with its value computed using the method introduced in \citet{JohannsenPsaltis} (circular markers).
The radii are shown as a function of observer inclination $\theta_{\rm o}$ for values of black hole spin $\spin=(0.01,0.25,0.5,0.75,0.99)$.
The two methods produce results that agree to within one percent for all values of black hole spin and observer inclination.
}
\label{fig:RadiusComparison}
\end{figure}

This definition for the average radius of the critical curve in terms of the image second moment differs from the definition introduced in \citet{JohannsenPsaltis}, Equation 4.
In \autoref{fig:RadiusComparison}, we compare the average radii of the critical curve for different black hole spins as a function of inclination angle as determined by these two methods.
The second moment method used here for calculating the average radius of the critical curve agrees with the results of the \citet{JohannsenPsaltis} method within one percent for all values of spin and inclination, with the only noticeable discrepancies occurring when both the spin and inclination are large (and the critical curve is at its least-circular).

\newpage

\bibliography{InnerShadow.bib}{}

\begin{thebibliography}{}
\expandafter\ifx\csname natexlab\endcsname\relax\def\natexlab#1{#1}\fi
\providecommand{\url}[1]{\href{#1}{#1}}
\providecommand{\dodoi}[1]{doi:~\href{http://doi.org/#1}{\nolinkurl{#1}}}
\providecommand{\doeprint}[1]{\href{http://ascl.net/#1}{\nolinkurl{http://ascl.net/#1}}}
\providecommand{\doarXiv}[1]{\href{https://arxiv.org/abs/#1}{\nolinkurl{https://arxiv.org/abs/#1}}}

\bibitem[{{Abramowski} {et~al.}(2012){Abramowski}, {Acero}, {Aharonian},
  {Akhperjanian}, {Anton}, {Balzer}, {Barnacka}, {Barres de Almeida},
  {Becherini}, {Becker}, \& et~al.}]{Abramowski2012}
{Abramowski}, A., {Acero}, F., {Aharonian}, F., {et~al.} 2012, \apj, 746, 151,
  \dodoi{10.1088/0004-637X/746/2/151}

\bibitem[{{Akiyama} {et~al.}(2015){Akiyama}, {Lu}, {Fish}, {Doeleman},
  {Broderick}, {Dexter}, {Hada}, {Kino}, {Nagai}, {Honma}, {Johnson}, {Algaba},
  {et~al.}}]{Akiyama15}
{Akiyama}, K., {Lu}, R.-S., {Fish}, V.~L., {et~al.} 2015, \apj, 807, 150,
  \dodoi{10.1088/0004-637X/807/2/150}

\bibitem[{Bardeen(1973)}]{Bardeen1973a}
Bardeen, J.~M. 1973, in Black Holes, ed. {C. DeWitt} \& {B. S. DeWitt} (New
  York: Gordon \& Breach), 215--239

\bibitem[{{Beckwith} \& {Done}(2005)}]{Beckwith}
{Beckwith}, K., \& {Done}, C. 2005, \mnras, 359, 1217,
  \dodoi{10.1111/j.1365-2966.2005.08980.x}

\bibitem[{{Beloborodov}(2002)}]{Beloborodov2002}
{Beloborodov}, A.~M. 2002, \apjl, 566, L85, \dodoi{10.1086/339511}

\bibitem[{{Broderick} \& {Loeb}(2006)}]{Broderick06}
{Broderick}, A.~E., \& {Loeb}, A. 2006, \apjl, 636, L109,
  \dodoi{10.1086/500008}

\bibitem[{{Broderick} {et~al.}(2020){Broderick}, {Pesce}, {Tiede}, {Pu}, \&
  {Gold}}]{Broderick20}
{Broderick}, A.~E., {Pesce}, D.~W., {Tiede}, P., {Pu}, H.-Y., \& {Gold}, R.
  2020, \apj, 898, 9, \dodoi{10.3847/1538-4357/ab9c1f}

\bibitem[{{Broderick} {et~al.}(2021){Broderick}, {Tiede}, {Pesce}, \&
  {Gold}}]{Broderick21}
{Broderick}, A.~E., {Tiede}, P., {Pesce}, D.~W., \& {Gold}, R. 2021, arXiv
  e-prints, arXiv:2105.09962.
\newblock \doarXiv{2105.09962}

\bibitem[{{Bronzwaer} {et~al.}(2021){Bronzwaer}, {Davelaar}, {Younsi},
  {Mo{\'s}cibrodzka}, {Olivares}, {Mizuno}, {Vos}, \&
  {Falcke}}]{Bronzwaer_2021}
{Bronzwaer}, T., {Davelaar}, J., {Younsi}, Z., {et~al.} 2021, \mnras, 501,
  4722, \dodoi{10.1093/mnras/staa3430}

\bibitem[{{Chael} {et~al.}(2018){Chael}, {Johnson}, {Bouman}, {Blackburn},
  {Akiyama}, \& {Narayan}}]{Chael_18}
{Chael}, A., {Johnson}, M.~D., {Bouman}, K.~L., {et~al.} 2018, \apj, 857, 23,
  \dodoi{10.3847/1538-4357/aab6a8}

\bibitem[{{Chael} {et~al.}(2019){Chael}, {Narayan}, \& {Johnson}}]{Chael_19}
{Chael}, A., {Narayan}, R., \& {Johnson}, M.~D. 2019, \mnras, 486, 2873,
  \dodoi{10.1093/mnras/stz988}

\bibitem[{{Chatterjee} {et~al.}(2020){Chatterjee}, {Younsi}, {Liska},
  {Tchekhovskoy}, {Markoff}, {Yoon}, {van Eijnatten}, {Hesp}, {Ingram}, \& {van
  der Klis}}]{Chaterjee20}
{Chatterjee}, K., {Younsi}, Z., {Liska}, M., {et~al.} 2020, \mnras, 499, 362,
  \dodoi{10.1093/mnras/staa2718}

\bibitem[{{Cunha} \& {Herdeiro}(2018)}]{Cunha_2018}
{Cunha}, P. V.~P., \& {Herdeiro}, C. A.~R. 2018, General Relativity and
  Gravitation, 50, 42, \dodoi{10.1007/s10714-018-2361-9}

\bibitem[{{Cunningham}(1975)}]{Cunningham_75}
{Cunningham}, C.~T. 1975, \apj, 202, 788, \dodoi{10.1086/154033}

\bibitem[{de~Vries(2003)}]{devries_2003}
de~Vries, A. 2003, Jahresschrift der Bochumer Interdisziplinären Gesellschaft.
\newblock \url{http://haegar.fh-swf.de/publikationen/pascal.pdf}

\bibitem[{{Dexter}(2016)}]{Dexter16}
{Dexter}, J. 2016, MNRAS, 462, 115, \dodoi{10.1093/mnras/stw1526}

\bibitem[{{Dexter} \& {Fragile}(2013)}]{DexFrag}
{Dexter}, J., \& {Fragile}, P. 2013, \mnras, 432, 2252,
  \dodoi{10.1093/mnras/stt583}

\bibitem[{{Dexter} {et~al.}(2012){Dexter}, {McKinney}, \& {Agol}}]{Dexter2012}
{Dexter}, J., {McKinney}, J.~C., \& {Agol}, E. 2012, \mnras, 421, 1517,
  \dodoi{10.1111/j.1365-2966.2012.20409.x}

\bibitem[{{Doeleman} {et~al.}(2019){Doeleman}, {Blackburn}, {Dexter}, {Gomez},
  {Johnson}, {Palumbo}, {Weintroub}, {Farah}, {Fish}, {Loinard}, {Lonsdale},
  {Narayanan}, {Patel}, {Pesce}, {Raymond}, {Tilanus}, {Wielgus}, {Akiyama},
  {Bower}, {Broderick}, {Deane}, {Fromm}, {Gammie}, {Gold}, {Janssen},
  {Kawashima}, {Krichbaum}, {Marrone}, {Matthews}, {Mizuno}, {Rezzolla},
  {Roelofs}, {Ros}, {Savolainen}, {Yuan}, {Zhao}, {Blackburn}, {Doeleman},
  {Dexter}, {Gomez}, {Johnson}, {Palumbo}, {Weintroub}, {Farah}, {Fish},
  {Loinard}, {Lonsdale}, {Narayanan}, {Patel}, {Pesce}, {Raymond}, {Tilanus},
  {Wielgus}, {Akiyama}, {Bower}, {Broderick}, {Deane}, {Fromm}, {Gammie},
  {Gold}, {Janssen}, {Kawashima}, {Krichbaum}, {Marrone}, {Matthews}, {Mizuno},
  {Rezzolla}, {Roelofs}, {Ros}, {Savolainen}, {Yuan}, \&
  {Zhao}}]{Astro2020Ground}
{Doeleman}, S., {Blackburn}, L., {Dexter}, J., {et~al.} 2019, in Bulletin of
  the American Astronomical Society, Vol.~51, 256.
\newblock \doarXiv{1909.01411}

\bibitem[{{Dokuchaev} \& {Nazarova}(2019)}]{DN19b}
{Dokuchaev}, V.~I., \& {Nazarova}, N.~O. 2019, Soviet Journal of Experimental
  and Theoretical Physics, 128, 578, \dodoi{10.1134/S1063776119030026}

\bibitem[{{Dokuchaev} \& {Nazarova}(2020{\natexlab{a}})}]{DN20}
---. 2020{\natexlab{a}}, Universe, 6, 154, \dodoi{10.3390/universe6090154}

\bibitem[{{Dokuchaev} \& {Nazarova}(2020{\natexlab{b}})}]{DN20b}
---. 2020{\natexlab{b}}, Physics Uspekhi, 63, 583,
  \dodoi{10.3367/UFNe.2020.01.038717}

\bibitem[{{EHT MWL Science Working Group} {et~al.}(2021){EHT MWL Science
  Working Group}, {Algaba}, {Anczarski}, {Asada}, {Balokovi{\'c}}, {Chandra},
  {Cui}, {Falcone}, {Giroletti}, {Goddi}, {Hada}, {Haggard}, {Jorstad}, {Kaur},
  {Kawashima}, {Keating}, {Kim}, {Kino}, {Komossa}, {Kravchenko}, {Krichbaum},
  {Lee}, {Lu}, {Lucchini}, {Markoff}, {Neilsen}, {Nowak}, {Park}, {Principe},
  {Ramakrishnan}, {Reynolds}, {Sasada}, {Savchenko}, {Williamson}, {Event
  Horizon Telescope Collaboration}, {Akiyama}, {Alberdi}, {Alef}, {Anantua},
  {Azulay}, {Baczko}, {Ball}, {Barrett}, {Bintley}, {Benson}, {Blackburn},
  {Blundell}, {Boland}, {Bouman}, {Bower}, {Boyce}, {Bremer}, {Brinkerink},
  {Brissenden}, {Britzen}, {Broderick}, {Broguiere}, {Bronzwaer}, {Byun},
  {Carlstrom}, {Chael}, {Chan}, {Chatterjee}, {Chatterjee}, {Chen}, {Chen},
  {Chesler}, {Cho}, {Christian}, {Conway}, {Cordes}, {Crawford}, {Crew},
  {Cruz-Osorio}, {Davelaar}, {de Laurentis}, {Deane}, {Dempsey}, {Desvignes},
  {Dexter}, {Doeleman}, {Eatough}, {Falcke}, {Farah}, {Fish}, {Fomalont},
  {Ford}, {Fraga-Encinas}, {Friberg}, {Fromm}, {Fuentes}, {Galison}, {Gammie},
  {Garc{\'\i}a}, {Gentaz}, {Georgiev}, {Gold}, {G{\'o}mez}, {G{\'o}mez-Ruiz},
  {Gu}, {Gurwell}, {Hecht}, {Hesper}, {Ho}, {Ho}, {Honma}, {Huang}, {Huang},
  {Hughes}, {Ikeda}, {Inoue}, {Issaoun}, {James}, {Jannuzi}, {Janssen},
  {Jeter}, {Jiang}, {Jim{\'e}nez-Rosales}, {Johnson}, {Jung}, {Karami},
  {Karuppusamy}, {Kettenis}, {Kim}, {Kim}, {Kim}, {Koay}, {Kofuji}, {Koch},
  {Koyama}, {Kramer}, {Kramer}, {Kuo}, {Lauer}, {Levis}, {Li}, {Li},
  {Lindqvist}, {Lico}, {Lindahl}, {Liu}, {Liu}, {Liuzzo}, {Lo}, {Lobanov},
  {Loinard}, {Lonsdale}, {MacDonald}, {Mao}, {Marchili}, {Marrone}, {Marscher},
  {Mart{\'\i}-Vidal}, {Matsushita}, {Matthews}, {Medeiros}, {Menten}, {Mizuno},
  {Mizuno}, {Moran}, {Moriyama}, {Moscibrodzka}, {M{\"u}ller}, {Musoke},
  {Mej{\'\i}as}, {Nagai}, {Nagar}, {Nakamura}, {Narayan}, {Narayanan},
  {Natarajan}, {Nathanail}, {Neri}, {Ni}, {Noutsos}, {Okino}, {Olivares},
  {Ortiz-Le{\'o}n}, {Oyama}, {{\"O}zel}, {Palumbo}, {Patel}, {Pen}, {Pesce},
  {Pi{\'e}tu}, {Plambeck}, {Popstefanija}, {Porth}, {P{\"o}tzl}, {Prather},
  {Preciado-L{\'o}pez}, {Psaltis}, {Pu}, {Rao}, {Rawlings}, {Raymond},
  {Rezzolla}, {Ricarte}, {Ripperda}, {Roelofs}, {Rogers}, {Ros}, {Rose},
  {Roshanineshat}, {Rottmann}, {Roy}, {Ruszczyk}, {Rygl}, {S{\'a}nchez},
  {S{\'a}nchez-Arguelles}, {Savolainen}, {Schloerb}, {Schuster}, {Shao},
  {Shen}, {Small}, {Sohn}, {Soohoo}, {Sun}, {Tazaki}, {Tetarenko}, {Tiede},
  {Tilanus}, {Titus}, {Toma}, {Torne}, {Trent}, {Traianou}, {Trippe}, {van
  Bemmel}, {van Langevelde}, {van Rossum}, {Wagner}, {Ward-Thompson}, {Wardle},
  {Weintroub}, {Wex}, {Wharton}, {Wielgus}, {Wong}, {Wu}, {Yoon}, {Young},
  {Young}, {Younsi}, {Yuan}, {Yuan}, {Zensus}, {Zhao}, {Zhao}, {Fermi Large
  Area Telescope Collaboration}, {Principe}, {Giroletti}, {D'Ammando},
  {Orienti}, {H.~E.~S.~S. Collaboration}, {Abdalla}, {Adam}, {Aharonian},
  {Benkhali}, {Ang{\"u}ner}, {Arcaro}, {Armand}, {Armstrong}, {Ashkar},
  {Backes}, {Baghmanyan}, {Barbosa Martins}, {Barnacka}, {Barnard},
  {Becherini}, {Berge}, {Bernl{\"o}hr}, {Bi}, {B{\"o}ttcher}, {Boisson},
  {Bolmont}, {de Lavergne}, {Breuhaus}, {Brun}, {Brun}, {Bryan}, {B{\"u}chele},
  {Bulik}, {Bylund}, {Caroff}, {Carosi}, {Casanova}, {Chand}, {Chen}, {Cotter},
  {Cury{\l}o}, {Damascene Mbarubucyeye}, {Davids}, {Davies}, {Deil}, {Devin},
  {Dewilt}, {Dirson}, {Djannati-Ata{\"\i}}, {Dmytriiev}, {Donath},
  {Doroshenko}, {Duffy}, {Dyks}, {Egberts}, {Eichhorn}, {Einecke}, {Emery},
  {Ernenwein}, {Feijen}, {Fegan}, {Fiasson}, {de Clairfontaine}, {Fontaine},
  {Funk}, {F{\"u}{\ss}ling}, {Gabici}, {Gallant}, {Giavitto}, {Giunti},
  {Glawion}, {Glicenstein}, {Gottschall}, {Grondin}, {Hahn}, {Haupt},
  {Hermann}, {Hinton}, {Hofmann}, {Hoischen}, {Holch}, {Holler}, {H{\"o}rbe},
  {Horns}, {Huber}, {Jamrozy}, {Jankowsky}, {Jankowsky}, {Jardin-Blicq},
  {Joshi}, {Jung-Richardt}, {Kasai}, {Kastendieck}, {Katarzy{\'n}ski}, {Katz},
  {Khangulyan}, {Kh{\'e}lifi}, {Klepser}, {Klu{\'z}niak}, {Komin}, {Konno},
  {Kosack}, {Kostunin}, {Kreter}, {Lamanna}, {Lemi{\`e}re}, {Lemoine-Goumard},
  {Lenain}, {Levy}, {Lohse}, {Lypova}, {Mackey}, {Majumdar}, {Malyshev},
  {Malyshev}, {Marandon}, {Marchegiani}, {Marcowith}, {Mares},
  {Mart{\'\i}-Devesa}, {Marx}, {Maurin}, {Meintjes}, {Meyer}, {Moderski},
  {Mohamed}, {Mohrmann}, {Montanari}, {Moore}, {Morris}, {Moulin}, {Muller},
  {Murach}, {Nakashima}, {Nayerhoda}, {de Naurois}, {Ndiyavala},
  {Niederwanger}, {Niemiec}, {Oakes}, {O'Brien}, {Odaka}, {Ohm},
  {Olivera-Nieto}, {de Ona Wilhelmi}, {Ostrowski}, {Panter}, {Panny},
  {Parsons}, {Peron}, {Peyaud}, {Piel}, {Pita}, {Poireau}, {Noel}, {Prokhorov},
  {Prokoph}, {P{\"u}hlhofer}, {Punch}, {Quirrenbach}, {Rauth}, {Reichherzer},
  {Reimer}, {Reimer}, {Remy}, {Renaud}, {Rieger}, {Rinchiuso}, {Romoli},
  {Rowell}, {Rudak}, {Ruiz-Velasco}, {Sahakian}, {Sailer}, {Sanchez},
  {Santangelo}, {Sasaki}, {Scalici}, {Schutte}, {Schwanke}, {Schwemmer},
  {Seglar-Arroyo}, {Senniappan}, {Seyffert}, {Shafi}, {Shiningayamwe},
  {Simoni}, {Sinha}, {Sol}, {Specovius}, {Spencer}, {Spir-Jacob}, {Stawarz},
  {Sun}, {Steenkamp}, {Stegmann}, {Steinmassl}, {Steppa}, {Takahashi},
  {Tavernier}, {Taylor}, {Terrier}, {Tiziani}, {Tluczykont}, {Tomankova},
  {Trichard}, {Tsirou}, {Tuffs}, {Uchiyama}, {van der Walt}, {van Eldik}, {van
  Rensburg}, {van Soelen}, {Vasileiadis}, {Veh}, {Venter}, {Vincent}, {Vink},
  {V{\"o}lk}, {Vuillaume}, {Wadiasingh}, {Wagner}, {Watson}, {Werner}, {White},
  {Wierzcholska}, {Wong}, {Yusafzai}, {Zacharias}, {Zanin}, {Zargaryan},
  {Zdziarski}, {Zech}, {Zhu}, {Zorn}, {Zouari}, {{\.Z}ywucka}, {MAGIC
  Collaboration}, {Acciari}, {Ansoldi}, {Antonelli}, {Engels}, {Artero},
  {Asano}, {Baack}, {Babi{\'c}}, {Baquero}, {de Almeida}, {Barrio}, {Becerra
  Gonz{\'a}lez}, {Bednarek}, {Bellizzi}, {Bernardini}, {Bernardos}, {Berti},
  {Besenrieder}, {Bhattacharyya}, {Bigongiari}, {Biland}, {Blanch}, {Bonnoli},
  {Bo{\v{s}}njak}, {Busetto}, {Carosi}, {Ceribella}, {Cerruti}, {Chai},
  {Chilingarian}, {Cikota}, {Colak}, {Colombo}, {Contreras}, {Cortina},
  {Covino}, {D'Amico}, {D'Elia}, {da Vela}, {Dazzi}, {de Angelis}, {de Lotto},
  {Delfino}, {Delgado}, {Delgado Mendez}, {Depaoli}, {di Pierro}, {di Venere},
  {Do Souto Espi{\~n}eira}, {Dominis Prester}, {Donini}, {Dorner}, {Doro},
  {Elsaesser}, {Ramazani}, {Fattorini}, {Ferrara}, {Fonseca}, {Font}, {Fruck},
  {Fukami}, {Garc{\'\i}a L{\'o}pez}, {Garczarczyk}, {Gasparyan}, {Gaug},
  {Giglietto}, {Giordano}, {Gliwny}, {Godinovi{\'c}}, {Green}, {Green},
  {Hadasch}, {Hahn}, {Heckmann}, {Herrera}, {Hoang}, {Hrupec}, {H{\"u}tten},
  {Inada}, {Inoue}, {Ishio}, {Iwamura}, {Jim{\'e}nez}, {Jormanainen}, {Jouvin},
  {Kajiwara}, {Karjalainen}, {Kerszberg}, {Kobayashi}, {Kubo}, {Kushida},
  {Lamastra}, {Lelas}, {Leone}, {Lindfors}, {Lombardi}, {Longo},
  {L{\'o}pez-Coto}, {L{\'o}pez-Moya}, {L{\'o}pez-Oramas}, {Loporchio}, {Machado
  de Oliveira Fraga}, {Maggio}, {Majumdar}, {Makariev}, {Mallamaci}, {Maneva},
  {Manganaro}, {Mannheim}, {Maraschi}, {Mariotti}, {Mart{\'\i}nez}, {Mazin},
  {Menchiari}, {Mender}, {Mi{\'c}anovi{\'c}}, {Miceli}, {Miener}, {Minev},
  {Miranda}, {Mirzoyan}, {Molina}, {Moralejo}, {Morcuende}, {Moreno},
  {Moretti}, {Neustroev}, {Nigro}, {Nilsson}, {Nishijima}, {Noda}, {Nozaki},
  {Ohtani}, {Oka}, {Otero-Santos}, {Paiano}, {Palatiello}, {Paneque},
  {Paoletti}, {Paredes}, {Pavleti{\'c}}, {Pe{\~n}il}, {Perennes}, {Persic},
  {Moroni}, {Prandini}, {Priyadarshi}, {Puljak}, {Rhode}, {Rib{\'o}}, {Rico},
  {Righi}, {Rugliancich}, {Saha}, {Sahakyan}, {Saito}, {Sakurai}, {Satalecka},
  {Saturni}, {Schleicher}, {Schmidt}, {Schweizer}, {Sitarek},
  {{\v{S}}nidari{\'c}}, {Sobczynska}, {Spolon}, {Stamerra}, {Strom}, {Strzys},
  {Suda}, {Suri{\'c}}, {Takahashi}, {Tavecchio}, {Temnikov}, {Terzi{\'c}},
  {Teshima}, {Tosti}, {Truzzi}, {Tutone}, {Ubach}, {van Scherpenberg}, {Vanzo},
  {Vazquez Acosta}, {Ventura}, {Verguilov}, {Vigorito}, {Vitale}, {Vovk},
  {Will}, {Wunderlich}, {Zari{\'c}}, {VERITAS Collaboration}, {Adams},
  {Benbow}, {Brill}, {Capasso}, {Christiansen}, {Chromey}, {Daniel}, {Errando},
  {Farrell}, {Feng}, {Finley}, {Fortson}, {Furniss}, {Gent}, {Giuri}, {Hassan},
  {Hervet}, {Holder}, {Hughes}, {Humensky}, {Jin}, {Kaaret}, {Kertzman},
  {Kieda}, {Kumar}, {Lang}, {Lundy}, {Maier}, {Moriarty}, {Mukherjee}, {Nieto},
  {Nievas-Rosillo}, {O'Brien}, {Ong}, {Otte}, {Patel}, {Pfrang}, {Pohl},
  {Prado}, {Pueschel}, {Quinn}, {Ragan}, {Reynolds}, {Ribeiro}, {Richards},
  {Roache}, {Rulten}, {Ryan}, {Santander}, {Sembroski}, {Shang}, {Weinstein},
  {Williams}, {Williamson}, {Eavn Collaboration}, {Hirota}, {Cui}, {Niinuma},
  {Ro}, {Sakai}, {Sawada-Satoh}, {Wajima}, {Wang}, {Liu}, \& {Yonekura}}]{Mwl}
{EHT MWL Science Working Group}, {Algaba}, J.~C., {Anczarski}, J., {et~al.}
  2021, \apjl, 911, L11, \dodoi{10.3847/2041-8213/abef71}

\bibitem[{{Falcke} {et~al.}(2000){Falcke}, {Melia}, \& {Agol}}]{Falcke2000}
{Falcke}, H., {Melia}, F., \& {Agol}, E. 2000, \apjl, 528, L13,
  \dodoi{10.1086/312423}

\bibitem[{{Farah} {et~al.}(2020){Farah}, {Pesce}, {Johnson}, \&
  {Blackburn}}]{Farah_2020}
{Farah}, J.~R., {Pesce}, D.~W., {Johnson}, M.~D., \& {Blackburn}, L. 2020,
  \apj, 900, 77, \dodoi{10.3847/1538-4357/aba59a}

\bibitem[{{Gammie} {et~al.}(2003){Gammie}, {McKinney}, \&
  {T{\'o}th}}]{Gammie03}
{Gammie}, C.~F., {McKinney}, J.~C., \& {T{\'o}th}, G. 2003, \apj, 589, 444,
  \dodoi{10.1086/374594}

\bibitem[{{Gates} {et~al.}(2020){Gates}, {Hadar}, \& {Lupsasca}}]{Gates2020}
{Gates}, D. E.~A., {Hadar}, S., \& {Lupsasca}, A. 2020, \prd, 102, 104041,
  \dodoi{10.1103/PhysRevD.102.104041}

\bibitem[{{Gralla} {et~al.}(2019){Gralla}, {Holz}, \& {Wald}}]{GHW}
{Gralla}, S.~E., {Holz}, D.~E., \& {Wald}, R.~M. 2019, \prd, 100, 024018,
  \dodoi{10.1103/PhysRevD.100.024018}

\bibitem[{{Gralla} \& {Lupsasca}(2020{\natexlab{a}})}]{GrallaLupsasca2020}
{Gralla}, S.~E., \& {Lupsasca}, A. 2020{\natexlab{a}}, \prd, 101, 044031,
  \dodoi{10.1103/PhysRevD.101.044031}

\bibitem[{{Gralla} \& {Lupsasca}(2020{\natexlab{b}})}]{GL20}
---. 2020{\natexlab{b}}, \prd, 102, 124003, \dodoi{10.1103/PhysRevD.102.124003}

\bibitem[{{Gralla} \& {Lupsasca}(2020{\natexlab{c}})}]{GrallaLupsasca2020b}
---. 2020{\natexlab{c}}, \prd, 101, 044032, \dodoi{10.1103/PhysRevD.101.044032}

\bibitem[{{Gralla} {et~al.}(2020){Gralla}, {Lupsasca}, \& {Marrone}}]{GLM_20}
{Gralla}, S.~E., {Lupsasca}, A., \& {Marrone}, D.~P. 2020, \prd, 102, 124004,
  \dodoi{10.1103/PhysRevD.102.124004}

\bibitem[{{Gravity Collaboration} {et~al.}(2019){Gravity Collaboration},
  {Abuter}, {Amorim}, {Baub{\"o}ck}, {Berger}, {Bonnet}, {Brandner},
  {Cl{\'e}net}, {Coud{\'e} Du Foresto}, {de Zeeuw}, {Dexter}, {Duvert},
  {Eckart}, {Eisenhauer}, {F{\"o}rster Schreiber}, {Garcia}, {Gao}, {Gendron},
  {Genzel}, {Gerhard}, {Gillessen}, {Habibi}, {Haubois}, {Henning}, {Hippler},
  {Horrobin}, {Jim{\'e}nez-Rosales}, {Jocou}, {Kervella}, {Lacour},
  {Lapeyr{\`e}re}, {Le Bouquin}, {L{\'e}na}, {Ott}, {Paumard}, {Perraut},
  {Perrin}, {Pfuhl}, {Rabien}, {Rodriguez Coira}, {Rousset}, {Scheithauer},
  {Sternberg}, {Straub}, {Straubmeier}, {Sturm}, {Tacconi}, {Vincent}, {von
  Fellenberg}, {Waisberg}, {Widmann}, {Wieprecht}, {Wiezorrek}, {Woillez}, \&
  {Yazici}}]{Gravity2019}
{Gravity Collaboration}, {Abuter}, R., {Amorim}, A., {et~al.} 2019, \aap, 625,
  L10, \dodoi{10.1051/0004-6361/201935656}

\bibitem[{{Hada} {et~al.}(2016){Hada}, {Kino}, {Doi}, {Nagai}, {Honma},
  {Akiyama}, {Tazaki}, {Lico}, {Giroletti}, {Giovannini}, {Orienti}, \&
  {Hagiwara}}]{Hada2016}
{Hada}, K., {Kino}, M., {Doi}, A., {et~al.} 2016, \apj, 817, 131,
  \dodoi{10.3847/0004-637X/817/2/131}

\bibitem[{{Hadar} {et~al.}(2021){Hadar}, {Johnson}, {Lupsasca}, \&
  {Wong}}]{Hadar2020}
{Hadar}, S., {Johnson}, M.~D., {Lupsasca}, A., \& {Wong}, G.~N. 2021, \prd,
  103, 104038, \dodoi{10.1103/PhysRevD.103.104038}

\bibitem[{{Igumenshchev} {et~al.}(2003){Igumenshchev}, {Narayan}, \&
  {Abramowicz}}]{Igumenschchev2003}
{Igumenshchev}, I.~V., {Narayan}, R., \& {Abramowicz}, M.~A. 2003, \apj, 592,
  1042, \dodoi{10.1086/375769}

\bibitem[{{Johannsen} \& {Psaltis}(2010)}]{JohannsenPsaltis}
{Johannsen}, T., \& {Psaltis}, D. 2010, \apj, 718, 446,
  \dodoi{10.1088/0004-637X/718/1/446}

\bibitem[{{Johnson} {et~al.}(2019){Johnson}, {Haworth}, {Pesce}, {Palumbo},
  {Blackburn}, {Akiyama}, {Boroson}, {Bouman}, {Farah}, {Fish}, {Honma},
  {Kawashima}, {Kino}, {Raymond}, {Silver}, {Weintroub}, {Wielgus}, {Doeleman},
  {Kauffmann}, {Keating}, {Krichbaum}, {Loinard}, {Narayanan}, {Doi}, {James},
  {Marrone}, {Mizuno}, \& {Nagai}}]{Astro2020Space}
{Johnson}, M., {Haworth}, K., {Pesce}, D.~W., {et~al.} 2019, in Bulletin of the
  American Astronomical Society, Vol.~51, 235.
\newblock \doarXiv{1909.01405}

\bibitem[{{Johnson} {et~al.}(2020){Johnson}, {Lupsasca}, {Strominger}, {Wong},
  {Hadar}, {Kapec}, {Narayan}, {Chael}, {Gammie}, {Galison}, {Palumbo},
  {Doeleman}, {Blackburn}, {Wielgus}, {Pesce}, {Farah}, \&
  {Moran}}]{Johnson_Ring}
{Johnson}, M.~D., {Lupsasca}, A., {Strominger}, A., {et~al.} 2020, Science
  Advances, 6, eaaz1310, \dodoi{10.1126/sciadv.aaz1310}

\bibitem[{{Junor} {et~al.}(1999){Junor}, {Biretta}, \& {Livio}}]{Junor99}
{Junor}, W., {Biretta}, J.~A., \& {Livio}, M. 1999, \nat, 401, 891,
  \dodoi{10.1038/44780}

\bibitem[{{Komissarov}(1999)}]{Komissarov99}
{Komissarov}, S.~S. 1999, \mnras, 303, 343,
  \dodoi{10.1046/j.1365-8711.1999.02244.x}

\bibitem[{{Leung} {et~al.}(2011){Leung}, {Gammie}, \& {Noble}}]{Leung11}
{Leung}, P.~K., {Gammie}, C.~F., \& {Noble}, S.~C. 2011, \apj, 737, 21,
  \dodoi{10.1088/0004-637X/737/1/21}

\bibitem[{{Luminet}(1979)}]{Luminet1979a}
{Luminet}, J.-P. 1979, \aap, 75, 228

\bibitem[{{Medeiros} {et~al.}(2020){Medeiros}, {Psaltis}, \&
  {{\"O}zel}}]{Medeiros20}
{Medeiros}, L., {Psaltis}, D., \& {{\"O}zel}, F. 2020, \apj, 896, 7,
  \dodoi{10.3847/1538-4357/ab8bd1}

\bibitem[{{Mertens} {et~al.}(2016){Mertens}, {Lobanov}, {Walker}, \&
  {Hardee}}]{Mertens2016}
{Mertens}, F., {Lobanov}, A.~P., {Walker}, R.~C., \& {Hardee}, P.~E. 2016,
  \aap, 595, A54, \dodoi{10.1051/0004-6361/201628829}

\bibitem[{{Mizuno} {et~al.}(2018){Mizuno}, {Younsi}, {Fromm}, {Porth}, {De
  Laurentis}, {Olivares}, {Falcke}, {Kramer}, \& {Rezzolla}}]{Mizuno_2018}
{Mizuno}, Y., {Younsi}, Z., {Fromm}, C.~M., {et~al.} 2018, Nature Astronomy, 2,
  585, \dodoi{10.1038/s41550-018-0449-5}

\bibitem[{{Mo{\'s}cibrodzka} {et~al.}(2016){Mo{\'s}cibrodzka}, {Falcke}, \&
  {Shiokawa}}]{Moscibrodzka_16}
{Mo{\'s}cibrodzka}, M., {Falcke}, H., \& {Shiokawa}, H. 2016, \aap, 586, A38,
  \dodoi{10.1051/0004-6361/201526630}

\bibitem[{{Mo{\'s}cibrodzka} \& {Gammie}(2018)}]{MoscibrodzkaIPole}
{Mo{\'s}cibrodzka}, M., \& {Gammie}, C.~F. 2018, \mnras, 475, 43,
  \dodoi{10.1093/mnras/stx3162}

\bibitem[{{Narayan} {et~al.}(2003){Narayan}, {Igumenshchev}, \&
  {Abramowicz}}]{NarayanMAD}
{Narayan}, R., {Igumenshchev}, I.~V., \& {Abramowicz}, M.~A. 2003, \pasj, 55,
  L69, \dodoi{10.1093/pasj/55.6.L69}

\bibitem[{{Narayan} {et~al.}(2019){Narayan}, {Johnson}, \&
  {Gammie}}]{Narayan_Shadow}
{Narayan}, R., {Johnson}, M.~D., \& {Gammie}, C.~F. 2019, \apjl, 885, L33,
  \dodoi{10.3847/2041-8213/ab518c}

\bibitem[{{Ohanian}(1987)}]{Ohanian_1987}
{Ohanian}, H.~C. 1987, American Journal of Physics, 55, 428,
  \dodoi{10.1119/1.15126}

\bibitem[{{Olivares} {et~al.}(2020){Olivares}, {Younsi}, {Fromm}, {De
  Laurentis}, {Porth}, {Mizuno}, {Falcke}, {Kramer}, \&
  {Rezzolla}}]{Olivares_2020}
{Olivares}, H., {Younsi}, Z., {Fromm}, C.~M., {et~al.} 2020, \mnras, 497, 521,
  \dodoi{10.1093/mnras/staa1878}

\bibitem[{{Pesce} {et~al.}(2019){Pesce}, {Haworth}, {Melnick}, {Blackburn},
  {Wielgus}, {Johnson}, {Raymond}, {Weintroub}, {Palumbo}, {Doeleman}, \&
  {James}}]{Astro2020Space2}
{Pesce}, D., {Haworth}, K., {Melnick}, G.~J., {et~al.} 2019, in Bulletin of the
  American Astronomical Society, Vol.~51, 176.
\newblock \doarXiv{1909.01408}

\bibitem[{{Porth} {et~al.}(2019){Porth}, {Chatterjee}, {Narayan}, {Gammie},
  {Mizuno}, {Anninos}, {Baker}, {Bugli}, {Chan}, {Davelaar}, {Del Zanna},
  {Etienne}, {Fragile}, {Kelly}, {Liska}, {Markoff}, {McKinney}, {Mishra},
  {Noble}, {Olivares}, {Prather}, {Rezzolla}, {Ryan}, {Stone}, {Tomei},
  {White}, {Younsi}, {Akiyama}, {Alberdi}, {Alef}, {Asada}, {Azulay}, {Baczko},
  {Ball}, {Balokovi{\'c}}, {Barrett}, {Bintley}, {Blackburn}, {Boland},
  {Bouman}, {Bower}, {Bremer}, {Brinkerink}, {Brissenden}, {Britzen},
  {Broderick}, {Broguiere}, {Bronzwaer}, {Byun}, {Carlstrom}, {Chael},
  {Chatterjee}, {Chen}, {Chen}, {Cho}, {Christian}, {Conway}, {Cordes},
  {Geoffrey}, {Crew}, {Cui}, {De Laurentis}, {Deane}, {Dempsey}, {Desvignes},
  {Doeleman}, {Eatough}, {Falcke}, {Fish}, {Fomalont}, {Fraga-Encinas},
  {Freeman}, {Friberg}, {Fromm}, {G{\'o}mez}, {Galison}, {Garc{\'\i}a},
  {Gentaz}, {Georgiev}, {Goddi}, {Gold}, {Gu}, {Gurwell}, {Hada}, {Hecht},
  {Hesper}, {Ho}, {Ho}, {Honma}, {Huang}, {Huang}, {Hughes}, {Ikeda}, {Inoue},
  {Issaoun}, {James}, {Jannuzi}, {Janssen}, {Jeter}, {Jiang}, {Johnson},
  {Jorstad}, {Jung}, {Karami}, {Karuppusamy}, {Kawashima}, {Keating},
  {Kettenis}, {Kim}, {Kim}, {Kim}, {Kino}, {Koay}, {Patrick}, {Koch}, {Koyama},
  {Kramer}, {Kramer}, {Krichbaum}, {Kuo}, {Lauer}, {Lee}, {Li}, {Li},
  {Lindqvist}, {Liu}, {Liuzzo}, {Lo}, {Lobanov}, {Loinard}, {Lonsdale}, {Lu},
  {MacDonald}, {Mao}, {Marrone}, {Marscher}, {Mart{\'\i}-Vidal}, {Matsushita},
  {Matthews}, {Medeiros}, {Menten}, {Mizuno}, {Moran}, {Moriyama},
  {Moscibrodzka}, {M{\"u}ller}, {Nagai}, {Nagar}, {Nakamura}, {Narayanan},
  {Natarajan}, {Neri}, {Ni}, {Noutsos}, {Okino}, {Oyama}, {{\"O}zel},
  {Palumbo}, {Patel}, {Pen}, {Pesce}, {Pi{\'e}tu}, {Plambeck}, {PopStefanija},
  {Preciado-L{\'o}pez}, {Psaltis}, {Pu}, {Ramakrishnan}, {Rao}, {Rawlings},
  {Raymond}, {Ripperda}, {Roelofs}, {Rogers}, {Ros}, {Rose}, {Roshanineshat},
  {Rottmann}, {Roy}, {Ruszczyk}, {Rygl}, {S{\'a}nchez},
  {S{\'a}nchez-Arguelles}, {Sasada}, {Savolainen}, {Schloerb}, {Schuster},
  {Shao}, {Shen}, {Small}, {Sohn}, {SooHoo}, {Tazaki}, {Tiede}, {Tilanus},
  {Titus}, {Toma}, {Torne}, {Trent}, {Trippe}, {Tsuda}, {van Bemmel}, {van
  Langevelde}, {van Rossum}, {Wagner}, {Wardle}, {Weintroub}, {Wex}, {Wharton},
  {Wielgus}, {Wong}, {Wu}, {Young}, {Young}, {Yuan}, {Yuan}, {Zensus}, {Zhao},
  {Zhao}, {Zhu}, \& {Event Horizon Telescope Collaboration}}]{CodeComparison}
{Porth}, O., {Chatterjee}, K., {Narayan}, R., {et~al.} 2019, \apjs, 243, 26,
  \dodoi{10.3847/1538-4365/ab29fd}

\bibitem[{{Psaltis} {et~al.}(2015){Psaltis}, {{\"O}zel}, {Chan}, \&
  {Marrone}}]{Psaltis_2015}
{Psaltis}, D., {{\"O}zel}, F., {Chan}, C.-K., \& {Marrone}, D.~P. 2015, \apj,
  814, 115, \dodoi{10.1088/0004-637X/814/2/115}

\bibitem[{{Raymond} {et~al.}(2021){Raymond}, {Palumbo}, {Paine}, {Blackburn},
  {C{\'o}rdova Rosado}, {Doeleman}, {Farah}, {Johnson}, {Roelofs}, {Tilanus},
  \& {Weintroub}}]{Raymond21}
{Raymond}, A.~W., {Palumbo}, D., {Paine}, S.~N., {et~al.} 2021, \apjs, 253, 5,
  \dodoi{10.3847/1538-3881/abc3c3}

\bibitem[{{Ripperda} {et~al.}(2020){Ripperda}, {Bacchini}, \&
  {Philippov}}]{Ripperda20}
{Ripperda}, B., {Bacchini}, F., \& {Philippov}, A.~A. 2020, \apj, 900, 100,
  \dodoi{10.3847/1538-4357/ababab}

\bibitem[{{Rowan} {et~al.}(2017){Rowan}, {Sironi}, \& {Narayan}}]{Rowan17}
{Rowan}, M., {Sironi}, L., \& {Narayan}, R. 2017, \apj, 850, 29,
  \dodoi{10.3847/1538-4357/aa9380}

\bibitem[{{S\k{a}dowski} {et~al.}(2014){S\k{a}dowski}, {Narayan}, {McKinney},
  \& {Tchekhovskoy}}]{KORAL14}
{S\k{a}dowski}, A., {Narayan}, R., {McKinney}, J.~C., \& {Tchekhovskoy}, A.
  2014, \mnras, 439, 503, \dodoi{10.1093/mnras/stt2479}

\bibitem[{{S\k{a}dowski} {et~al.}(2013){S\k{a}dowski}, {Narayan},
  {Tchekhovskoy}, \& {Zhu}}]{KORAL13}
{S\k{a}dowski}, A., {Narayan}, R., {Tchekhovskoy}, A., \& {Zhu}, Y. 2013,
  \mnras, 429, 3533, \dodoi{10.1093/mnras/sts632}

\bibitem[{{S\k{a}dowski} {et~al.}(2017){S\k{a}dowski}, {Wielgus}, {Narayan},
  {Abarca}, {McKinney}, \& {Chael}}]{KORAL16}
{S\k{a}dowski}, A., {Wielgus}, M., {Narayan}, R., {et~al.} 2017, \mnras, 466,
  705, \dodoi{10.1093/mnras/stw3116}

\bibitem[{{Stawarz} {et~al.}(2006){Stawarz}, {Aharonian}, {Kataoka},
  {Ostrowski}, {Siemiginowska}, \& {Sikora}}]{Stawarz06}
{Stawarz}, L., {Aharonian}, F., {Kataoka}, J., {et~al.} 2006, \mnras, 370, 981,
  \dodoi{10.1111/j.1365-2966.2006.10525.x}

\bibitem[{{Takahashi}(2004)}]{Takahashi}
{Takahashi}, R. 2004, \apj, 611, 996, \dodoi{10.1086/422403}

\bibitem[{{Tchekhovskoy} {et~al.}(2011){Tchekhovskoy}, {Narayan}, \&
  {McKinney}}]{Tchekhovskoy11}
{Tchekhovskoy}, A., {Narayan}, R., \& {McKinney}, J.~C. 2011, \mnras, 418, L79,
  \dodoi{10.1111/j.1745-3933.2011.01147.x}

\bibitem[{{Teo}(2003)}]{Teo_2003}
{Teo}, E. 2003, General Relativity and Gravitation, 35, 1909,
  \dodoi{10.1023/A:1026286607562}

\bibitem[{{The Event Horizon Telescope Collaboration}
  {et~al.}(2019{\natexlab{a}})}]{PaperI}
{The Event Horizon Telescope Collaboration}, {et~al.} 2019{\natexlab{a}},
  \apjl, 875, L1, \dodoi{10.3847/2041-8213/ab0ec7}

\bibitem[{{The Event Horizon Telescope Collaboration}
  {et~al.}(2019{\natexlab{b}})}]{PaperII}
---. 2019{\natexlab{b}}, \apjl, 875, L2, \dodoi{10.3847/2041-8213/ab0c96}

\bibitem[{{The Event Horizon Telescope Collaboration}
  {et~al.}(2019{\natexlab{c}})}]{PaperIII}
---. 2019{\natexlab{c}}, \apjl, 875, L2, \dodoi{10.3847/2041-8213/ab0c96}

\bibitem[{{The Event Horizon Telescope Collaboration}
  {et~al.}(2019{\natexlab{d}})}]{PaperIV}
---. 2019{\natexlab{d}}, \apjl, 875, L4, \dodoi{10.3847/2041-8213/ab0e85}

\bibitem[{{The Event Horizon Telescope Collaboration}
  {et~al.}(2019{\natexlab{e}})}]{PaperV}
---. 2019{\natexlab{e}}, \apjl, 875, L5, \dodoi{10.3847/2041-8213/ab0f43}

\bibitem[{{The Event Horizon Telescope Collaboration}
  {et~al.}(2019{\natexlab{f}})}]{PaperVI}
---. 2019{\natexlab{f}}, \apjl, 875, L6, \dodoi{10.3847/2041-8213/ab1141}

\bibitem[{{The Event Horizon Telescope Collaboration}
  {et~al.}(2021{\natexlab{a}})}]{PaperVII}
---. 2021{\natexlab{a}}, \apjl, 910, L12, \dodoi{10.3847/2041-8213/abe71d}

\bibitem[{{The Event Horizon Telescope Collaboration}
  {et~al.}(2021{\natexlab{b}})}]{PaperVIII}
---. 2021{\natexlab{b}}, \apjl, 910, L13, \dodoi{10.3847/2041-8213/abe4de}

\bibitem[{{Vincent} {et~al.}(2016){Vincent}, {Meliani}, {Grandcl{\'e}ment},
  {Gourgoulhon}, \& {Straub}}]{Vincent_2016}
{Vincent}, F.~H., {Meliani}, Z., {Grandcl{\'e}ment}, P., {Gourgoulhon}, E., \&
  {Straub}, O. 2016, Classical and Quantum Gravity, 33, 105015,
  \dodoi{10.1088/0264-9381/33/10/105015}

\bibitem[{{Walker} {et~al.}(2018){Walker}, {Hardee}, {Davies}, {Ly}, \&
  {Junor}}]{Walker18}
{Walker}, R.~C., {Hardee}, P.~E., {Davies}, F.~B., {Ly}, C., \& {Junor}, W.
  2018, \apj, 855, 128, \dodoi{10.3847/1538-4357/aaafcc}

\bibitem[{{White} {et~al.}(2019){White}, {Quataert}, \& {Blaes}}]{White19}
{White}, C.~J., {Quataert}, E., \& {Blaes}, O. 2019, \apj, 878, 51,
  \dodoi{10.3847/1538-4357/ab089e}

\end{thebibliography}
\bibliographystyle{aasjournal}

\end{document}